\documentclass[12pt]{article}
\usepackage[english]{babel}

\usepackage[margin=1in]{geometry}
\usepackage{setspace}
\usepackage{amsmath}
\usepackage{rotating, graphicx}
\usepackage[colorlinks=true, allcolors=blue]{hyperref}
\usepackage{amssymb}
\usepackage{amsmath}
\usepackage{amsfonts}
\usepackage{threeparttable}
\usepackage{booktabs, makecell, longtable}
\usepackage{threeparttablex}
\usepackage{dirtytalk}
\usepackage{bbm}
\usepackage{siunitx}
\usepackage{xcolor}
\usepackage{subcaption}
\usepackage{chngcntr}
\usepackage[authoryear]{natbib}
\usepackage[super]{nth}
\usepackage{afterpage}
\usepackage{epigraph}
\usepackage{authblk}
\usepackage{caption}
\usepackage{afterpage}
\usepackage{datetime}
\usepackage{subcaption}
\usepackage{makecell}
\usepackage{chapterbib}
\usepackage{multirow}
\usepackage[page,toc,titletoc,title]{appendix}
\usepackage[section]{placeins}
\usepackage{titletoc}
\usepackage{minitoc}
\usepackage{rotating}
\usepackage{tabularx}
\usepackage{array}
\usepackage{adjustbox}
\usepackage{verbatim}

\providecommand{\keywords}[1]
{
  \small	
  \textbf{Keywords:} #1
}

\title{Sacred Ecology}
\author[1]{Neha Deopa\thanks{Corresponding author: n.deopa@exeter.ac.uk. \\ University of Exeter, Department of Economics, Rennes Drive, EX4 4PU Exeter, United Kingdom}} \author[1]{Daniele Rinaldo\thanks{d.rinaldo@exeter.ac.uk}}
\affil[1]{University of Exeter and Land, Environment, Economics and Policy Institute (LEEP)} 
\date{}
\begin{document}

\maketitle  \pagenumbering{gobble}


\begin{abstract}
\noindent \small 
Can religious beliefs sustain ecosystems? We study the case of Benin, the cradle of Vodun, where African Traditional Religions (ATR) place forests within a sacred sphere and where adherence declined sharply during the expansion of Charismatic Pentecostalism.  
Using census data between 1992 and 2013, we leverage exogenous variation in proximity to the precolonial Adja migration corridor, where Vodun institutions were historically concentrated, 
together with changes in the national expansion of Pentecostal broadcasting.
We find that declining ATR adherence reduces forest--savanna cover and increases subsequent forest loss.
We find that ATR-adherent households are less likely than non-adherent ones to clear surrounding forest resources as a response to idiosyncratic shocks and livelihood losses.
We embed this evidence in a model of forest use with adherence-dependent environmental preferences and show how the decline of ATR beliefs can generate substantial and potentially unsustainable losses in tree cover. 
More broadly, our findings show that traditional beliefs can function as informal conservation institutions, and that their erosion can have persistent environmental consequences.
\\
\\
\begin{jel}
\footnotesize Z12, Z13, Q5, O13
\end{jel}
\\[5pt] 
\noindent\keywords{African Traditional Religions, Deforestation, Environmental Preferences, Benin}

\end{abstract}

\newpage
\newgeometry{margin=0.8in} 
\onehalfspacing
\pagenumbering{arabic}
\section{Introduction}
\setlength\epigraphwidth{\textwidth}
\setlength\epigraphrule{0pt}
\epigraph{ 
\textit{Open your eyes, stranger, and know where to step.
Here, a tree is not a tree, a spring is not a spring.
Everything is mysterious and mystical.}}{-- Eustache Prudencio, \itshape Vents du Lac}

Religion is a a powerful means by which societies shape worldviews, knowledge systems and ultimately human behaviour. The influence religious beliefs can have on individual attitudes such as prosocial behaviour, interpersonal relationships and economic outcomes is well established \citep{guiso2006does,iannaccone2009economics, iyer2016new}. What is less known, however, is how religion can shape human interactions with nature. The ecological and anthropological literatures provide several examples of systems of indigenous people and local communities whose ecological knowledge is embedded within spiritual or religious beliefs and where nature is imbued with sacredness \citep{taylor2016lynn, berkes2017sacred}. This phenomenon is especially conspicuous on the African continent where sacredness of common pool resources is prevalent among communities that adhere to African Traditional Religions (ATR) 
\citep{sheridan2008african}. 

ATR are religions that are indigenous to Africa and are based on the fundamental belief that the sacred and the everyday are inseparable, making religious life indistinguishable from daily experience \citep{amanze2024african}.
An important feature of the ATR worldview is that divinities are the personification of natural phenomena and spirits are known to inhabit rivers, streams, forests, mountains, and other natural occurrences on the African landscape. 
Forests, distinctly, have defined the physical and spiritual environment of ATR communities.
Sacred forests and groves feature in all aspects of ATR culture: history, art, worship, medicine, and social structure 
\citep{food1990major, aderibigbe2022palgrave}. 
Despite the significant role belief systems play in shaping human behaviour, there is limited evidence and understanding of how  religious beliefs drive human-ecosystem interactions, particularly for non-Abrahamic religions such as ATR. 

This knowledge gap has important implications for global conservation policy and decision-making as there is a burgeoning interest in the conservation discourse to reformulate the people–nature relationship by recognizing the heterogeneous values of traditional belief systems \citep{ives2024role}. 
Current environmental governance overwhelmingly ignores the cultural and spiritual ways in which people around the world understand, value, and relate to nature. According to \cite{ipbes2019intergovernmental}, this narrowness underpins the current biodiversity crisis. Religion, therefore, provides a relevant lens to understand the socio-ecological process that shapes nature values at both individual and collective levels. Leveraging these values will therefore allow global policy to achieve sustainable and more equitable conservation outcomes.

Our study explores how adherence to ATR, which embeds forests within a sacred cosmology, affects forest-related environmental outcomes.\footnote{Although ATR refers to African Traditional Religions, we use the acronym in the singular throughout.} This question is especially salient given the long-run decline of ATR across sub-Saharan Africa, alongside the considerable expansion of Christianity and Islam \citep{lugo2010tolerance}. 
Within this context, we are interested in understanding what happens to a landscape when the belief systems that sacralised it recede. 
We examine this question by means of an instrumental variable approach, an exploration of heterogeneity and channels, and a theoretical framework that formalises ATR-driven pro-environmental attitudes, fit to data. 

Studying traditional religions typically poses an empirical challenge because ATR adherence is widely under-reported in national surveys, reflecting the long-run legacy of colonial missionary efforts that stigmatized these belief systems \citep{claffey2007christian, stoop2019voodoo}. Self-reported affiliation may therefore be a poor proxy for actual beliefs and practices and often exhibits limited variation across communities. The Republic of Benin provides a particularly informative setting in which these concerns are substantially reduced. As the historic home of the Kingdom of Dahomey and the cradle of Vodun, the country’s most prominent ATR, Benin formally recognises traditional religions in its constitution, grants them a national holiday, and accords them the same legal status as Christianity and Islam. ATR adherence is therefore less stigmatised and more freely reported, generating substantial local variation consistent with Benin’s exceptionally high degree of religious diversity \citep{pewresearchcenter_2014}. Consistent with the broader continental trend, ATR adherence in Benin also declined sharply between 1992 and 2013, with most of the decline absorbed by Christian denominations rather than Islam or non-religion.
Benin is also well suited to studying environmental consequences because land use change remains strongly shaped by local household and community decisions, as in much of sub-Saharan Africa. Forests in Benin continue to serve important social, cultural, and religious functions \citep{world2020benin}, and the country contains an estimated 3,000 sacred forests and groves used for traditional rituals. These features make Benin a particularly informative setting for testing whether ATR beliefs impact forest outcomes.

We use the latest three waves of the population census from 1992, 2002, and 2013, providing us with arrondissement-level (administrative division 3) information on socio-economic and demographic characteristics, including religion. The resulting dataset is a panel of 546 arrondissements observed over a period of 3 census years. We define ATR adherence as the percentage of individuals living in an arrondissement that self-report as following ATR. 
We identify the impact of ATR on two remotely sensed environmental outcomes: the share of arrondissement area under forest--savanna cover in the European Space Agency's annual land-cover maps, and \citet{hansen2013high} stand-replacing forest loss in the year following each census. 

Our data allow the econometric specification to incorporate a comprehensive set of environmental and socio-economic covariates as well as a rich set of fixed effects. We include arrondissement fixed effects, isolating the effect of ATR adherence from any time-invariant characteristics potentially associated with forests, such as proximity to rivers and baseline ecological endowments. Crucially, we include ethnic homeland$\times$year fixed effects, where homelands are the \citet{murdock1967ethnographic} ethnic polygons, so that identification comes from comparing arrondissements within the same homeland in the same census year. These fixed effects absorb nonparametrically any time-varying legacy shared at the ethnic group, specifically the slave-trade exposure that \citet{nunn2011slave} define at exactly this level, cultural norms and social structure as well as any group-specific trend in conversion or in land use. 
Finally, we allow the arrondissement's latitude and longitude and its soil suitability for maize, cassava and cotton to have census-year-specific effects, so that smooth spatial trends and suitability-driven commodity cycles do not confound as religious change.

The OLS results support our hypothesis that ATR adherence sustains forest outcomes. A fundamental concern is endogeneity, as the decline of ATR is not random. Conversion may be selective as charismatic churches may recruit fastest where markets are integrating and forests are already being cleared, or, conversely, ATR adherents may resist longest in places whose forests were never under pressure. OLS estimates, therefore, would conflate the effect of religious change with the process that produced it. 
We address these issues using a novel instrument that combines a sixteenth-century sacred geography with a twentieth-century shock to the religious market.

The spatial component of the instrument exploits the sacred geography created by the migrations that founded the Kingdom of Dahomey. Historical accounts trace the Adja royal lineage from Tado through Allada to Abomey, a corridor that became closely associated with the political and religious institutions of Vodun and was established by the mid-seventeenth century, several centuries before our study period \citep{cornevin1981republique, law1991slave, bay2012wives}. We measure each arrondissement’s exposure to this geography relative to the contemporaneous migration corridors of the Yoruba, whose religious trajectory subsequently diverged sharply from that of the Adja-Dahomey heartland. Christianity reached Yoruba areas early in the nineteenth century and was well established there by the early twentieth century, whereas missions remained largely confined to the coast in Dahomey and Vodun institutions persisted in its interior. 
The two corridors therefore trace sacred geographies of similar historical depth but with markedly different persistence of traditional religion. 
Specifically, relative Adja proximity (REP) is the log distance to the nearest Yoruba corridor minus the log distance to the Adja corridor, so that higher values indicate greater historical exposure to the Adja--Vodun heartland.
The temporal component exploits the liberalisation of Benin’s broadcasting sector in 1997, which gave Charismatic Pentecostal churches a national platform for the first time. We therefore interact REP with the national number of licensed Pentecostal radio stations at each census wave, which rose from zero in 1992 to one in 2002 and three in 2013 \citep{gratz2011paroles}. 
The interaction captures differential exposure to the national Pentecostal expansion. 

The intuition behind the instrument is the following: Charismatic churches across Africa portray traditional religion as a source of evil forces and direct their proselytising at ATR strongholds. 
When the national religious market opened, the places with the most to lose were those along the corridor with deeper and more persistent ATR institutions, where adherence then fell fastest. The first stage confirms this, and within ethnic homeland and conditional on controls, each licensed Pentecostal station lowered the ATR share by 1.3 percentage points more per unit of relative proximity to the Adja corridor, with an effective $F$-statistic above 60. Reassuringly, the corridor's outflow went to Christian denominations and not to Islam, which is what a conversion shock, rather than migration or secularisation, should look like.

Our 2SLS estimates imply that the loss of ATR caused substantial increases in deforestation. 
Each percentage point of adherence lost reduced forest--savanna cover by 0.185 percentage points of arrondissement area and raised forest loss in the following year by 0.068 percentage points. Relative to the sample means, the cover effect is small per point of adherence but large in aggregate: scaled by the 21-point national decline in adherence between 1992 and 2013, it implies about 3.9 percentage points less cover, or one-seventh of the mean, in an arrondissement experiencing the average decline. The loss effect is large in proportional terms because annual clearing is rare, averaging 0.4 percent of land: each point of adherence lost raises it by 17 percent. Both estimates are stable across fixed-effect structures and hold in the southern sub-sample where the two corridors actually compete for religious space.

We address threats to identification through a broad set of robustness checks. Using annual land cover maps, we find forest--savanna cover along the corridor shows no differential trajectory between 1992 and 1996, before the first Pentecostal station was licensed, and diverges only once broadcasting begins, intensifying after the 2003 station entries. 
The results are not driven by the particular 0/1/3 station count and alternative national shifters, including the national Pentecostal share, national ATR outflow, and a linear trend, yield similar first-stage and 2SLS estimates, while unrestricted year coefficients closely reproduce the timing of broadcaster expansion. 
Historical population and cropland reconstructions show that REP did not predict settlement prior to the migrations, suggesting that it is not proxying for underlying agricultural suitability or accessibility. Following \citet{borusyak2023nonrandom}, we also recenter the instrument using 500 counterfactual Tado-anchored routes, with similar results. 
The estimates are robust to controlling for distance to precolonial state capitals, accounting for land-formalisation programme, migration-driven demographic change, and using alternative clustering and Conley standard errors. 
Finally, estimates are stable across alternative instrument constructions, including different distance floors, linear distances, and migration waypoint-based and absolute proximity measures.

Given this evidence, we then explore the mechanisms through which declining ATR adherence translates into forest loss. The aggregate effect does not appear to operate through other land pressures correlated with religious change. Interactions of instrumented adherence with baseline ethnic fractionalization, drought exposure, poverty, and non-literacy are small and statistically insignificant, while the main effect remains stable. 
Household evidence instead points towards individual behaviour. 
Using the 2018 and 2021 Enquête Harmonisée sur le Conditions de Vie des Ménages (EHCVM) surveys matched to surrounding forest outcomes, we investigate responses to recent household shocks and related losses. 
We find that 
ATR-headed households clear less forest following idiosyncratic shocks and retain more forest cover, compared to non-ATR households. 
This pattern is similar across agricultural, income, and livestock losses, but does not appear for climate shocks, which increase clearing broadly across all households. 
We observe the opposite result for Christian households.
The results are not explained by difference in exposure or resources as within survey cluster, ATR-headed households report the same number of shocks as their neighbours but more losses from them, have lower salaried income and own fewer of the plots they farm, all of which would predict more clearing, not less.
Reported coping strategies provide little evidence of substitution into other margins, ATR-headed households are no more likely to draw on savings, sell assets, receive transfers, or intensify religious activity following shocks. Instead, they appear to absorb idiosyncratic losses without liquidating either physical or natural capital. The household evidence therefore identifies ATR as a state-contingent restraint on forest use as adherents hold back from drawing on the forest especially when need is greatest. 
We formalise this behavioural response in a theoretical framework.

What the household evidence lacks is disentangling what is it that adherents value when they hold back from forest use. ATR may sacralise an adherent's own environment, a grove or a tree where spirits reside, so that restraint protects a local stock; or it may carry a concern for the forest at large, so that welfare depends on how much forest is left standing across the landscape. 
The two readings have different aggregate implications, and we build a model in which both can coexist.
A population of individuals extract from a stochastically growing forest; adherence lowers the utility of consuming one's own forest i.e. the \emph{sacred} channel, and assigns positive utility to the forest stock across the population, the \emph{ecology} channel. 
Since the spatial distribution of the forest stock depends on everyone's deforestation activity, the equilibrium is a mean-field game. It delivers three results. Extraction falls with individual adherence, as in the reduced form. When adherence declines in the population, aggregate clearing rises because there are fewer restrained individuals, while those who remain adherent restrain further in response to the thinning forest they observe, which reconciles the arrondissement-level and household-level evidence. Finally, only the ecology channel can shift a population's deforestation activity from unsustainable to sustainable: the sacred channel alone can reduce individual clearing rates, but never below the forest's natural growth rate.

We estimate the model on Benin, recovering the adherence distribution from the census, the forest's natural growth rate from the protected W--Arly--Pendjari complex, and the preference structure from the relationship between adherence and canopy growth across arrondissements. 
The fitted model reproduces the observed 2013 canopy distribution. 
Setting adherence to zero everywhere lowers tree canopy by about 10 percent and pushes national deforestation above the forest's growth rate, so that without ATR the resource is depleted in the long run.
At the observed spatial forest distribution, only the high-adherence departments along the Adja corridor and in the Atakora extract sustainably. We read the decline of ATR, then, not as a change in who clears the forest but as the erosion of a preference that kept clearing within ecological limits.

Our results contribute to a burgeoning literature in economics that focuses on African Traditional Religions \citep{alonso2016voodoo, alidou2019only, stoop2019voodoo, alidou2021beliefs, le2022social, butinda2023importance, bonikowski2026nature} and refines our understanding of the impact of religions beyond the Abrahamic faiths and within the developing world context \citep{iyer2016new, lowes2026religion}.
There is also a parallel growing literature on the social impact of witchcraft and supernatural beliefs \citep{gershman2016witchcraft,nunn2017being}. 
Such beliefs are not mutually exclusive to traditional religions in Africa, and can cut across social status, education, gender and ethnic and religious affiliations \citep{falen2018african}. Our paper also broadly contributes to the larger work on the economics of religion, both empirical and theoretical 
\citep{mccleary2006religion,becker2009weber,carvalho2019advances,alesina2023religion,ciscato2024astrology,montero2025price}, 
the role played by culture in public goods provision \citep{owen2007culture,alesina1999public,alesina2019public} and the political economy of deforestation \citep{pfaff1999drives, burgess2012political}. 
Our research pushes this literature towards the environment, which the economics of religion has not studied systematically either empirically or theoretically. To our knowledge, we provide the first causal evidence that the decline of a religious belief system changes land use, and the first model in which religious preferences over a shared ecosystem are formalised and estimated.

Our findings have substantive real-world policy implications. First, the conversion of ATR adherents, which continues across sub-Saharan Africa, has an ecological cost that is absent from current accounts of religious change. In the case we focus on, the Republic of Benin, this decrease has meant measurably less standing cover and faster clearing. Second, because the mechanism acts via individual preferences rather than group-wide characteristics, the loss is not one that stronger community institutions or formal monitoring can readily replace. 
A conservation policy that treats sacred landscapes only as a governance arrangement would omit a core dimension that sustains them. 
Third, the restraint is strongest when households are hit by shocks, which is precisely when poverty-driven clearing accelerates. Where adherence has receded, policies that protect against idiosyncratic losses, such as insurance or transfers, become a direct instrument of forest conservation. 
More broadly, our results support the view that environmental valuation which ignores spiritual and cultural dimensions to nature will systematically understate what is lost when these aspects disappear.

The paper proceeds as follows. Section \ref{back+data} provides background on African Traditional Religions, and the historical and contemporary overview of Benin. Section \ref{data} introduces the data. Section \ref{empirics} presents our empirical strategy, our results and discusses threats to identification. Section \ref{sox} discusses heterogeneity and shock responses.
Section \ref{theory} presents the model, the structural estimates and counterfactuals. 

\section{Background}\label{back+data}

\subsection{African Traditional Religions \& Forests}\label{back: atr}

African Traditional Religions refer to the indigenous religions of the African continent and encapsulate a significant belief system, thought patterns, and ritual practices reflective of the culture and geography unique to Africa. 
The term ``African Traditional Religions'' has been used as a nomenclature by scholars such as \cite{idowu1973african} and \cite{mbiti1970concepts,mbiti1990african} to distinguish it from other \emph{imported} religions practiced in Africa today such as Christianity and Islam. 
Although ATR varies widely from region to region and does not have sacred scriptures, and is without an all-encompassing founder or central historical figure, there are several common features that unify the traditional religions on the continent under the term ATR \citep{aderibigbe2022palgrave}. 

One such characteristic is the belief and veneration of divinities with expression in natural phenomena. These divinities are usually spirits that are associated with natural forces like rivers, lakes, trees, mountains and forests, which become central to communal rituals and are regarded as sacred. This is especially true for West Africa where nature's sacred element has led to large pantheons of spirits and divinities. For example, in the coastal regions of West Africa, the water deity Mami Wata (Mother of Water) is highly revered and celebrated. 
Of particular interest within the ATR cosmology are trees and forests: from the Congo basin to the forests of Western Africa, they are thought of as a sacred place which is revered in the traditional belief systems as they are intimately linked with ancestry and cultural heritage. The value of forests go beyond utilitarian purposes and they are often seen as a natural boundary between the living and the spirit world. Forest trees can represent the links between the sky and earth, often symbolising links between the spiritual world of ancestors and people. For example, the \emph{Iroko} tree (Milicia excelsa), also known as African teak, holds a significant place in West African culture and for Vodun adherents is one of the most important trees in Dahomey’s history with significant natural and traditional wealth. The deciduous characteristic of the trees give it an ambiguous image which reflects the tree’s power to give life and rebirth as well as to bring about death. In many African myths and stories, the tree is portrayed as an ancestral symbol of wisdom, authority and custom, providing a bond between the dead and the living  \citep{juhe2010forets}. 

The inseparable link between religion and nature pervades traditional African religious life, and ATR worldview often advocates for preserving the harmony between humans and nature. 
This insight alludes to the possibility that ATR cosmology can represent a blueprint for ecological principles and adaptation that endeavours to maintain an equilibrium between natural resources and consumption, and is in line with studies in ecology and anthropology that demonstrate this result \citep{reichel1976cosmology, taylor2016lynn}. Considering the salient nature of forests and trees in ATR beliefs, it is of interest to further explore this relationship.
    


\subsection{Benin: The Cradle of Vodun}\label{back:vodun}

Present-day Benin is an amalgam of several distinct historical entities whose legacies continue to shape the country’s social and political geography \citep{claffey2007christian}. In the north-east, the former Bariba kingdom of Borgou lay along the trade routes linking the Sahel and included a substantial Fulani presence, while the north-west was inhabited by several smaller acephalous groups, including the Otamari, that did not form centralised states. The south-east is predominantly Yoruba-speaking and historically oriented toward Porto-Novo and, more broadly, the Yoruba polities of present-day Nigeria. The south-west, which is primarily inhabited by Adja and Fon people, is the heartland of the hegemonic Kingdom of Dahomey, founded at Abomey in the seventeenth century by the Adja royal lineage of Allada. 

These historical formations remained closely associated with particular ethnic groups and also structured the geography of religious change. Islam spread primarily through the northern trade routes, Christianity reached the Yoruba speaking groups in south-east relatively early through connections with Nigeria, while Vodun became deeply institutionalised in the Adja heartland in south-west. Dahomey's expansion and participation in the Atlantic slave trade contributed to the incorporation of religious practices from conquered populations, while the royal ancestral traditions embedded Vodun within the institutions of the state. Religious change in Benin has therefore never been a national process but one that runs along the boundaries of historical polities and the ethnic groups they contained \citep{manning2004slavery}, a structure that Section~\ref{iv} exploits directly and that our homeland$\times$year fixed effects absorb.

Post 1872, under French colonial rule and the Marxist-Leninist regime of Mathieu Kérékou (1972--1991), traditional religions were marginalised and in 1976 anti-witchcraft laws exposed ATR adherents to persecution, leading to a sharp decline in reported practice \citep{kahn2011policing}. The democratic transition reversed this move and the 1990 constitution recognised ATR alongside Christianity and Islam. Concurrently, the new leadership promoted Vodun as part of a Béninois national identity, and the ``Ouidah~'92'' festival, still celebrated today, inaugurated a public revival \citep{tall1995democratie}. 
This official standing is why self-reported ATR adherence in Benin, unlike in most of the region, can be taken as a credible measure of belief. The persistence of ATR is also visible in the landscapes through which these beliefs are practiced. Forests in Benin today serve as a place for social, cultural and religious activities. For example, in For\^et Sacr\'ee de Kpass\`e (Sacred Forest of Kpass\`e) in Ouidah, adherents care for the forest understanding that they are caring for their ancestors, whose spirits are believed to be reincarnated in Loko (Iroko) trees. The government estimates some 3,000 sacred forests and groves covering 0.16 percent of the national territory. These sites persist under heavy clearing pressure from smallholder and shifting agriculture.

\section{Data}\label{data}

\subsection{ATR Adherence}
\begin{figure}
    \centering
    \includegraphics[scale=0.7]{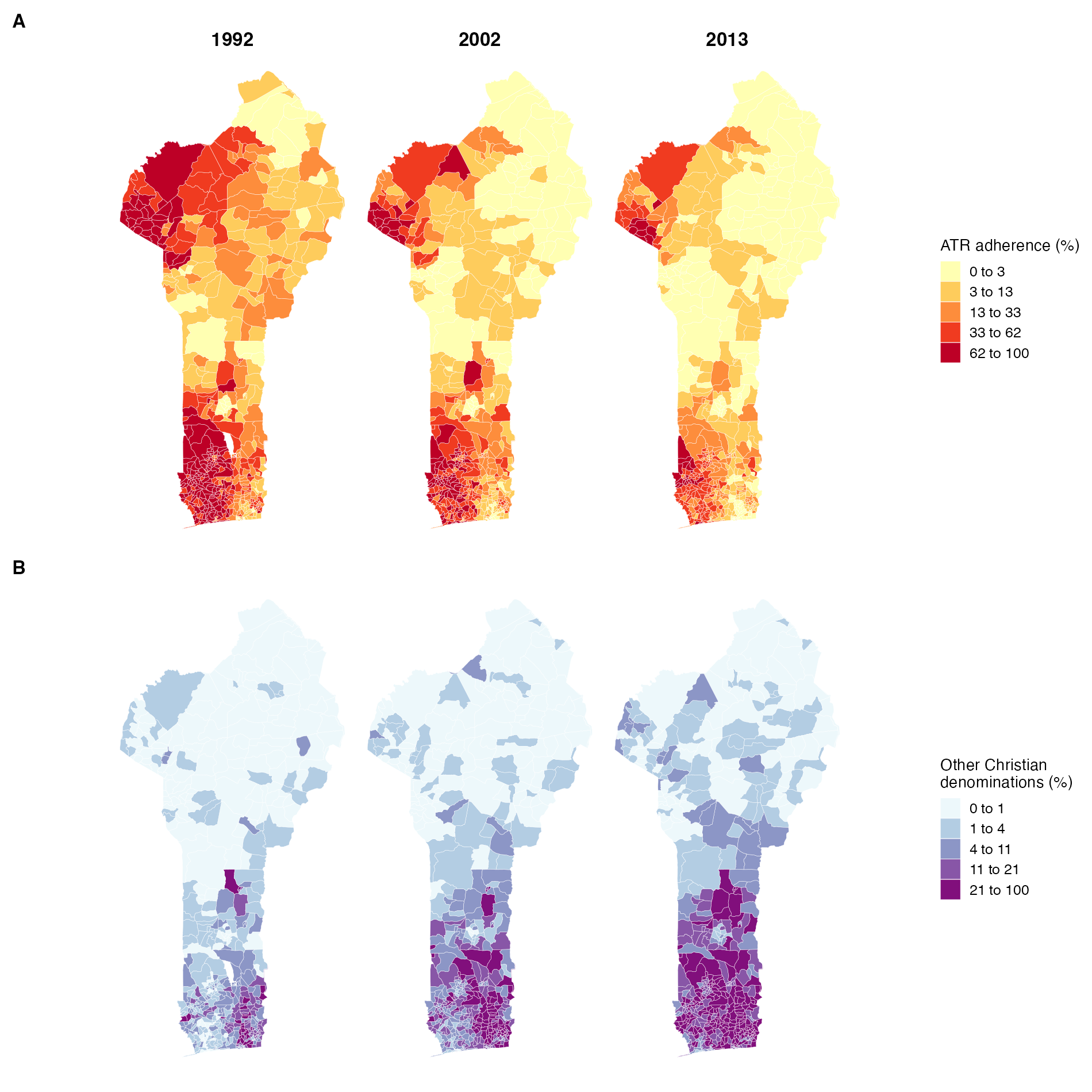}
    \caption{Evolution of  (A) ATR adherence and (B) other Christianity across arrondissements over census waves.}
    \label{fig:atr_other_evolution}
\end{figure}

\begin{figure}
    \centering
\includegraphics[scale=0.9]{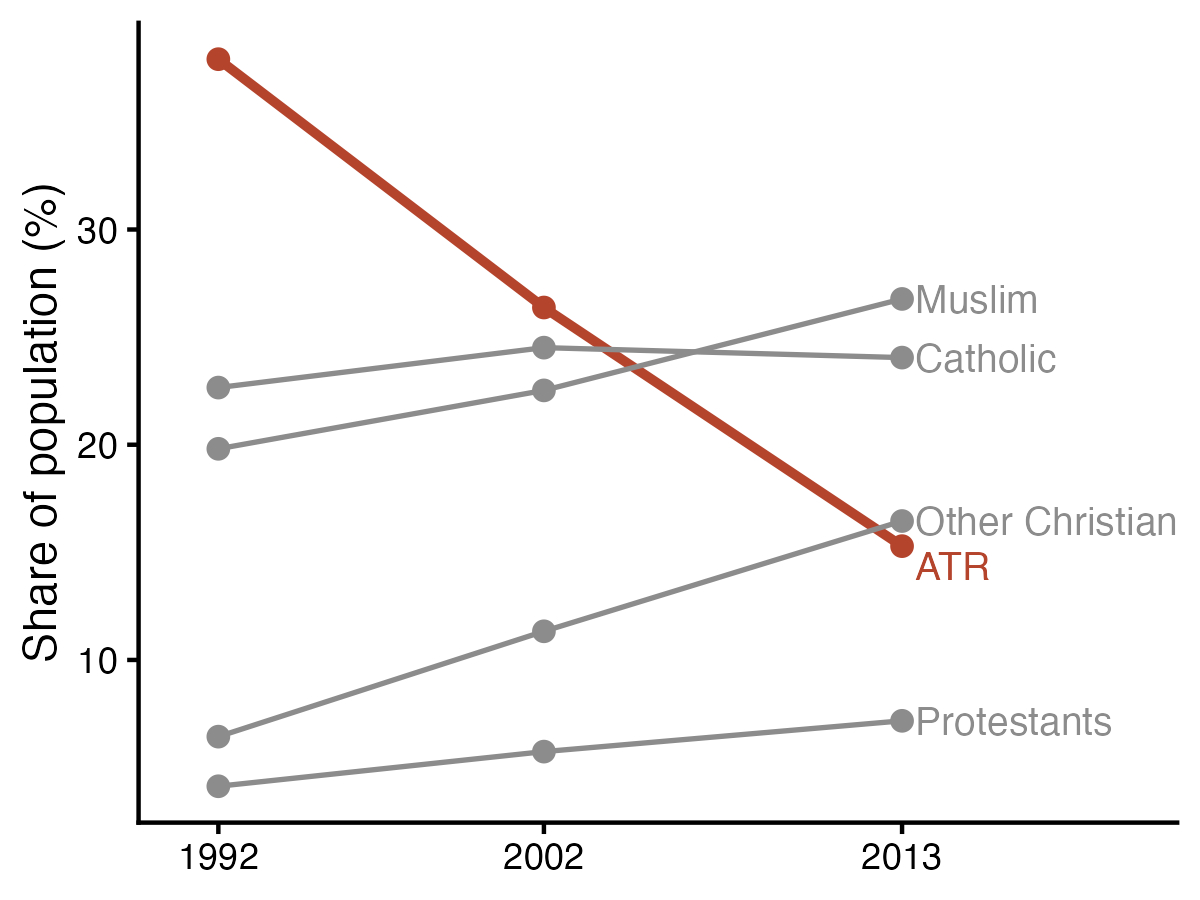}
 \caption{Share of adherents of major religions across the three censuses. }
\label{fig: census_summ}
    \end{figure}

We leverage data from the three latest waves of the Republic of Benin population census conducted in 1992, 2002 and 2013.\footnote{The individual and household microdata consists of a random sample of 10\% of person and household records from the censuses, allowing for linkage between individuals in the same family and the same household.} The census records individual and household socio-demographic characteristics, including religious affiliation. Benin is divided into 12 departments, 77 communes, 546
arrondissements and 5,295 villages and urban quarters; our unit of analysis is
the arrondissement, the third administrative tier. We measure ATR adherence as the percentage or share of individuals living in an arrondissement \emph{i} in census year \emph{t} that self-report as following ATR.\footnote{Whilst the 1992 census has a single category representing Traditional religions, the remaining two waves make a distinction between Vodun and \say{other Traditional religions}.We combine the two
into a single ATR category so that the measure is comparable across waves.} Figure~\ref{fig:atr_other_evolution} maps the two religious categories that drive
the change over our sample period. Panel A shows ATR adherence by
arrondissement in each census year. In 1992, high-adherence arrondissements
form two distinct blocs: a southern one that overlaps the historical boundaries
of the Dahomey kingdom, and a north-western one in the Atakora mountain range, home to
the Otamari people. The two blocs have different origins. The
southern bloc is the Vodun heartland of the Adja and Fon people, tied to the royalty of  
of Allada and Abomey. The Otamari are an acephalous population whose
traditional religion centres on earth and ancestor veneration, and whose settlement behind the mountain range reflects a long history of resisting incorporation by the
neighbouring Bariba kingdoms, Dahomean states and by colonial administration
\citep{cornevin1981republique}. By 2013 both blocs have
thinned, but the retreat is far steeper in the south. Panel B shows the mirror
image: the rise of \say{other Christian} denominations which is predominantly
Charismatic and Pentecostal churches that emphasize a \say{born again}
identity, and is concentrated almost entirely in the southern arrondissements
where ATR was losing ground, and remains marginal in the north. Figure~\ref{fig: census_summ} reports the full national composition by
religion over the three waves. Despite these regional patterns, every major religion is practiced throughout
the country, and Benin ranks among the most religiously diverse countries in
the world \citep{pewresearchcenter_2014}. 

\subsection{Environmental Data} 

We focus on two environmental outcomes, forest cover
and forest loss, drawn from two separate remote-sensing products. In 2009 the European Space Agency (ESA) launched its Climate Change Initiative
(CCI) to provide high-quality satellite-derived records of Essential Climate
Variables, including global land cover (LC). The ESA-CCI-LC dataset provides
consistent annual global LC maps at 300m resolution from 1992 to 2015. The
data are categorical: each pixel is assigned to one of 37 LC types under the
FAO Land Cover Classification System, with high overall global accuracy \citep{defourny2009accuracy}. 
Supplemental Appendix Table~\ref{lctype} lists the classes relevant to our study, as defined in \cite{esa2017land}.
Benin lies within the Dahomey Gap, the dry corridor of Guinean
forest-savanna mosaic that interrupts the West African rainforest belt between
Ghana and Nigeria \citep{natta2003riparian}. Outside a few remaining patches of semi-deciduous forest in the south, the country's woody vegetation consists of dry deciduous forest, woodland and tree savanna that merge into one another along a continuous gradient of canopy density rather than forming separate groups \citep{world2020benin} 
Since Benin's forests are mainly represented as a savanna-mosaic ecosystem, our primary outcome combines tree cover with shrublands and we define \emph{forest-savanna cover} as the union of the ESA forest classes (broadleaved and needleleaved, evergreen and deciduous,
mixed, flooded, and mosaic tree-and-shrub with tree cover above 50\%
) and deciduous shrubland. 
For each annual map we compute the area of every pixel and aggregate to arrondissements, such that the forest-savanna share of arrondissement $i$
in year $t$ is the summed area of pixels in these classes divided by the
arrondissement's total pixel area, expressed in percent. 

For our second outcome, forest loss, we use the Global Forest Change (GFC) product of \cite{hansen2013high} which maps annual gross tree-cover loss at
30~m resolution from 2001 to 2024 using Landsat imagery. Every pixel is
assigned a canopy-cover percentage in the year 2000 and, the first year in which a clearing or complete canopy removal is subsequently detected. 
Unlike the ESA-CCI-LC classification, which is a stock that reflects the accumulated history of clearing and regrowth up to the census, the GFC data is a flow that indicates where and when tree cover is removed.\footnote{Technical documentation is available at \url{https://glad.earthengine.app/view/global-forest-change}.}
We define the baseline forest cover as the pixels with at least 10\% tree canopy cover in 2000, a standard threshold for countries with forest-savanna ecosystems \citep{mermoz2018impacts}.
Forest loss is then computed as the area of baseline forest pixels in arrondissement $i$ flagged as lost in year $t+1$, divided by the arrondissement's total area and expressed in percent. We use the one-year lead rather than the contemporaneous year for two reasons. 
First, the one-year lead allows us to see how ATR adherence observed at the census predicts the forest clearing that follows it. Second, census enumeration takes place during the year, so 
using the lead ensures that forest loss is measured strictly afterwards and removes any mechanical overlap.
Because the GFC record begins in 2001, loss is observed for the 2002 and 2013 census waves only, so the loss regressions use a two-wave panel. 

\subsection{Time varying controls, heterogeneity and shocks}\label{data_sox}

The census provides us with arrondissement level socio-economic variables which are used as time varying controls: share of informal employment, Catholic adherence, Muslim adherence and other Christian adherence. As the census does not record income or consumption, we construct an asset-based wealth
index from the household questionnaire. We harmonise nine binary indicators of
dwelling quality and services that are recorded in all three waves (roof, walls, floor, electricity, drinking water, toilet, wastewater disposal, garbage collection and cooking fuel). For each arrondissement $i$ and census year $t$ our measure of
local living standards is the mean of this index across resident households;
higher values indicate a larger share of households with improved dwellings and access to services.
We also control for population density (Gridded Population of the World, Version 4.11 and 3), nighttime lights (version 4 DMSP-OLS), yearly precipitation and yearly mean temperature (CRU TS v4.07), all extracted from AidData \citep{goodman2019geoquery}. We also use the Annual Average Daily Traffic (AADT) on all extra-urban highways, primary roads, and secondary roads, which proxies for changes in accessibility and the built environment \citep{van2024global}. Geographic controls such as maize, cotton and cassava soil suitability are taken from FAO GAEZ.\footnote{\url{https://gaez.fao.org/}}

For the heterogeneity exercises we gather and harmonise additional datasets. 
We first gather the Climate Hazards Group InfraRed Precipitation with Station data (CHIRPS, version 3.0), a gridded rainfall product at 0.05$^\circ$
resolution which is relevant for long-run drought exposure. For each year from 1981 to 2013 we sum monthly rainfall over the April-October growing season and average the resulting seasonal total within each arrondissement, weighting grid cells by their coverage fraction. 
We calculate drought frequency as the share of the 33 seasons in which the arrondissement's growing-season rainfall fell at least 20\% below its own 1981-2013 mean.


Second, the census also provides the ethnic group of each respondent and following \cite{alesina2003fractionalization}, we compute ethnic fractionalization for arrondissement $i$ in census year $t$
as $\mathrm{ELF}_{it} = 1 - \sum_{k} s_{kit}^{2}$, where $s_{kit}$ is the
population share of ethnic group $k$. The index equals the probability that
two randomly drawn residents belong to different groups and ranges from zero (complete homogeneity) to one (extreme fragmentation). Finally, while the census does not provide consistent information on education, it does have a question on non-literacy across the three waves, which allows us to look at the share of non-literacy in each arrondissement. 

For exploring channels, we complement our census analysis using the 2018 and 2021 rollouts of the Enqu\^ete Harmonis\'ee sur le Conditions de Vie des M\'enages (EHCVM) for Benin, which is a representative survey of 8012 households done in two waves each year. 
We then merge the household-level information with our two main forest outcomes. We measure both forest cover and loss around each EHCVM survey cluster using the GPS coordinates recorded in the 2018/19 survey year, which are identical for the 2021/22 year. The two years cover 474 and 466 clusters respectively, 422 of them in both waves.
We first compute the share of area classified in the ESA-CCI-LC forest–savanna aggregate within a 10 km radius around each cluster, using exact area-weighted extraction over geodesic buffers. 
For the GFC Hansen forest loss, we now aggregate the original pixels to 10 km grid cells.
For each cell and year we compute 
forest loss as a percentage of the cell's total land area, which is directly comparable to the arrondissement-level loss measure described earlier. 
Each EHCVM cluster is assigned the value of the cell containing its coordinates. 
We then create an indicator for whether the head of the household is an ATR adherent, reported as ``Animiste'' in the survey,
and drop households that do not report their religion. We are thus left with 4288 unique households for a total of 6032 observations over two years. 
We then construct a set of relevant covariates (number of members, education, ratio of plot ownership, total plot surface, total assets, salary). Table \ref{tab:sumstats} presents the summary statistics for the arrondissement panel and Supplemental Appendix Table \ref{tab:sumstats_lsms} presents the summary statistics for the EHCVM panel. 

\section{Empirical Strategy}\label{empirics}

This section first presents the baseline OLS specification. Section \ref{iv} presents our identification strategy, section \ref{ivresult} shows our results and section \ref{threats} discusses threats to identification and robustness. 

Our baseline specification is the following:
\begin{equation}
\text{Y}_{iht} = \beta\,\text{ATR}_{it} + \boldsymbol{\theta}\textbf{X}_{it}
 + \boldsymbol{\gamma}\,(\textbf{G}_i \times \boldsymbol{\tau}_t)
 + \alpha_i + \phi_{ht} + \epsilon_{iht},
\label{eq: main_ols}
\end{equation}

\begin{table}
\centering
\begin{threeparttable}
\caption{Summary statistics, arrondissement panel}
\label{tab:sumstats}
\begin{tabular}{l rr rrr r}
\toprule
 & \multicolumn{2}{c}{Pooled} & \multicolumn{3}{c}{Mean by census year} & \\
\cmidrule(lr){2-3} \cmidrule(lr){4-6}
Variable & Mean & SD & 1992 & 2002 & 2013 & $N$ \\
    \midrule
    \multicolumn{7}{l}{\textit{Panel A: Forest outcomes}} \\
    \quad Forest--savanna share (\%) & 27.80 & 30.65 & 28.86 & 27.70 & 26.85 & 1,637 \\
    \quad Forest loss in $t{+}1$ (\% of area) & 0.39 & 0.79 & -- & 0.21 & 0.57 & 1,084 \\
    \addlinespace
    \multicolumn{7}{l}{\textit{Panel B: Religious composition}} \\
    \quad ATR share (\%) & 31.05 & 28.91 & 43.00 & 30.75 & 19.41 & 1,637 \\
    \quad Catholic share (\%) & 22.50 & 17.83 & 20.92 & 23.21 & 23.35 & 1,637 \\
    \quad Muslim share (\%) & 20.19 & 28.85 & 18.06 & 20.46 & 22.05 & 1,637 \\
    \quad Other Christian share (\%) & 11.22 & 11.78 & 5.69 & 10.65 & 17.32 & 1,637 \\
\addlinespace
\multicolumn{7}{l}{\textit{Panel C: Instrument}} \\
\quad Relative Adja proximity (REP) & 0.17 & 1.79 & 0.17 & 0.17 & 0.17 & 1,638 \\
\quad Pentecostal radio stations  & 1.33 & 1.25 & 0.00 & 1.00 & 3.00 & 1,637 \\
\addlinespace
\multicolumn{7}{l}{\textit{Panel D: Time varying controls}} \\
\quad Log population density & 4.73 & 1.44 & 4.41 & 4.71 & 5.08 & 1,637 \\
\quad Night-time lights & 2.77 & 9.04 & 1.73 & 2.68 & 3.91 & 1,637 \\
\quad Informal employment (\%) & 39.62 & 10.24 & 43.06 & 44.94 & 30.86 & 1,637 \\
\quad Poorest wealth quintile (\%) & 31.03 & 27.64 & 54.04 & 26.29 & 12.81 & 1,637 \\
\quad Mean temperature (\textdegree C) & 27.60 & 0.49 & 27.18 & 27.86 & 27.78 & 1,637 \\
\quad Precipitation (mm/month) & 84.85 & 11.53 & 76.93 & 91.73 & 85.87 & 1,637 \\
\quad Road traffic (AADT) & 1,640.59 & 1,867.03 & 381.33 & 898.04 & 3,640.10 & 1,637 \\
\addlinespace
\multicolumn{7}{l}{\textit{Panel E: Geographic controls}} \\
\quad GAEZ suitability: maize & 2,598.18 & 2,129.99 & 2,596.07 & 2,598.18 & 2,598.18 & 1,638 \\
\quad GAEZ suitability: cassava & 2,947.53 & 2,110.30 & 2,946.04 & 2,947.53 & 2,947.53 & 1,638 \\
\quad GAEZ suitability: cotton & 2,386.77 & 1,645.19 & 2,384.48 & 2,386.77 & 2,386.77 & 1,638 \\
\addlinespace
\multicolumn{7}{l}{\textit{Panel F: Heterogeneity moderators}} \\
\quad Drought frequency (1992) & 0.11 & 0.06 & 0.11 & 0.11 & 0.11 & 1,638 \\
\quad Ethnic fractionalization (1992) & 0.37 & 0.26 & 0.37 & 0.37 & 0.37 & 1,635 \\
\quad Non-literacy (\%) (1992) & 72.41 & 11.52 & 72.41 & 72.41 & 72.41 & 1,635 \\
\addlinespace
\bottomrule
\end{tabular}
\begin{tablenotes}[flushleft]\footnotesize
\item \textit{Notes:} Unit of observation is the arrondissement--census year (546 arrondissements; 1992, 2002, 2013). Forest--savanna share is the ESA-CCI share of arrondissement area in forest and savanna classes; forest loss is Hansen stand-replacing loss in the year following the census as a share of arrondissement area (2002 and 2013 only); Religious shares are from the census microdata. REP is $\ln(\text{dist. to Yoruba corridor}) - \ln(\text{dist. to Adja corridor})$; Pentecostal radio stations is the national count of licensed Charismatic Pentecostal stations at each census year (0, 1, 3). GAEZ suitability indices are time-invariant and enter the regressions interacted with census-year indicators. One arrondissement (Hou\'eko, Cov\`e) is absent from the 1992 census, giving 1,637 observations.
\end{tablenotes}
\end{threeparttable}
\end{table}

where $\text{Y}_{iht}$ denotes the share of forest-savanna cover in arrondissement $i$,
located in ethnic homeland $h$, in census year $t\in\{1992,2002,2013\}$, or the
share of the arrondissement subject to forest loss in the year following
the census (available for 2002 and 2013), as defined in Section~\ref{data}. $\textbf{X}_{it}$ contains time-varying arrondissement
characteristics: log population density, night-time light intensity, the share
of employment in the informal sector, mean wealth index, mean temperature and precipitation, and annual average daily traffic on the nearest road. Finally,
$\textbf{G}_i\times\boldsymbol{\tau}_t$ denotes 
denotes period-specific geographic controls covering spatial and soil suitability: the arrondissement centroid's latitude and longitude and the FAO GAEZ soil suitability for Benin's three main crops i.e. maize, cassava and cotton. These time invariant controls are interacted with indicators for the 2002 and 2013 waves, 
to account for any period-specific shock that varies smoothly with location or crop suitability such as rainfall and climate trends that follow the north-south gradient, and the expansion of cultivation onto land suited to particular crops as prices and agricultural extension programmes change. 
Specification~\eqref{eq: main_ols} includes two sets of fixed effects.
Arrondissement fixed effects $\alpha_i$ absorb any time-invariant characteristic
correlated with both forest and ATR adherence such as the historical or initial stock of forest and protected areas, distance to rivers and the Atlantic coastline, as well as baseline market and road access. 
Importantly, homeland$\times$year fixed effects $\phi_{ht}$ absorb
all shocks common to arrondissements of the same ethnic homeland in the same
year, and the homeland boundaries follow \cite{murdock1967ethnographic}. 
As discussed in Section \ref{back:vodun} since religious change in Benin unfolded along ethnic lines, $\beta$ is identified from variation in ATR adherence between arrondissements of the same ethnic homeland, in the same census year, that are comparable in location and agricultural potential. Finally, standard errors are clustered at the arrondissement level and shown within parentheses.


Table~\ref{tab:ols_atr_combined} reports the OLS estimates.
In the pooled cross-section, ATR adherence is negatively associated with forest cover, as high-ATR arrondissements are concentrated in the densely settled south-west of Benin. The sign reverses once arrondissement fixed effects are added and the coefficient increases under homeland$\times$year fixed effects with geographic controls, such that within an arrondissement over the sample period, a fall in the ATR share correlates with a fall in forest cover and a rise in subsequent forest loss. This contrast between the cross-sectional and the within-arrondissement correlations is the motivation for the fixed effects structure, and column (6) will remain our preferred specification in what follows. The OLS results, however, remain exposed to time-varying confounders and measurement error in census religious affiliation, which we discuss in the next section.

\begin{table}[!htbp]
\centering
\caption{ATR adherence, forest--savanna cover, and subsequent forest loss: OLS}
\label{tab:ols_atr_combined}

\begin{threeparttable}

\small
\setlength{\tabcolsep}{4.5pt}
\renewcommand{\arraystretch}{1.05}

\begin{tabular}{l*{6}{c}}
\toprule
Model:
& (1) & (2) & (3) & (4) & (5) & (6) \\
\midrule

\multicolumn{7}{c}{
\textit{Panel A: Dependent variable: forest--savanna cover (\%)}
} \\[2pt]

ATR share (\%)
& -0.292$^{***}$
& -0.122$^{***}$
& 0.016
& 0.023$^{*}$
& 0.044$^{***}$
& 0.051$^{***}$ \\

& (0.037)
& (0.041)
& (0.012)
& (0.012)
& (0.013)
& (0.012) \\[2pt]

Observations
& 1,637
& 1,637
& 1,637
& 1,637
& 1,634
& 1,634 \\

\midrule

\multicolumn{7}{c}{
\textit{Panel B: Dependent variable: subsequent forest loss (\%)}
} \\[2pt]

ATR share (\%)
& 0.002$^{**}$
& 0.001
& -0.020$^{***}$
& -0.020$^{***}$
& -0.019$^{***}$
& -0.016$^{***}$ \\

& (0.001)
& (0.001)
& (0.004)
& (0.004)
& (0.004)
& (0.005) \\[2pt]

Observations
& 1,084
& 1,084
& 1,084
& 1,084
& 1,082
& 1,082 \\

\midrule

Time-varying controls
&  & Yes & Yes & Yes & Yes & Yes \\

Geographic controls
&  &  &  &  &  & Yes \\

Arrondissement FE
&  &  & Yes & Yes & Yes & Yes \\

Year FE
&  &  &  & Yes &  &  \\

Homeland $\times$ Year FE
&  &  &  &  & Yes & Yes \\

\bottomrule
\end{tabular}

\begin{tablenotes}[flushleft]
\scriptsize
\item \textit{Notes:}
Time-varying controls include log population density, nighttime-light
intensity, informal-employment share, mean wealth index, mean temperature,
precipitation, and annual average daily traffic on the nearest road.
Geographic controls comprise the arrondissement centroid's latitude and
longitude and FAO-GAEZ soil suitability for maize, cassava, and cotton,
each interacted with indicators for the 2002 and 2013 waves. The forest-loss outcome is only available for the 2002 and 2013 waves.
Standard errors, clustered by arrondissement, are in parentheses.
$^{*}p<0.10$; $^{**}p<0.05$; $^{***}p<0.01$.
\end{tablenotes}

\end{threeparttable}
\end{table}

\subsection{Identification and relevance} \label{iv}

Let us now consider the main potential sources of bias in the OLS estimates. The first source is attenuation bias, since ATR may be measured with error, particularly due to syncretism in the population. The choice of Republic of Benin mitigates this issue to a large extent, given the comparatively explicit recognition of Vodun and other traditional religions. Nevertheless, evidence from across sub-Saharan Africa suggests that adherence to traditional religions may be under-reported. 
This phenomenon would therefore cause us to understate the effect of ATR by estimating coefficients smaller in magnitude than the true underlying effect.

A second set of concerns arises from time-varying processes of modernisation. Road construction, market integration, and urbanisation may simultaneously reduce ATR adherence through greater exposure to alternative religions and reduce forest cover by lowering the cost of accessing and clearing land. Such processes, if unaccounted for, would generate a spurious positive within-arrondissement association between ATR adherence and forest cover even in the absence of a causal effect of ATR. Bias in the opposite direction is also possible. In peri-urban arrondissements, particularly in southern Benin, ATR adherence may decline rapidly as Pentecostal and other Christian denominations expand first in urban areas, while agricultural land is simultaneously abandoned as labour shifts out of agriculture. 
The resulting secondary forest regrowth, a phenomenon commonly observed in West Africa \citep{schroeder2010secondary}, could generate a negative relationship between changes in ATR adherence and forest loss. 
Lastly, reverse causality is possible. Sacred groves constitute part of the physical and institutional infrastructure through which ATR practices are maintained. Their destruction may therefore weaken the institutions sustaining ATR adherence, potentially generating causal links from environmental outcomes to religious composition rather than the reverse.

Any of these factors can bias the Table \ref{tab:ols_atr_combined} estimates in either direction.
To address these concerns, we employ an instrumental variable strategy of the exposure design kind. Our instrument $Z_{it}$ is obtained by interacting a historically determined measure of local exposure with a national time-varying shifter:


\begin{equation}
    Z_{it} = \underbrace{\text{Relative Adja Proximity}_{i}}_{\text{Local exposure}}
        \times \underbrace{\text{Pentecostal radio stations}_{t}}_{\text{National shifter}}.
        \label{eq: instrument} 
\end{equation}
The RElative Adja Proximity (REP) is computed as 
\begin{equation}
\text{REP}_{i}
       = \ln\!\big(\text{Dist. from Yoruba corridor}_{i}\big)
        - \ln\!\big(\text{Dist. from Adja corridor}_{i}\big)
        \label{eq: relprox},  
\end{equation}
where higher values of $\text{REP}_i$ indicate arrondissements closer to the historical migration corridor of the Adja, the ethnic group of the founders of the Dahomey kingdom. 
Our design exploits how national changes in religious broadcasting generate differential changes in ATR adherence based on each arrondissment $i$'s exposure to historical migration corridors. 
We first describe the local exposure, then the national shifter, and finally how the two combine to deliver identification.

\begin{figure}
    \centering
    \includegraphics[scale=0.8]{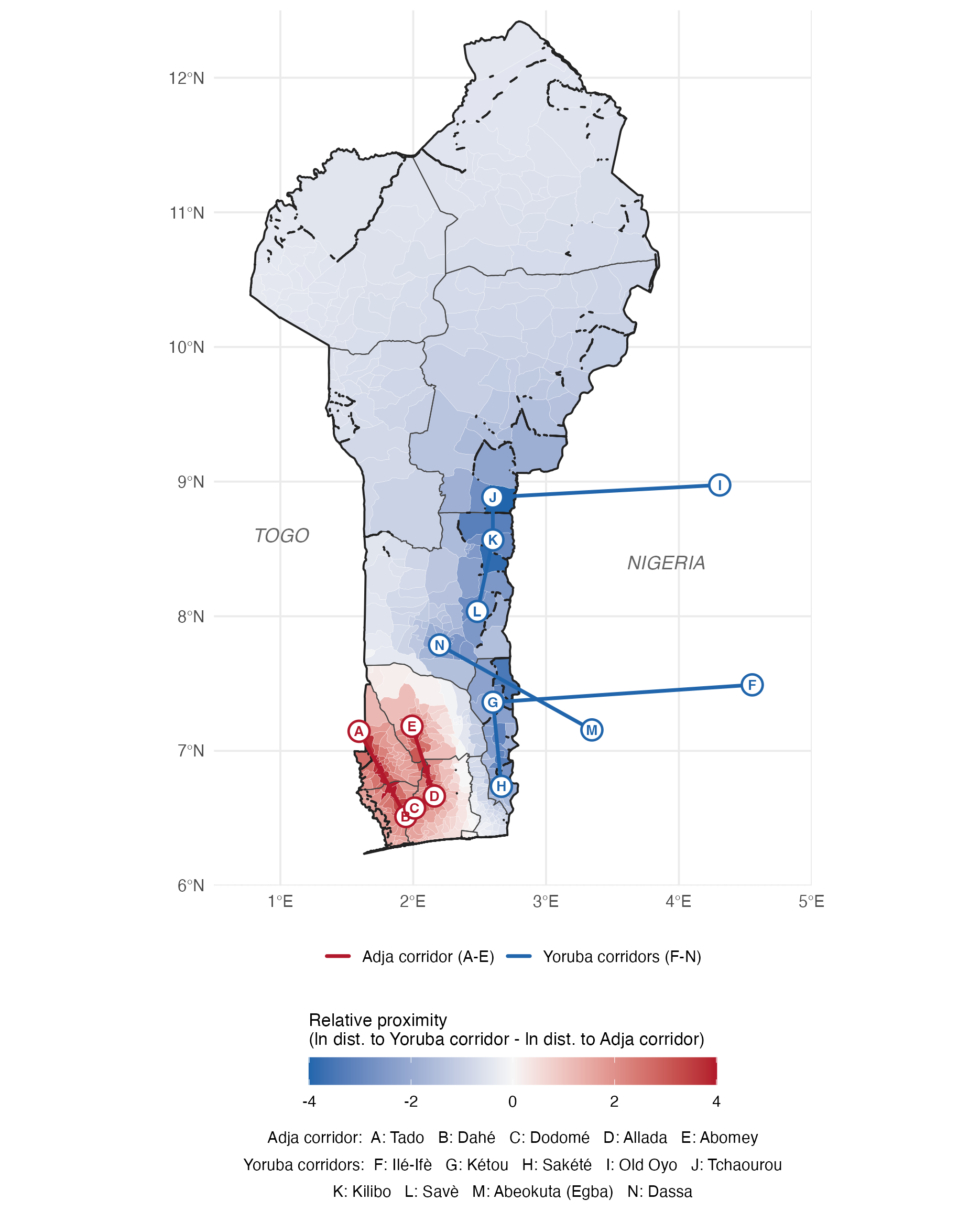}
    \caption{Migration waypoints and corridor.}
    \label{fig:corridors}
\end{figure}

\textbf{Local exposure:} 
We reconstruct the migration corridors using Robert Cornevin's \emph{La R\'epublique Populaire du B\'enin} \citep{cornevin1981republique}. In this book Cornevin synthesised a range of written and oral traditions collected during his work in Togo and Dahomey and subsequently as head of documentation at the Acad\'emie des Sciences d'Outre-Mer (The Academy of Overseas Sciences). 
Although historical accounts differ on particular episodes and etymologies, Cornevin emphasises that the traditions are \emph{``relativement homog\`enes concernant l'itin\'eraire''} that is, relatively homogeneous with respect to the migration path. 
Our reconstruction therefore relies only on the elements that are common across the historical sources, which are the name, geographical position and ordering of the principal waypoints. 
Figure \ref{fig:corridors} maps these reconstructed waypoints and the resulting migration corridors.

Several historical sources such as \cite{cornevin1981republique}, \cite{law1991slave} and \cite{bay2012wives} identify Tado (Point A in Figure \ref{fig:corridors}) as the ancestral homeland of the Adja. According to traditions, the royal Adja lineage of Tado were forced out after a failed bid for the throne.
Carrying the ancestors' relics and regalia, they reached first the shores of Lake Ah\'em\'e, where they founded Dah\'e (Point B), before moving to Dodom\'e on the D\'ekam\'e plateau (Point C) and subsequently to Allada (Point D). The
sources date these stages only relatively, however, the Allada waypoint has a definitive timeline, as 
Zoz\'e, the earliest remembered king of Allada, reigned
at the close of the 16th century, and the waypoint already appears on the Dutch cartographer Arnold von Langren's map of 1596. Allada became an important royal and religious centre, and the sacred remains of the ancestor Panther Prince Agassou were ultimately buried there. A further dynastic dispute in the early seventeenth century divided the lineage into three branches. One remained in Allada, a second established the kingdom of Porto-Novo, and a third, led by the Prince Do-Aklin, moved northwards to Abomey (Point E), which subsequently became the political centre and capital of the Kingdom of Dahomey. The corridor was thus
well established by the mid-17th century. 
The kings of Abomey were Adja by descent and maintained the corridor as a ritual axis as deceased kings were carried south for burial beside Agassou at Allada. 
The king was not himself a god, but royal authority had a sacral character:
at death each king entered the order of ancestors and, in time, the Vodun pantheon.
Religion was organised as an hierarchy that mirrored the state's own. The
deified royal ancestors stood at its head, served by a priesthood under
royal control \citep{claffey2007christian}. 
Since royal ancestral rites were institutions of the state rather
than simply lineage practices, Vodun practices persisted across generations and remained anchored to particular places. 
The corridor is therefore not simply a settlement route but the spine of Vodun's sacred geography.

The Yoruba provide the natural historical comparison underlying our measure of relative exposure. Alongside the Adja, the Yoruba constitute one of the major historical populations of southern and central Benin, and their spatial distribution can likewise be traced to a well-documented sequence of migrations originating from a sacred centre. Yoruba traditions identify Il\'e-If\'e (Point F) as the ancestral and ritual origin of the Yoruba world. Cornevin's account traces several branches of migration from Il\'e-If\'e into the territory of present-day Benin. One branch established K\'etou (Point G), which became an important centre of the Beninese Yoruba world and the point of origin of several related communities. From K\'etou, a southern branch settled at Sak\'et\'e (Point H), founded in the 16th century. A second branch originated in Old Oyo (Point I) and moved westward through Tchaourou (Point J) and Kilibo (Point K) before establishing the Tchab\'e kingdom at Sav\`e (Point L) in the 17th century. And the third were Egba, a subgroup of Yoruba, who migrated from Abeokuta (Point M) and settled on the hills of Dassa-Zoum\`e (Point N). Our Yoruba corridor follows these three branches.  Both corridors were therefore in place by 1650, more than three centuries before our census panel begins.

The Yoruba polities and Dahomey were longstanding regional competitors, repeatedly contesting territory, political influence, and control over slave-raiding and trading networks. The comparison between the groups matter because the two geographies experienced markedly different trajectories of religious change. In the Yoruba kingdoms of Benin,
traditional religion was polycentric, organised around towns and
lineages rather than a royal pantheon fused with a centralised state, and these areas encountered Christianity very early on. The Christian missions reached the Egba people in Abeokuta in the 1840s; Yoruba returnees from Sierra Leone and Brazil, many
of them former captives of Dahomey, carried the new faith inland
\citep{claffey2007christian}, and K\'etou's king Ad\'ebia received the pastor
Gollmer, the first European to visit, as early as 1835. Dahomey, by contrast, kept missions under royal control at
the coast; \emph{``no mission ever moved beyond coastal quarantine to penetrate the
kingdom''} \citep{claffey2007christian}. By the early twentieth century
Christianity was established in the Yoruba areas while Vodun institutions
persisted in the Adja-Dahomey heartland. Measuring proximity to the Adja corridor \emph{relative} to the Yoruba corridor thus contrasts two sacred geographies of the same age and standing, the first still holding ATR beliefs and institutions when Charismatic Pentecostalism became popular in the late nineties.

We build the exposure measure in four steps. First, we georeference the documented waypoints/stations of the Adja migration chronologically (points A-E) and connect consecutive stations with straight-line segments to form a single polyline, the Adja corridor. Second, we construct the comparison corridor in the same way from the three documented streams of Yoruba settlement (points F-H, I-L and M-N) and thus the Yoruba corridor is the
union of these three polylines. Third, we compute, for each
arrondissement centroid, the Euclidean distance in kilometres to the nearest
point on each corridor, taking the minimum over all segments of that
corridor. Distance is therefore measured to the route itself, not to the
nearest station, so as to not understate exposure for the arrondissements that lie on the actual migration line.
Fourth, relative exposure is computed as the
difference in log distances as shown in Eq. \eqref{eq: relprox} so that positive values indicate an arrondissement relatively closer to the
Adja corridor than to the Yoruba corridor. 
Because the measure is a ratio, it is invariant to the
distance unit and does not depend on where along the corridor an
arrondissement lies.

The resulting historical geography has substantial predictive power for contemporary religious affiliation. Proximity to the Adja corridor captures the local depth and persistence of Vodun institutions centuries after the migrations themselves. Panel A of Figure~\ref{fig:atr_other_evolution} shows that present-day ATR adherence in the south continues to follow the Tado-Allada-Abomey axis. 
Therefore, arrondissements close to the corridor entered our sample in 1992 with the largest stock of adherents exposed to conversion.

\textbf{The national shifter:} The instrument's time variation comes from a national shifter which is the number of licensed Charismatic Pentecostal
radio stations in Benin for each census year. It captures the national change in Benin's religious composition as the 1990s saw an unprecedented growth of new
churches, and by 2001 the census conducted by the Christian project ARCEB (Action pour la recherche et la croissance des Eglises au Benin) counted nearly 4000 local congregations of new Pentecostal denominations \citep{mayrargue20059}. 
Charismatic Pentecostalism became a highly visible presence in the public sphere which was characterized by a radical concept of salvation and considering the born-again individual as already belonging to the \say{Kingdom of God}. 
A key part of the proselytising discourse of the Charismatic churches was to
vehemently reject traditional religions, which they diabolised as harbingers of
evil spirits \citep{hackett1998charismatic, asamoah2009african}. Benin illustrated the force of this discourse at the highest level: Mathieu K\'er\'ekou, the
military-Marxist ruler of 1972-1991, returned to power through the 1996
elections presenting himself as a Born-Again Christian, distancing himself
from a past associated with the ``occult forces'' of Vodun
\citep{camilla2005nouveaux}. A large part of this Pentecostal expansion ran through the media. 
Benin's 1990 constitutional transition liberalised the media sector, and from 1997 the licensing of independent radio permitted religious broadcasting in all its forms, from dedicated stations to contracted hours on local ones. By 2013 the main licensed Christian broadcasters were Radio Immacul\'ee Conception (1998), the broadcaster of the Roman Catholic Church, and the two Charismatic Pentecostal broadcasters, Radio Maranatha (first station 1998, second 2003) and Radio All\'eluia (2003) \citep{gratz2011paroles}.
The Pentecostal broadcasters' strategy was to Christianise popular culture
for Born-Again consumption, which involved persistent efforts to evangelize
and portray a negative view of ATRs in Christian dominated media \citep{hackett1998charismatic, asamoah2009african}. 
Our shifter counts these Pentecostal stations at each census wave: 0 in 1992, 1 in 2002, 3 in 2013. We show robustness in Section \ref{threats} using the national share of Pentecostals instead, as well as other functional forms. 

We deliberately exclude the Catholic station as we exploit the Post 1990s \say{Pentecostalism explosion} in Benin \citep{claffey2007christian}. 
This choice is validated by Appendix Figure~\ref{fig: census_summ} which shows that ATR was the largest single affiliation in 1992
and had fallen to well under half that share by 2013, and the growth in Christian adherence
over our period is almost entirely charismatic: the \say{other Christian}
census category rose from 6.4 to 16.5 percent of the population between
1992 and 2013 while the Catholic and Protestant shares were flat. Moreover, Supplemental Appendix Table \ref{tab:destination} decomposes the corridor's ATR outflow by destination and more than three-quarters went to Christian denominations and
none to Islam, so the compositional change the instrument picks up is
consistent with a Christian proselytising shock and inconsistent with
the corridor simply modernising or receiving Muslim migrants. In essence, our time variation reflects the overall decrease of ATR adherence across Benin due to conversion pressure. We remark that by counting nationally the number of Pentecostal stations and not considering their location, we avoid using their spatial placement as part of our identifying variation, which would then inherit part of the endogeneity we are trying to remove. 


Identification comes from the interaction of the two components: when national broadcasting expanded, the places with the largest stock of ATR adherents, and the places that Charismatic Pentecostalism named as its target of evangelization, had the most to lose. 
These channels predict a negative first stage:
as Pentecostal broadcasting intensified, arrondissements closer to the Adja corridor shed ATR adherence the most,  within ethnic homeland and accounting for the time-varying and geographic controls described previously. 
All temporal variation in the instrument is national and all spatial variation is precolonial; neither is chosen by households, churches or governments during the sample period.
The exclusion restriction requires that, conditional on the covariates, trends and fixed effects, the corridor position affects forest outcomes on the broadcast timeline only through religious composition. 
We now present the main results of the paper. 

\subsection{Results}\label{ivresult}

Table~\ref{tab:iv_esa_atr} presents the 2SLS estimates for forest-savanna
cover, together with the first stage and reduced form. The first stage in
Panel B is strong and stable: a one-unit increase in relative Adja proximity
reduces the ATR share by 1.3-1.8 percentage points per radio station, with
effective $F$-statistics of 62-129 in the full panel and 59 in the southern
sub-sample. The reduced form in Panel C consistently shows that
forest-savanna cover falls along the corridor as broadcasting expands.
Panel A presents the 2SLS estimates. In our preferred specification, which includes arrondissement and homeland$\times$year fixed effects with geographic controls (column 3), the coefficient is 0.185 so each percentage point of ATR adherence lost reduces forest--savanna cover by 0.19 percentage points of arrondissement area. Per point this is 0.7 percent of the sample mean, but adherence did not fall by a point. Scaled by the 21-point national decline between 1992 and 2013, the estimate implies 3.9 percentage points less cover, one-seventh of the mean, in an arrondissement experiencing the average decline. The estimate is stable across all specifications: moving from two-way to
homeland$\times$year fixed effects raises the coefficient from 0.084 to
0.123, adding the geographic controls yields 0.185, and restricting to the
southern departments yields 0.158. The
Anderson-Rubin and tF \citep{lee2022valid} confidence intervals exclude
zero in every column of Table~\ref{tab:iv_esa_atr}. The 2SLS estimates exceed the OLS estimates from
Table~\ref{tab:ols_atr_combined} by a factor of about four, consistent with
attenuation from measurement error, with negatively signed confounders such
as peri-urban conversion accompanied by land abandonment, and with 2SLS
recovering a local average treatment effect: the instrument identifies the
effect among corridor arrondissements, where Vodun institutions are densest.

It is relevant to report the southern sub-sample in column (4) as it is the effective domain of our LATE. Relative Adja proximity contrasts two geographies the Adja corridor in the south-west and the Yoruba corridors in the
south-east, and it is only where these corridors compete for religious
space that the measure captures differential exposure to the institutions
that resisted Christianization. In the far north, arrondissements are
distant from both corridors, and variation in the measure reduces to a
generic east-west gradient with no corresponding historical contest. As
the southern estimate sits close to the full-panel
estimate, it indicates that the full-panel results are driven by
the corridor contrast itself rather than by out-of-domain variation.

Table~\ref{tab:iv_loss_atr} shows the corresponding estimates for forest loss in $t+1$. The first stage on this two-wave sample is of
similar strength (coefficients of 1.0-1.4, effective $F$ of 75--103 in
the full panel), and the reduced form changes sign as expected.
The 2SLS estimate in our preferred specification is $-0.068$ i.e. a one-percentage-point decline in ATR adherence raises subsequent annual forest loss by 0.07 percentage points which is 17.5 percent of the sample mean. The large proportional magnitude reflects the fact that annual stand-replacing loss averages only about 0.4 percent of arrondissement area, so an absolute effect of 0.07 percentage points of land area per year per point of ATR decline is substantial relative to this low base rate.

Both cover and loss results provide parallel evidence of our core finding, which is that for the areas where Pentecostal conversion eroded ATR adherence, standing cover is lower and ongoing clearing is faster. 
The southern loss estimate in column~(4) is negative but imprecise ($-0.024$, s.e.\ 0.030, effective $F = 20.1$),
so we rely on the full panel for the loss outcome throughout. 
Across both outcomes the first stage, reduced form, and all point estimates agree with those expected from the Adja corridor channel discussed in Section \ref{iv}.
\begin{table}[!htb]
\centering
\caption{ATR adherence and forest--savanna cover:
2SLS, first stage, and reduced form}
\label{tab:iv_esa_atr}

\begin{threeparttable}

\small
\setlength{\tabcolsep}{6pt}
\renewcommand{\arraystretch}{1.08}

\begin{tabular}{lcccc}
\toprule
& \multicolumn{3}{c}{Full sample}
& \multicolumn{1}{c}{South} \\
\cmidrule(lr){2-4}\cmidrule(lr){5-5}
& (1) & (2) & (3) & (4) \\
\midrule

\multicolumn{5}{l}{
\textit{Panel A: 2SLS. Dependent variable: forest--savanna share (\%)}
} \\[2pt]

ATR share (\%)
& 0.084$^{***}$
& 0.123$^{***}$
& 0.185$^{***}$
& 0.158$^{***}$ \\

& (0.029)
& (0.037)
& (0.056)
& (0.045) \\[2pt]

Effect rel.\ to mean
& +0.3\%
& +0.4\%
& +0.7\%
& +1.0\% \\

AR 95\% CI
& [0.03, 0.14]
& [0.05, 0.20]
& [0.08, 0.31]
& [0.08, 0.26] \\

tF 95\% CI
& [0.03, 0.14]
& [0.05, 0.20]
& [0.07, 0.30]
& [0.06, 0.25] \\

Critical value ($c_F$)
& 1.96
& 1.96
& 2.08
& 2.10 \\

\midrule

\multicolumn{5}{l}{
\textit{Panel B: First stage. Dependent variable: ATR share (\%)}
} \\[2pt]

$\text{REP}_{i} \times \text{Pentecostal radio stations}_{t}$
& -1.257$^{***}$
& -1.285$^{***}$
& -1.341$^{***}$
& -1.797$^{***}$ \\

& (0.111)
& (0.124)
& (0.171)
& (0.234) \\[2pt]

Effective $F$
& 128.6
& 107.9
& 61.5
& 59.1 \\

\midrule

\multicolumn{5}{l}{
\textit{Panel C: Reduced form. Dependent variable: forest--savanna share (\%)}
} \\[2pt]

$\text{REP}_{i} \times \text{Pentecostal radio stations}_{t}$
& -0.105$^{***}$
& -0.158$^{***}$
& -0.248$^{***}$
& -0.284$^{***}$ \\

& (0.036)
& (0.046)
& (0.069)
& (0.076) \\

\midrule


Time-varying controls
& Yes & Yes & Yes & Yes \\

Geographic controls
&  &  & Yes & Yes \\

Arrondissement FE
& Yes & Yes & Yes & Yes \\

Year FE
& Yes &  &  &  \\

Homeland $\times$ Year FE
&  & Yes & Yes & Yes \\

\midrule

Observations
& 1,637
& 1,634
& 1,634
& 1,166 \\

\bottomrule
\end{tabular}

\begin{tablenotes}[flushleft]
\scriptsize
\item \textit{Notes:}
Time-varying controls include log population density, nighttime-light
intensity, informal-employment share, mean wealth index, mean temperature,
precipitation, and annual average daily traffic on the nearest road.
Geographic controls comprise the arrondissement centroid's latitude and
longitude and FAO-GAEZ soil suitability for maize, cassava, and cotton,
each interacted with indicators for the 2002 and 2013 waves.
AR denotes the Anderson--Rubin 95\% confidence interval.
tF denotes the Lee et al.\ (2022) adjusted 95\% confidence interval;
$c_F$ is the corresponding $F$-dependent critical value.
Effective $F$ is the first-stage effective $F$-statistic.
Standard errors, clustered by arrondissement, in parentheses.
$^{*}p<0.10$; $^{**}p<0.05$; $^{***}p<0.01$.
\end{tablenotes}

\end{threeparttable}
\end{table}

\begin{table}[!htb]
\centering
\caption{ATR adherence and subsequent forest loss (Hansen):
2SLS, first stage, and reduced form}
\label{tab:iv_loss_atr}

\begin{threeparttable}

\small
\setlength{\tabcolsep}{6pt}
\renewcommand{\arraystretch}{1.08}

\begin{tabular}{lcccc}
\toprule
& \multicolumn{3}{c}{Full sample}
& \multicolumn{1}{c}{South} \\
\cmidrule(lr){2-4}\cmidrule(lr){5-5}
& (1) & (2) & (3) & (4) \\
\midrule

\multicolumn{5}{l}{
\textit{Panel A: 2SLS. Dependent variable: subsequent forest loss (\%)}
} \\[2pt]

ATR share (\%)
& -0.094$^{***}$
& -0.072$^{***}$
& -0.068$^{***}$
& -0.024 \\

& (0.015)
& (0.013)
& (0.015)
& (0.030) \\[2pt]

Effect rel.\ to mean
& -24.1\%
& -18.4\%
& -17.5\%
& -5.3\% \\

AR 95\% CI
& [-0.13, -0.07]
& [-0.10, -0.05]
& [-0.10, -0.04]
& [-0.09, 0.04] \\

tF 95\% CI
& [-0.12, -0.06]
& [-0.10, -0.05]
& [-0.10, -0.04]
& [-0.10, 0.05] \\

Critical value ($c_F$)
& 2.03
& 1.96
& 2.03
& 2.59 \\

\midrule

\multicolumn{5}{l}{
\textit{Panel B: First stage. Dependent variable: ATR share (\%)}
} \\[2pt]

$\text{REP}_{i} \times \text{Pentecostal radio stations}_{t}$
& -1.040$^{***}$
& -1.427$^{***}$
& -1.431$^{***}$
& -1.096$^{***}$ \\

& (0.118)
& (0.141)
& (0.165)
& (0.244) \\[2pt]

Effective $F$
& 78.0
& 102.6
& 74.9
& 20.1 \\

\midrule

\multicolumn{5}{l}{
\textit{Panel C: Reduced form. Dependent variable: subsequent forest loss (\%)}
} \\[2pt]

$\text{REP}_{i} \times \text{Pentecostal radio stations}_{t}$
& 0.098$^{***}$
& 0.103$^{***}$
& 0.098$^{***}$
& 0.026 \\

& (0.014)
& (0.018)
& (0.021)
& (0.033) \\

\midrule


Time-varying controls
& Yes & Yes & Yes & Yes \\

Geographic controls
&  &  & Yes & Yes \\

Arrondissement FE
& Yes & Yes & Yes & Yes \\

Year FE
& Yes &  &  &  \\

Homeland $\times$ Year FE
&  & Yes & Yes & Yes \\

\midrule

Observations
& 1,084
& 1,082
& 1,082
& 770 \\

\bottomrule
\end{tabular}

\begin{tablenotes}[flushleft]
\scriptsize
\item \textit{Notes:}

The Hansen forest loss outcome is only available for the 2002 and 2013 waves.
Time-varying controls include log population density, nighttime-light
intensity, informal-employment share, mean wealth index, mean temperature,
precipitation, and annual average daily traffic on the nearest road.
Geographic controls comprise the arrondissement centroid's latitude and
longitude and FAO-GAEZ soil suitability for maize, cassava, and cotton,
each interacted with indicators for the 2002 and 2013 waves.
AR denotes the Anderson--Rubin 95\% confidence interval.
tF denotes the Lee et al.\ (2022) adjusted 95\% confidence interval;
$c_F$ is the corresponding $F$-dependent critical value.
Effective $F$ is the first-stage effective $F$-statistic.
Standard errors, clustered by arrondissement, in parentheses.
$^{*}p<0.10$; $^{**}p<0.05$; $^{***}p<0.01$.
\end{tablenotes}

\end{threeparttable}
\end{table}

\subsection{Threats to identification and robustness checks}\label{threats}

\begin{figure}
    \centering
    \includegraphics[scale=0.7]{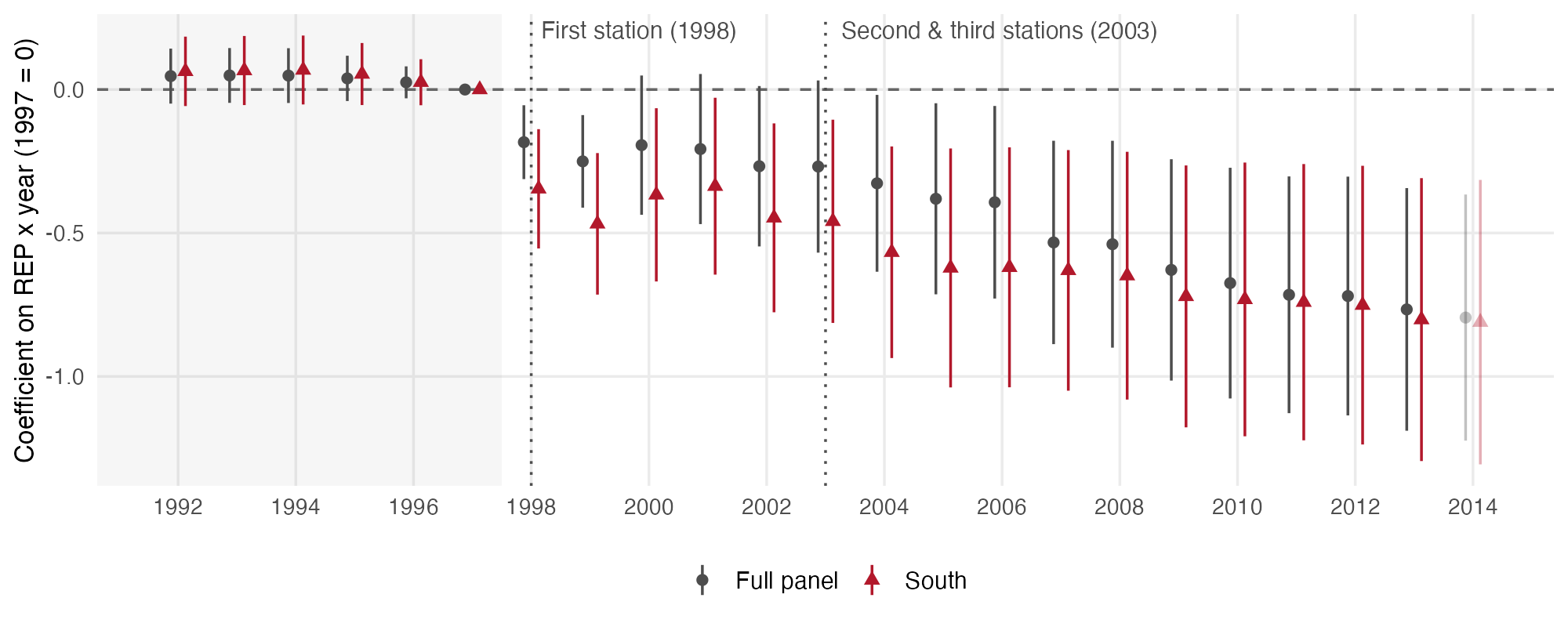}
    \caption{Forest-savanna cover along the corridor: annual event study. ESA CCI annual land cover, 1991-2014; arrondissement, homeland x year FE and geographic controls; 95\% CI, clustered by arrondissement.}. 
    \label{fig:es_esa_annual}
\end{figure}

The key tenet behind the causal validity of our results is whether anything other than religious change could produce the observed pattern of timing and geography in ATR adherence, since the exclusion restriction requires for our instrument to affect forest outcomes only through ATR adherence, conditional on fixed effects and controls. 
In this section we first show how the corridor exhibits no pre-broadcast forest trends and that ATR decline follows the broadcasting timeline; we then show how the corridor's geography is unrelated to settlement history, generic origin proximity, or state centralisation; finally we address potential confounders such as a contemporaneous land reform and migration-related population change, and discuss robustness checks on spatial inference and instrument construction choices.

Although we only have three census waves that report ATR adherence, the ESA outcome is observed annually, and therefore we can check if the Adja migration corridor was already generating a different forest trajectory before the broadcasting sector was liberalized in 1997 and a Pentecostal boom took place in Benin, confounding our estimates. 
We regress the annual ESA CCI forest-savanna share on REP$_i\times$year dummies over 1992-2014 and arrondissement and homeland-year fixed effects, to test the presence of corridor pre-trends on forest dynamics, as shown in Figure~\ref{fig:es_esa_annual}. 
We first reassuringly observe there is no corridor-driven pre-trend: the 1992-1996 coefficients are indistinguishable from zero in both samples and are jointly insignificant (Wald $p=0.59$ south,
$0.88$ full panel). Second, the divergence starts with the broadcast era and intensifies after the 2003 station entries, reproducing the dynamics implied by our national shifter in the first stage.

Our instrument imposes a specific form on the national shifter based on the national count of Pentecostal radio stations (0/1/3), therefore, it is important to check whether our results are robust to alternative temporal specifications. 
In Table~\ref{tab:shifter_esa_atr_full} we investigate what form the corridor's ATR decline actually follows, by comparing our shifter to alternative national series over the same period, and to unrestricted year coefficients. 
In order to directly compare the first stage coefficients we rescale every shifter candidate to run from 0 in 1992 to 1 in 2013, such that each one is the implied 1992–2013 change in ATR share per unit of REP$_i$. Across the radio station count, the national share of other Christianity, the national ATR outflow and a linear time trend, this coefficient is stable at about $-4$ and the 2SLS estimate moves only between 0.185 and 0.191. A post-1992 dummy, which would identify the democratic transition rather than conversion dynamics, imposes a 2013/2002 ratio of 1.0 and fits worse as its effective $F$ is 30.9 against 53--61 for the other shifters, and its ratio is incompatible with the unrestricted year coefficients in the final row, which put the corridor's ATR decline at a 2013/2002 ratio of 3.9, closest to the radio-station count's imposed 3 and well above the linear trend's 2.1. The corridor's religious realignment therefore follows the path of
Charismatic broadcasting, not the democratic transition of 1991--92 nor a linear secular modernisation trend. Supplemental Appendix
Table~\ref{tab:shifter_esa_atr_south} repeats the exercise for the
southern sub-sample with the same conclusions.

\begin{table}[!htb]
\centering
\caption{Alternative temporal profiles of the national shifter: full sample}
\label{tab:shifter_esa_atr_full}

\begin{threeparttable}

\small
\setlength{\tabcolsep}{3.0pt}
\renewcommand{\arraystretch}{1.05}

\begin{tabular}{
@{}
p{3.4cm}
c
c
c
c
c
c
@{}
}
\toprule
Shifter $S_t$
& \makecell{$S_t$ path\\(92/02/13)}
& \makecell{2013/2002\\ratio}
& \makecell{REP $\times$ 2013\\coef.\ ($t$)}
& \makecell{Effective\\$F$}
& \makecell{2SLS\\$\hat{\beta}$}
& \makecell{Sargan\\$p$} \\
\midrule

Pentecostal radio stations
& 0 / 1 / 3
& 3.0
& -4.02 (-7.8)
& 61.4
& \makecell{0.185$^{***}$\\(0.056)}
& -- \\

Other Christianity share
& 0 / 4.96 / 11.6
& 2.3
& -4.03 (-7.6)
& 57.3
& \makecell{0.188$^{***}$\\(0.058)}
& -- \\

ATR outflow since 1992
& 0 / 12.2 / 23.6
& 1.9
& -3.94 (-7.3)
& 53.0
& \makecell{0.191$^{***}$\\(0.061)}
& -- \\

Linear time trend
& 0 / 10 / 21
& 2.1
& -3.99 (-7.4)
& 55.0
& \makecell{0.190$^{***}$\\(0.060)}
& -- \\

Post-1992 dummy
& 0 / 1 / 1
& 1.0
& -2.47 (-5.6)
& 30.9
& \makecell{0.212$^{***}$\\(0.081)}
& -- \\

Unrestricted year coefficients
& --
& 3.9
& -3.96 (-7.3)
& 38.4
& \makecell{0.182$^{***}$\\(0.054)}
& 0.438 \\

\bottomrule
\end{tabular}

\begin{tablenotes}[flushleft]
\scriptsize
\item \textit{Notes:}
$N=1{,}637$.
Each row instruments ATR share with
REP$_i \times \widetilde{S}_t$, where
$\widetilde{S}_t=(S_t-S_{1992})/(S_{2013}-S_{1992})$,
so each national series is normalized to zero in 1992 and one in 2013.
The coefficient on REP $\times$ 2013 therefore gives the implied
1992--2013 change in ATR share per unit of REP and is comparable across
candidate shifters; the associated $t$-statistic is in parentheses.
The 2013/2002 ratio describes the temporal profile imposed by each
series. 
The final row instead instruments ATR share with REP interacted separately
with the 2002 and 2013 indicators, leaving the first-stage time profile
unrestricted.
All specifications include arrondissement and
homeland $\times$ year fixed effects, time-varying and
geographic controls.
Time-varying controls include log population density, nighttime-light
intensity, informal-employment share, mean wealth index, mean temperature,
precipitation, and annual average daily traffic on the nearest road.
Geographic controls comprise latitude, longitude, and FAO-GAEZ soil
suitability for maize, cassava, and cotton, each interacted with census-year
indicators.
Standard errors, clustered by arrondissement, are reported in parentheses
beneath the 2SLS estimates.
$^{*}p<0.10$; $^{**}p<0.05$; $^{***}p<0.01$.
\end{tablenotes}

\end{threeparttable}
\end{table} 

Another possible concern is that REP$_{i}$ might simply be an index of fundamental suitability and accessibility such as fertile land and rivers. In this case, the south-western gradient that REP$_{i}$ traces would show differential land use change in previous periods or centuries, not because of ATR but because that's where people historically settled and cleared. To address this potential issue we use HYDE (History Database of the Global Environment) which provides spatially explicit reconstructions of historical population and land use over very long time periods \citep{klein2011hyde}. HYDE's
historical maps are reconstructed rather than observed, and population density and
cropland are allocated across cells based on factors that make settlement and land use change more likely such as soil quality, climate and terrain. This allows HYDE to be
a direct test of our concern since if REP$_i$ were simply a proxy for underlying
geographic advantage, it would predict HYDE's allocations. For each HYDE period $t \in \{1100, 1200, \ldots, 2000\}$ we estimate 
\begin{equation}\label{eq:hyde}
\operatorname{ihs}(Y_{ct}) = \beta_t \, \text{REP}_c
  + \delta_{1t}\,\text{lat}_c + \delta_{2t}\,\text{lon}_c + \varepsilon_{ct},
\end{equation}
where $Y_{ct}$ is either population density or the cropland share of HYDE grid
cell $c$ in year $t$, and REP$_c$ is the relative Adja proximity of
the cell's centroid, constructed exactly as in Eq.~(\ref{eq: relprox}).
Standard errors allow for spatial correlation following
\citet{conley1999gmm} with a 50~km cutoff. A nonzero $\hat\beta_t$ in any period would indicate that corridor position predicts settlement or cultivation conditional on coarse location.

\begin{figure}
    \centering
    \includegraphics[scale=0.75]{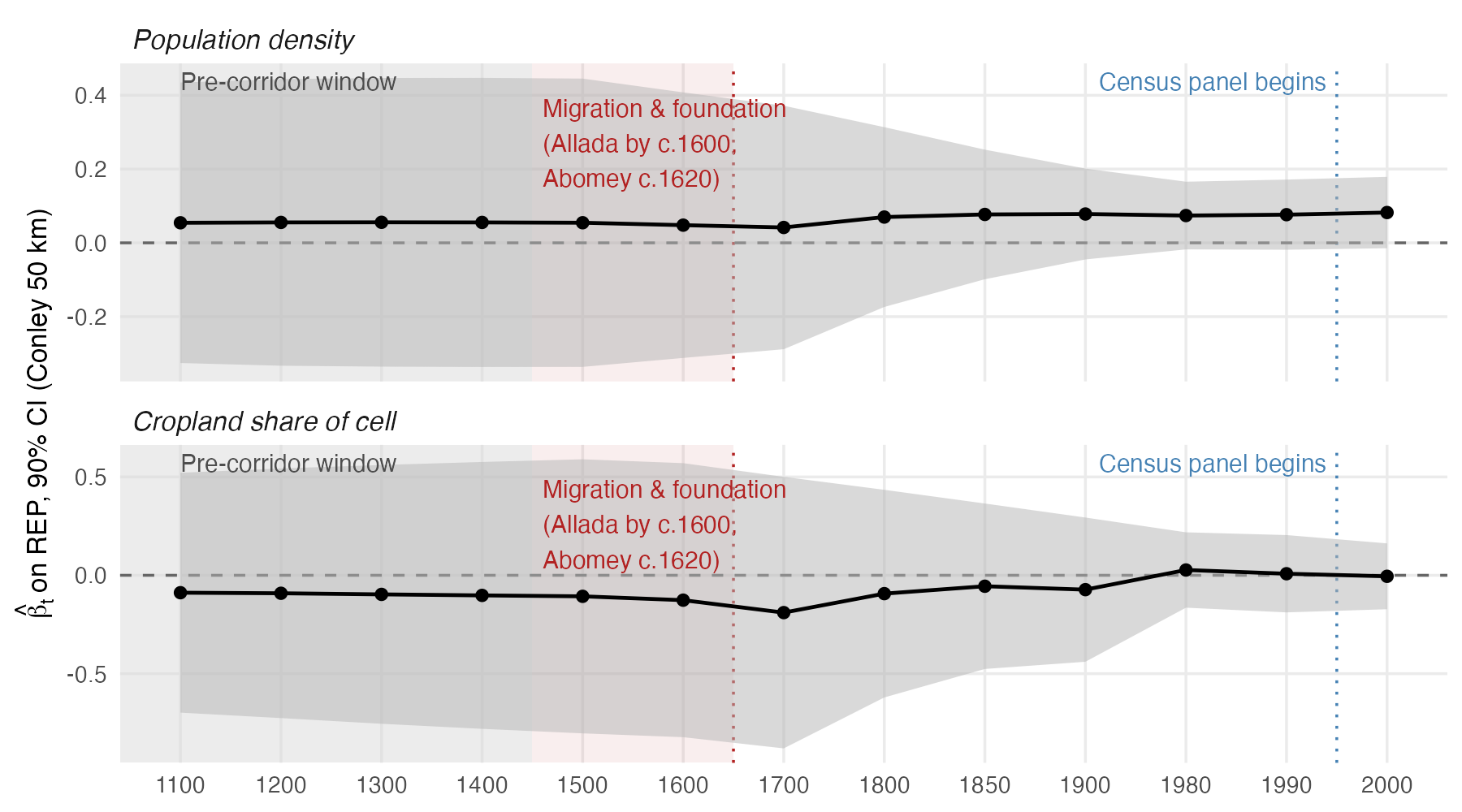}
    \caption{Anthropogenic footprint along the Adja corridor, 1100-2000 CE. HYDE 3.3 event study with latitude and longitude controls and Conley standard errors.}
    \label{fig:hyde}
\end{figure}

Figure~\ref{fig:hyde} plots the $\hat\beta_t$ series and shows that corridor position predicts settlement in no period. This is especially relevant for the pre-migration period (1100--1400~CE) as HYDE data shows substantial settlement variation across cells in this period (we find one standard deviation of log population density corresponds to roughly a four-fold difference in density), showing changes on their suitability and accessibility characteristics, yet none of it correlates with the migration corridor. 
This absence of correlation persists through the migration era and into the modern period, where HYDE is from observed census and satellite data and the confidence intervals tighten accordingly, therefore, the corridor position does not seem to have any impact on historic settlement or cultivation conditional on coarse position.

\begin{table}[!htb]
\centering
\caption{Corridor exogeneity: recentering}
\label{tab:bh_recentering}
\begin{threeparttable}

\small
\setlength{\tabcolsep}{6pt}
\renewcommand{\arraystretch}{1.06}

\begin{tabular}{lcccc}
\toprule
& \multicolumn{2}{c}{Forest--savanna cover}
& \multicolumn{2}{c}{Subsequent forest loss} \\
\cmidrule(lr){2-3}\cmidrule(lr){4-5}
& Baseline & Recentered & Baseline & Recentered \\
& (1) & (2) & (3) & (4) \\
\midrule

ATR share (\%)
& 0.185$^{***}$
& 0.159$^{***}$
& -0.068$^{***}$
& -0.100$^{***}$ \\

& (0.056)
& (0.047)
& (0.015)
& (0.023) \\

\midrule

Observations
& 1,637
& 1,637
& 1,084
& 1,084 \\

Effective $F$
& 61.4
& 52.6
& 74.8
& 55.2 \\

\bottomrule
\end{tabular}

\begin{tablenotes}[flushleft]
\scriptsize
\item \textit{Notes:}
All specifications include arrondissement and
homeland $\times$ year fixed effects, time-varying and geographic
controls.
Time-varying controls include log population density, nighttime-light
intensity, informal-employment share, mean wealth index, mean temperature,
precipitation, and annual average daily traffic on the nearest road.
Geographic controls comprise the arrondissement centroid's latitude and
longitude and FAO-GAEZ soil suitability for maize, cassava, and cotton,
each interacted with indicators for the 2002 and 2013 waves.
Standard errors, clustered by arrondissement, are in parentheses.
$^{*}p<0.10$; $^{**}p<0.05$; $^{***}p<0.01$.
\end{tablenotes}

\end{threeparttable}
\end{table}
Another concern is in the way that REP$_i$ is built, since any arrondissement near Tado, or anywhere in the south-west, is by construction close to any migration corridor the Adja might have taken. A part of REP$_i$ therefore reflects general proximity to the origin region and the country's south-western quadrant, rather than its position relative to the realised route. This region is also close to the coast, home to important economic centers and markets, so even if the migration route is as good as random among the plausible ones, exposure built from it might not be randomly assigned across space. Places near the origin may thus be systematically exposed to these effects under every other potential corridor. Following \citet{borusyak2023nonrandom}, we purge this generic component by recentering the exposure against its expectation over randomized counterfactual corridors. We draw 500 alternative routes anchored at Tado, matching the historical spine's segment lengths and rotating the path angle at random, and then compute each arrondissement's expected relative proximity $\overline{\text{REP}}_i$ over these draws. 
The expected exposure is highly correlated with the realised one and absorbs about two-fifths of its variation. 
The recentered instrument $(\text{REP}_i-\overline{\text{REP}}_i)\times \text{radio stations}_t$ thus isolates the component due
to the particular path through Dah\'e, Dodom\'e and Allada that the tradition records. Table \ref{tab:bh_recentering} shows that the estimates are
robust, confirming that the identifying variation is the realised Adja migration route and not proximity to Tado or the south-west.

\begin{table}[!htb]
\centering
\caption{The precolonial state: truncated corridor and state distance controls}
\label{tab:state_legacy}

\begin{threeparttable}

\small
\setlength{\tabcolsep}{5.5pt}
\renewcommand{\arraystretch}{1.06}

\begin{tabular}{lcccc}
\toprule
& \multicolumn{2}{c}{Truncated corridor}
& \multicolumn{2}{c}{State distance controls} \\
\cmidrule(lr){2-3}\cmidrule(lr){4-5}
& Cover & Loss & Cover & Loss \\
& (1) & (2) & (3) & (4) \\
\midrule

ATR share (\%)
& 0.247$^{**}$
& -0.131$^{***}$
& 0.203$^{**}$
& -0.132$^{***}$ \\

& (0.118)
& (0.024)
& (0.089)
& (0.026) \\

\midrule

State distance $\times$ Year
&  &  & Yes & Yes \\
\midrule

Observations
& 1,637
& 1,084
& 1,637
& 1,084 \\

Effective $F$
& 17.1
& 35.8
& 26.6
& 45.7 \\

\bottomrule
\end{tabular}

\begin{tablenotes}[flushleft]
\scriptsize
\item \textit{Notes:}
All specifications include arrondissement and
homeland $\times$ year fixed effects, time-varying  and
geographic controls.
Time-varying controls include log population density, nighttime-light
intensity, informal-employment share, mean wealth index, mean temperature,
precipitation, and annual average daily traffic on the nearest road.
Geographic controls comprise latitude, longitude, and FAO-GAEZ soil
suitability for maize, cassava, and cotton, each interacted with the
2002 and 2013 census-wave indicators.
Standard errors, clustered by arrondissement, are in parentheses.
$^{*}p<0.10$; $^{**}p<0.05$; $^{***}p<0.01$.
\end{tablenotes}

\end{threeparttable}
\end{table}
As our instrument exploits the presence of Dahomey Kingdom, it raises the concern that REP$_i$ might simply be a proxy for the kingdom's political centralisation elements, which have a strong positive association with contemporary regional economic development \citep{michalopoulos2013pre}, rather than capturing the sacred geography. We tackle this issue in two ways. In column (1) and (2) of Table \ref{tab:state_legacy} we show the results using a truncated migration Adja corridor where we drop Abomey, the final waypoint, from the corridor spine. This is because once the Adja settled in Abomey, the capital of Dahomey, a centralised state was formed and removing the city from our corridor allows us to focus entirely on the pre-state migration corridor (Tado, Dah\'e, Dodom\'e and Allada) and avoid capturing centralisation in our instrument. Results are unchanged and show equivalent results to our main specification albeit with a weakened instrument due to the removal of a key exposure predictor.
As a further check, in column (3) and (4) in Table \ref{tab:state_legacy} we augment our 2SLS specification with 
the log distance to both Abomey and Porto-Novo,
the latter being a important pre-colonial kingdom in the southwest near the coast which might be influencing centralisation, each interacted with indicator variables for 2002 and 2013. The results are again essentially unchanged.

We also address additional threats that may be confounding the outcomes. 
First, is the rolling out of the Plans Fonciers Ruraux (PFR) programme in 2009 by the government of Benin. PFR was a bundle of
interventions formalising customary land rights, with \emph{randomised}
assignment of treated villages, and \citet{wren2020formalizing} show that implementation of this programme reduced
forest loss in Benin. Supplemental Appendix Table~\ref{tab:mig_pfr_combined} shows that controlling for PFR treatment leaves both estimates essentially unchanged (0.182 for cover, $-0.068$ for loss), and instrumented interaction estimates show no differential effect of ATR in PFR arrondissements. Therefore, the forest dynamics due to ATR change do not differ between where land rights were and were not formalised.

Second, between 1991 and 2013 arrondissements may have undergone significant population change through migration. The ATR share of an arrondissement changes through conversion to Pentecostalism but may also change through in- and out-migration. In-migrants are often from the north such as cotton farmers, Fulani pastoralists along with urban spillover, and they are disproportionately Christian or Muslim and often clear land on arrival. Our fixed effects can not account for this, however,  migration simply adds an additional channel through which the religious composition of an arrondissement changes over our sample period. Supplemental Appendix Table~\ref{tab:mig_pfr_combined} shows
this compositional channel is quantitatively small. Adding the
born-here share i.e. the share of the population in an arrondissement that has been living there since birth, as a control moves the estimates only from 0.185 to 0.190 for cover and from $-0.068$ to $-0.072$ for loss; and interacting instrumented ATR adherence with an indicator for above-median residential stability in 1992, a predetermined measure of exposure to migration, yields small and insignificant interactions for both outcomes. The corridor's religious change thus operates through conversion rather than population dynamics.

Third, since REP$_i$ is a function of location, neighbouring arrondissements share exposure and plausibly share shocks, so residuals may be spatially correlated beyond the arrondissement. Clustering at the commune level (77 communes, the tier at which
decentralised land governance operates) raises the standard errors moderately to 0.065 for cover and 0.026 for loss, without altering inference. Supplemental Appendix Figure~\ref{fig:conley} reports \citet{conley1999gmm} standard errors across distance cutoffs from 25 to 200~km. For both outcomes the 90\% confidence interval excludes zero at every cutoff. 

Lastly, we show robustness of our results to alternative choices for constructing the instrument: the distance metric, the treatment of near-zero distances, anchoring exposure on the corridor polylines or on their waypoints, the relative form, and the composition of the Yoruba reference corridor. Supplemental Appendix Table~\ref{tab:spec_curve}
re-estimates the preferred specification with the instrument rebuilt
under each alternative 5 and 10~km distance floors, linear rather
than log distances, distances to the nearest waypoint rather than to
the corridor polyline, absolute Adja proximity ($-\ln
d_{\text{Adja}}$) in place of the relative measure, and dropping each
of the three Yoruba branches in turn. Across all variants the cover
coefficient ranges from 0.137 to 0.247 and the loss coefficient from
$-0.047$ to $-0.082$, with every estimate significant at the one
percent level. Anchoring exposure on the waypoints strengthens the
first stage further (effective $F \approx 142$), consistent with the
sacred geography being concentrated at the historical settlement
sites, while replacing the relative measure with absolute Adja
proximity roughly halves the effective $F$ at similar point estimates,
which is the practical case for the relative form. 
Dropping any single Yoruba branch leaves every estimate within one standard error of the baseline, so no construction choice or individual branch drives the results. 

The last robustness check replaces the ATR share with the Vodun share, which the census records separately from other traditional religions in the 2002 and 2013 waves, yields coefficients of the same sign and significance (0.122 for cover, $-0.056$ for loss, both $p<0.01$). Since the reduced form is common to both, the smaller coefficient implies a steeper first stage for Vodun than for total ATR, which is consistent with our story: the corridor's religious decline is concentrated in the Vodun category at the core of the sacred geography mechanism, rather than in traditional affiliation generically.

\section{Heterogeneity and response to shocks}\label{sox}

Before turning to household-level evidence on mechanisms, we investigate if the aggregate effect simply captures other types of
pressure on land. Ethnic fragmentation undermines the collective action
that forest commons often require. \citet{alesina2019public} show that
deforestation in Indonesia rises with local ethnic diversity as
communities fail to coordinate on conservation, and religious change
along the migration corridor could proxy for such fragmentation. Similarly, adverse climate realisations push farming households to expand cultivated
area, and recent work documents deforestation as a direct response to
climate-related livelihood shocks, so a drought-prone corridor could generate our estimates without any religious channel \citep{ordonez2025crops}. 
Poverty lowers the shadow price of forest relative to immediate consumption, making natural capital the asset of last resort for credit-constrained households 
\citep{ferraro2020conditional},
and low literacy can limit both off-farm alternatives and the reach of
conservation policies. 
If any of these pressures drove our estimates instead of the erosion of religious beliefs, their effect should be concentrated in the arrondissements most exposed to them.

\begin{figure}
    \centering
   \includegraphics[scale=0.65]{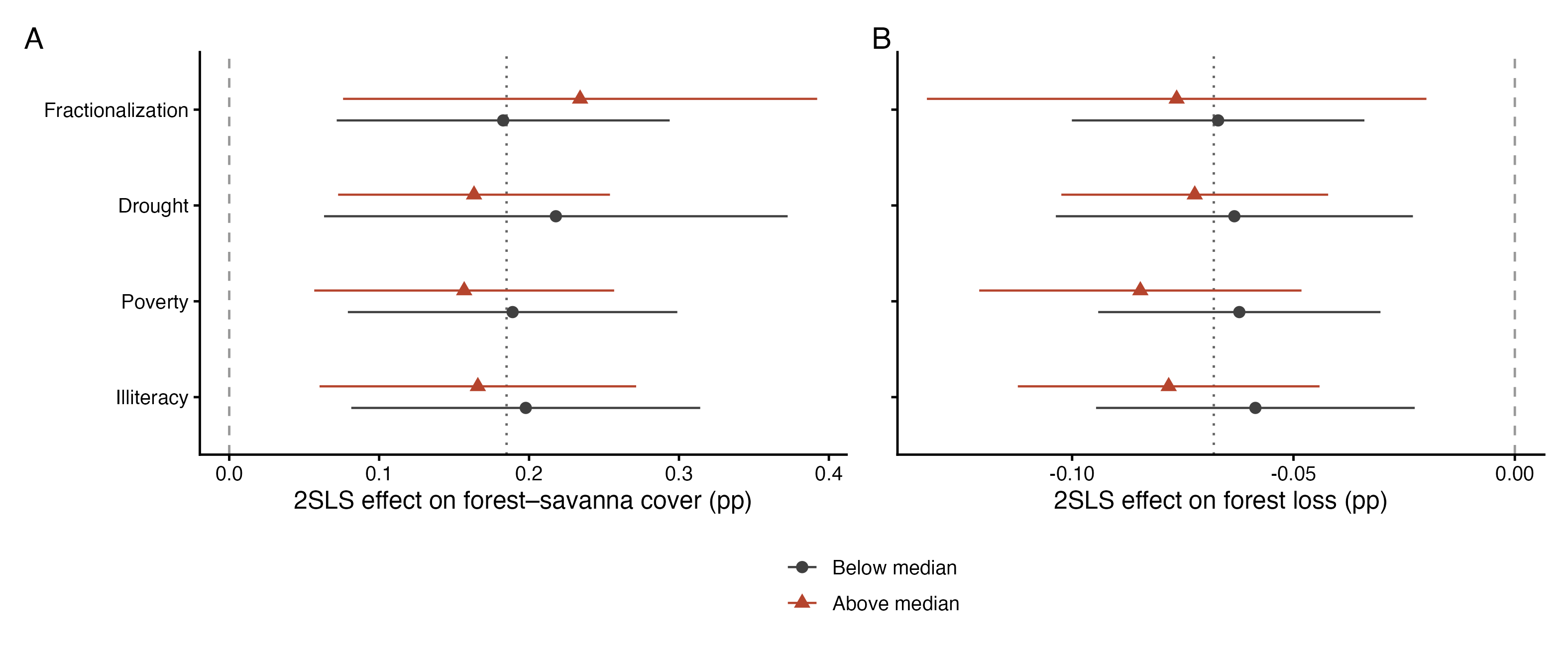}
    \caption{Marginal effect of ATR adherence below and above the median
of each baseline moderator with 95\% confidence intervals. Dotted vertical lines mark the baseline
2SLS estimates (0.185 for cover, $-0.068$ for loss).}
  \label{fig:het_marginal}
\end{figure}

We interact instrumented ATR adherence with indicators for above-median baseline values of four moderators: ethnic fractionalization (1992), drought frequency, poverty (below-median 1992 mean wealth index), and non-literacy.
Every interaction is small and statistically insignificant for both outcomes, and the main effect is stable across all columns (0.18--0.22 for cover, $-0.06$ to $-0.07$ for loss). 
Figure~\ref{fig:het_marginal} plots the marginal effects below and above each median, and we report the full results in Supplemental Appendix Table ~\ref{tab:het_combined}. 
The effect of religious change on deforestation is thus not driven by the corridor's poorer, more drought-prone, less literate, or more fractionalized arrondissements clearing land at different rates.

The mechanism via which ATR drives land use change seems therefore to stem from individual actions and their preferences.
ATR adherence can thus be seen as an individual-level primitive that disciplines resource use. This mechanism should be most visible when individual constraints are more binding and when the shadow cost of not deforesting is the highest, that is when households are subject to adverse shocks. 
Such shocks can be household-specific, such as diseases or family member deaths, or shared, such as droughts and wildfires.
Such a margin is invisible in census aggregates but is directly observable in survey data, and it is where religious beliefs that discipline resource use will be most visible.

We therefore want to test whether ATR-headed households hit by shocks do not liquidate natural capital (forests) as a response, and whether non-ATR households behave differently. The EHCVM dataset described in Section \ref{data_sox} is a perfect fit for our purpose since the survey asks which shocks have been suffered by each household in the previous three years. This information is key for our analysis as it describes \emph{locally relevant} shocks as perceived by the households. The information is self-reported, so the data has the limitation of effectively mixing real losses with reporting propensity. We include cluster-level rainfall data averaged over the same shock recall 3-year periods as a weather control for each survey year, 2016-2018 for the first survey wave and 2019-2021 for the second wave. 

We consider two main categories of shocks, which we define as climate (droughts, floodings, fires, crop/livestock disease) and idiosyncratic shocks, the latter being household-specific (disease/deaths of household members, divorce, sudden end of transfers, bankruptcies, salary losses and thefts). We count for each household and wave the number of shocks reported in each category.
However, the presence of a shock does not necessarily tell us whether the household was worse off. The survey also records, for each shock, whether it caused a loss to agricultural production, to income or assets, or to livestock. From these we build four further counts per household and wave: the number of agricultural losses, of income and asset losses, of livestock losses, and any of these three. Supplemental Appendix Table~\ref{tab:balance_lsms} reports the balance of shocks and losses between households with ATR and non-ATR heads. The table provides us with some key insights. Within survey clusters, both categories of shocks are balanced across households, but losses are not. ATR households report more losses of every type, have roughly half the income and own a smaller share of the plots they farm compared to the non-ATR ones. The poverty-deforestation link should then imply that less secure, harder-hit households should be the ones liquidating natural capital when a shock hits.



We estimate the same specification separately for each of these six measures and for each of the two forest outcomes:
\begin{equation}
    Y_{ct} = \beta_1 \text{S}_{it}+ \beta_2 \mathbbm{1}[ \text{ATR} ]_{it} + \beta_3 (\text{S}_{it}  \times \mathbbm{1}[ \text{ATR} ]_{it}) + \boldsymbol{ \theta} \mathbf{X}_{ict} + \boldsymbol{\gamma} (\mathbf G_{ic} \times \mathbf \tau_t ) + \alpha_t + \phi_{c} + \eta_{ict}, \label{sox2}    
\end{equation}
where $Y_{ct}$ is the ESA forest--savanna cover share or the Hansen forest loss share around survey cluster $c$ in year $t$, as discussed in Section \ref{data_sox}. $S_{it}$ is the number of idiosyncratic shocks, the number of climate shocks, or one of the four loss counts reported by household $i$ for the three years preceding year $t$. $\mathbbm{1}[\text{ATR}]_{it}$ indicates that the head of household $i$ reports ATR in year $t$. $\mathbf{X}_{it}$ is a flexible set of household covariates (household size, education, plot ownership and area, assets, and salary) together with wave-year dummies and a cubic spline in cluster-level rainfall over the same three-year recall window; $\mathbf{G}_c \times \boldsymbol{\tau}_t$ interacts the cluster's latitude and longitude with year indicators; and $\alpha_t$ and $\phi_c$ are year and survey-cluster fixed effects. The coefficient of interest is $\beta_3$ which captures the differential for ATR-headed households on the change in forest outcomes per reported shock or loss. 
Since the rainfall spline absorbs the direct effect of weather on the landscape, $\beta_1$ and $\beta_3$ capture the response of forest dynamics to shocks and losses as \emph{perceived and reported} by households, rather than the effect of climate itself. 
All landscape variables are measured at the cluster centroid, so standard errors are clustered by survey cluster throughout; Conley standard errors with 25 km cutoffs give equivalent results.

\begin{table}[!htb]
\centering
\caption{Household shocks, losses and forest outcomes by ATR adherence}
\label{tab:shocks_full}
\begin{threeparttable}
\small
\setlength{\tabcolsep}{3.5pt}
\renewcommand{\arraystretch}{1.05}
\begin{tabular}{lcccccc}
\toprule
 & \multicolumn{2}{c}{Number of shocks} & \multicolumn{4}{c}{Number of losses caused by shocks} \\
\cmidrule(lr){2-3}\cmidrule(lr){4-7}
 & Idiosyncratic & Climate & Any & Agri- & Income/ & Livestock \\
  & &  &  & cultural & assets & \\
 & (1) & (2) & (3) & (4) & (5) & (6) \\
\midrule
\multicolumn{7}{l}{\textit{Panel A: ESA forest--savanna cover share, 10 km (\%)}} \\[2pt]
Shock / loss $\times$ ATR head ($\beta_3$) & 0.193$^{***}$ & -0.011 & 0.143$^{***}$ & 0.142$^{***}$ & 0.139$^{***}$ & 0.154$^{***}$ \\
 & (0.054) & (0.061) & (0.041) & (0.045) & (0.041) & (0.042) \\
Effect for ATR households ($\beta_1+\beta_3$) & 0.155$^{***}$ & 0.105$^{**}$ & 0.169$^{***}$ & 0.163$^{***}$ & 0.162$^{***}$ & 0.148$^{***}$ \\
 & (0.041) & (0.046) & (0.040) & (0.043) & (0.040) & (0.036) \\
Shock / loss count ($\beta_1$) & -0.037 & 0.116$^{**}$ & 0.026 & 0.020 & 0.023 & -0.006 \\
 & (0.032) & (0.050) & (0.031) & (0.031) & (0.032) & (0.028) \\
ATR head ($\beta_2$) & 0.033 & 0.029 & 0.011 & 0.017 & 0.014 & 0.022 \\
 & (0.045) & (0.045) & (0.046) & (0.046) & (0.046) & (0.045) \\
\addlinespace
Mean of dependent variable & 30.01 & 30.01 & 30.01 & 30.01 & 30.01 & 30.01 \\
$\beta_3$ relative to mean & +0.6\% & -0.0\% & +0.5\% & +0.5\% & +0.5\% & +0.5\% \\
\midrule
\multicolumn{7}{l}{\textit{Panel B: Hansen forest loss over the 3-year recall window, 10 km (\% of land)}} \\[2pt]
Shock / loss $\times$ ATR head ($\beta_3$) & -0.049$^{*}$ & -0.001 & -0.068$^{***}$ & -0.072$^{***}$ & -0.077$^{***}$ & -0.068$^{***}$ \\
 & (0.028) & (0.027) & (0.025) & (0.027) & (0.025) & (0.025) \\
Effect for ATR households ($\beta_1+\beta_3$) & -0.057$^{**}$ & 0.034 & -0.051$^{**}$ & -0.055$^{**}$ & -0.060$^{**}$ & -0.054$^{**}$ \\
 & (0.024) & (0.030) & (0.024) & (0.027) & (0.024) & (0.024) \\
Shock / loss count ($\beta_1$) & -0.008 & 0.035$^{**}$ & 0.017 & 0.017 & 0.017 & 0.014 \\
 & (0.012) & (0.014) & (0.011) & (0.010) & (0.011) & (0.009) \\
ATR head ($\beta_2$) & -0.007 & -0.007 & 0.001 & -0.001 & 0.002 & -0.003 \\
 & (0.025) & (0.025) & (0.025) & (0.025) & (0.025) & (0.025) \\
\addlinespace
Mean of dependent variable & 1.28 & 1.28 & 1.28 & 1.28 & 1.28 & 1.28 \\
$\beta_3$ relative to mean & -3.8\% & -0.1\% & -5.3\% & -5.6\% & -6.0\% & -5.3\% \\
\midrule
Mean shock / loss count & 0.60 & 0.66 & 1.45 & 1.26 & 1.38 & 1.00 \\
Observations & 5,997 & 5,997 & 5,997 & 5,997 & 5,997 & 5,997 \\
\bottomrule
\end{tabular}
\begin{tablenotes}[flushleft]\scriptsize
\item \textit{Notes:} Each column reports OLS estimates of equation~\eqref{sox2} with the column's shock or loss count as $S_{it}$, centered at its sample mean (reported in the penultimate row), so that $\beta_2$ is the ATR versus non-ATR difference for a household reporting the average number of shocks or losses; $\beta_1$ and $\beta_3$ are unaffected by centering. Effect for ATR households is $\beta_1+\beta_3$ with a delta-method standard error. $\beta_3$ relative to mean divides the interaction by the sample mean of the outcome. All columns include survey-cluster and year fixed effects, household controls (household size, share of plots owned, plot area, assets, salary), a natural cubic spline (3 d.f.) in cluster rainfall over the survey year (Panel A) or the recall window (Panel B), wave $\times$ year indicators, and the cluster's latitude and longitude interacted with year. The sample is 6,032 household-years, of which 2 lack a control and 33 are the sole observation in their survey cluster and are absorbed by the cluster fixed effect. Standard errors clustered by survey cluster in parentheses. $^{*}p<0.10$; $^{**}p<0.05$; $^{***}p<0.01$.
\end{tablenotes}
\end{threeparttable}
\end{table}

Table~\ref{tab:shocks_full} reports the results.
For idiosyncratic shocks and for every category of loss, $\hat\beta_3$ is positive and significant for forest cover and negative and significant for forest clearing. We observe that per reported loss, the forest--savanna share around an ATR-headed household is 0.14--0.15 pp higher, and clearing over the recall window is 0.07--0.08 pp lower, than around a non-ATR household in the same cluster. The total response of ATR-headed households, $\beta_1+\beta_3$, is a \emph{reduction} in clearing of 0.05--0.06 pp of land per reported loss, irrespective of whether the loss is to agricultural production, to income and assets, or to livestock.
Households with non-ATR heads, by contrast, do not change their use of the surrounding landscape in response to the shocks they report. The direct effect $\beta_1$ of idiosyncratic shocks and of every category of loss is small and statistically indistinguishable from zero for both outcomes. The exception is climate shocks, which raise subsequent clearing for all households, with $0.035$ pp of land per reported shock over the three-year recall window, about 2.8\% of the 1.28 pp sample mean.\footnote{The direct effect of climate shocks on the forest--savanna share is positive ($\hat\beta_1 = 0.116$, s.e.\ 0.050). Hansen forest loss is a flow of stand-replacing clearing over the recall window, whereas cover is a stock measured in the survey year.} For climate shocks $\hat\beta_3$ is essentially zero for both outcomes indicating that climate shocks raise clearing for all households alike, without any ATR differential. This is consistent with no drought heterogeneity at the arrondissement level, and reflects the fact that climate shocks are highly spatially correlated for households within the same cluster, and individual households
cannot compensate for such a shock.
Finally, at the average number of reported shocks or losses, ATR-headed households and their non-ATR neighbours sit on landscapes with the same cover and the same clearing: $\hat\beta_2$ is below 0.04 pp for cover and below 0.01 pp for loss in every column, none significant. 
As a key placebo test, we estimate \eqref{sox2} with an indicator for a Christian household head, a comparison that mirrors our main identification strategy, and find the opposite results. For Christian households $\hat\beta_3$ is significant and positive for forest clearing and negative, though not significant, for forest--savanna cover, for all categories of losses. We illustrate these results in Supplementary Appendix Figure \ref{fig:christians_coping}.


These results are robust to replacing survey cluster with household fixed effects, where the magnitudes are larger. While the shocks are balanced across both types of households, there maybe concern that ATR households reporting more losses could partly be due to reporting propensity rather than material hardship. We therefore instrument losses with idiosyncratic shock counts and find our results continue to hold. Results are reported in Supplemental Appendix Table \ref{tab:shocks_iv}. We also explore whether ATR-adherent households compensate by means of other coping strategies. The survey records up to three strategies the household adopted for each of the three most severe shocks. Among households reporting at least one idiosyncratic shock, ATR-headed households are no more likely than their neighbours to draw on savings, sell livestock or other assets, pawn land, receive help from relatives or engage in religious activity; they are however more likely to reduce consumption and less likely to expand cultivation into the off-season i.e. less likely to intensify land use. We show these results in Supplementary Appendix Figure \ref{fig:coping_atr}.
The restraint we observe in deforestation is thus because ATR households absorb idiosyncratic shocks and losses rather than liquidating natural capital.

\section{Combining the Sacred and the Ecology}\label{theory}

The evidence we have obtained so far establishes three facts. Where ATR receded, forest-savanna cover fell and clearing accelerated, and this effect is not driven by ethnic fragmentation, drought, poverty or literacy. Second, at the household level ATR adherence appears as a restraint on resource use, with ATR-headed households absorbing idiosyncratic losses without drawing on the forest and without substituting into any other reported coping strategy. 
What these facts cannot tell us is \emph{what} adherents value when they hold back. Two readings are consistent with the evidence. In the first, ATR sacralises an adherent's own environment, a grove, a tree, a stretch of forest where ancestors and spirits reside, and restraint is the protection of that local stock. 
In the second, ATR carries a concern for the state of the forest at large, so that an adherent's welfare depends on how much forest is left standing across the landscape, not only on what they themselves can use. 
Local sacralisation protects patches, whereas a concern for the overall forest can make an entire population's resource use sustainable or not. This section builds a model in which both interpretations can be present. 

The model is deliberately parsimonious in its ingredients. A population of individuals, each of whom we will identify in the data with an arrondissement, extracts from a forest stock that grows naturally and fluctuates randomly. Each individual has a level of ATR adherence between zero and one, drawn from the distribution we observe in the census. Adherence enters preferences in two places. It lowers the utility an adherent draws from consuming their own forest, which is the \emph{sacred} channel, and it raises the weight they place on the median forest cover across the whole population, which is the \emph{ecology} channel. Because the median forest depends on what everyone else extracts, each individual's decision depends on the aggregate behaviour of all others. 
In this equilibrium, a representative individual of each adherence level best-responds to the distribution of forest that the population's own choices generate. 
The model therefore formalises the stock-valuing preferences that the household evidence reveals, calibrated by ATR adherence. 


\subsection{Theoretical framework}

Let us present a framework where $N$ individuals indexed by $i \in [1,\dots,N]$ draw utility continuously in $t\in [0,T]$ from extracting / consuming a quantity $q_{it}$ of the forest cover $X_{it}$ available for individual $i$, assumed to evolve according to the following dynamics:   

\begin{equation}
 dX_{it}= \left ( \mu X_{it}  - q_{it} \right ) dt + \sigma X_{it}  d W_{it},
 \label{gbm}
\end{equation}
where every $i$-th $W_{it}$ is an independent Brownian motion.
The ecological parameters $\mu$ and $\sigma$ are respectively the growth rate of the forest cover and the standard deviation of the increments of each Brownian motion\footnote{We assume the technical requirements of a triple $(\mathbb{R}^+, P, \mathcal{F}_t)$ comprised by the positive real domain - forest cover cannot be negative - a probability measure $P$ and a filtration (an information set) $\mathcal{F}_t$. We could extend the model, adding some complexity, to one in which each individual has their own $\mu_i$ and $\sigma_i$. However, such a framework would not allow us to identify $\mu_i$ from the Benin data we have, a point we will discuss at length in the following section, and thus we leave such a model for future extensions.}.
We retain the same index $i$ for individuals that we have used in the previous sections to identify Benin arrondissements, since in estimating the model parameters we will associate an ``individual'' as a representative agent of each arrondissement in the data. 

Individuals maximize the following objective function:


\begin{equation}
J^i(X_{it},q_{it},t) = \mathbb{E}_t\int_t^T e^{-\rho s}  u( q_{is},    p^a(x,s) ; a_i )ds + b(X_{iT}), \label{jcrit}
\end{equation} 
where $a_i \in A \subseteq [0,1], a_i \sim p_0(a)$ is the individual ATR adherence, normalized  between 0 (``not believing at all'') and 1 (``fully adherent''). This term parametrises the role played by heterogeneous ATR beliefs, defined within a bounded set $A$ and is distributed across the population according to a measure $p_0(a)$, assumed to be known by the individuals. 
The utility function includes the spatial (``cross-sectional'')  empirical distribution of the resource \emph{across} individuals $p(x, t)$ where $x = [ X_{1t}, \dots, X_{Nt}] $, defined as $p(dx,t)=  \frac{1}{N} \sum_{i=1}^N \delta_{X_{it}}(dx)$, where $\delta_x$ is a unit mass function at each $X_{it}$. 
This distribution maps the state of the resource for all the $N$ individuals at any time $t$  into a real-valued utility quantity, 
a signal that individuals are able to observe which captures all available information on the overall state of the resource. 
This term represents the utility individuals draw due to the ecology, due to \emph{pro-environmental attitudes}\footnote{We note that the model setup is general, in which $x$ can be any renewable resource and $a$ can represent a wide variety of preferences over the overall resource distribution $p^a(x,t)$. For the purpose of this paper and to understand the implications within the context of religion and ecology we restrict our attention to $a_i$ as individual ATR adherence, $X_{it}$ as the forest area managed by individual $i$ at time $t$ and $p^a(x,t)$ is the (empirical) forest cover distribution. }. 
The distribution captures the interactions across all individuals, since it is continuously determined for each $t$ by everyone's resource use policies, and hence indirectly dependent on the distribution of $a$. Each adherence level $a_i$ calibrates how both $q$ and $p$ generate utility for each individual $i$.
We consider utility functions $u(q, p ; a)$ such that $u_q\geq 0, u_{qq}\leq 0$. 
    Lastly, the function $b$ is a general bequest function, for simplicity assumed independent of $a_i$. 

We now assume that the signal that individuals observe to gauge the overall state of the forest cover is the \emph{geometric mean} of the forest distribution. 
This quantity is formalised naturally by 
$\exp\mathbb{E}^p [ \log X_{it}^{q}] := \bar{X_t}$, 
where the expectation is with respect to the joint measure $p:= p^a(x,t)$ of forest cover and ATR beliefs induced by $X_{it}^q$ (the forest cover $X_{it}$ associated with any feasible
extraction policy $q$) . We thus augment our information set to include the information available to all individuals at time 0 over the initial state of both forest and adherence\footnote{In more technical terms, we augment the filtration $\mathcal{F}_t$ to $\mathcal{F}_t^a$, which includes all information on ATR beliefs across the population.}. Note that for a discrete number $j=1,\dots,n$ of observations $x_j$, $p^a(x,t)$ is the empirical measure associated to the realizations $x_j$ and $\bar{X}_t$ becomes the familiar expression for the geometric mean $(\Pi_j x_j)^{1/n}$. Individuals, therefore, observe the \emph{median} forest cover across the population. 
The choice of the geometric mean is one of convenience, as one can derive explicitly the dynamics of the median forest cover and characterize precisely the model solution. We now assume $u_{\bar{x}} \geq 0, u_{\bar{x}\bar{x}}\leq 0$. 
In Appendix Section \ref{prob_setup} we show all the formal details, where we also show how the median state of the forest depends not only on the ecological parameters $\mu$ and $\sigma$, but also on the heterogeneity in beliefs within the population.

The presence of $\bar{X}_t$ in the utility function of each individual represents the role that the state of the environment, and in particular its \emph{scarcity}, plays in determining pro-environmental attitudes, and ultimately shaping their use policies. The term shows how populations may be concerned about a collapsing ecosystem, a consequence of global, rather than individual, actions. 
What matters for individual welfare is the overall state of forests, beyond what is available for one to exploit, as decreasing aggregate levels of forest reduce individual utility. 
The term is endogenous as it depends on the probability measure $p^a(x,t)$ which identifies the spatial forest distribution at any time $t$. 

With our reduced form results we have uncovered signs of a substitutability mechanism at play, such that deforestation is at odds with ATR adherents' value associated to nature. 
In our framework, we will therefore formalize ATR beliefs as the way adherents adjust the utility drawn from using their available forest stock $q_{it}$ and the existence value of the forest at large $\bar{X}_t$.  We will assume the scarcity term to enter the utility function as $\bar{X}_t^{g_2(a)}$ and the forest consumption term as $q^{g_1(a)}$, where we impose $g'_1  \leq 0,  g'_2 \geq 0 $.\footnote{We note that by using $\bar{X}_t$ as the signal of the global state of the environment observed by individuals we implicitly choose not to include peer effects in our framework, since pro-environmental attitudes involve drawing utility from observing an aggregate state, rather than observing an aggregate quantity which is chosen by other individuals.  
In other words, individuals do not react directly to what others \emph{do} (cutting trees), but rather observe a wider outcome.
We however note that including peer effects is something that would be entirely possible using our framework, for example by using $1/\bar{q}_t$ (i.e. the median deforestation policy) instead of $\bar{X}_t$. We leave this aspect to potential future extensions.} In our framework, individual $i$ draws utility from a composite good given by $q_{it}^{g_1(a_i)} \bar{X}_t^{g_2(a_i)}$, which is expressed in forest cover units and is a combination of individual consumption $q_{it}$ and overall median
forest state $\bar{X}_t$. This combination is calibrated by one's own individual adherence $a_i$.

The intuition of this formalisation is as follows: $g_1$ (the sacred) determines how ATR adherence affects individual deforestation utility, which defines the value individuals give to \emph{their} forest cover. 
This local environment can be specific trees or groves considered sacred, where spirits may lie. The function $g_1$  is decreasing in ATR adherence, so the more individuals are adherent, the less utility they will draw from consuming forest products from their own local stock. The pro-environmental attitudes are calibrated by $g_2$ (the ecology), which controls how individuals care about the \emph{global} environment, rather than individual / specific parts of forest, via its scarcity.  
Individual ATR adherence levels $a_i$, therefore, calibrate the \emph{degree of substitutability} between individual consumption $q$ and the state of the resource across all individuals. 
If $g_1 + g_2 = 1 , \forall a \in A$, then individual resource use and pro-environmental attitudes are perfect substitutes.\footnote{We do not mandate that local forest consumption and global forest scarcity should be perfect substitutes: all we require is $g_1' \leq 0 , g_2' \geq 0$, both continuous.}

We now assume a continuum of individuals, which is equivalent to assuming a large number $N$ of individuals, and yields an instance of a mean-field game.
It has been proven \citep{lasry2007mean} how under fairly general assumptions the infinite-population / continuum limit can effectively approximate the equilibrium of a finite-$N$ game, and the system can be fully characterized by a single representative agent interacting with a distribution of states (the mean field). In short, the large $N$ limit allows us to switch from solving for \emph{all} deforestation interactions $q_{it}$ in the formation of $\bar{X}_t$, to the interaction between the deforestation  $q_{t}$ by a representative individual with adherence $a$, the endogenous spatial forest distribution $p^a(x,t)$ and the adherence distribution $p_0(a)$ across the population.\footnote{On the technical side, the large-$N$ limit problem is now equipped with the augmented filtration $\mathcal{F}^a_t$ which now has to be set as the smallest filtration such that the beliefs ``type'' vector $[a_i]$ drawn from $a \sim p_0(a)$ is $\mathcal{F}_0$-measurable and to which every Brownian motion $W_{it}$ remains adapted.}

The equilibrium deforestation policy $q^{MFE}_t$ is obtained by solving the following problem:
\begin{align}
q_t^{MFE} \to \sup_{q\in Q} \ \ &  \mathbb{E}_{t} \int_t^T e^{-\rho s}  u \bigg( q^{g_1(a)} \bar{X}_t^{g_2(a)} \bigg ) ds + b(X_T) \label{max} \\
    & dX _t =  \left ( \mu X _t - q \right ) dt + \sigma X _t d W_t, \quad X_0 = x_0 \nonumber \\
     & \bar{X}_t = \exp \int_{[0,\infty)} \int_A \log x \ p^a( dx,t) p_0(da).  \label{coupl_prob}
\end{align}
%
For all details on deriving the setup of the problem \eqref{max} we refer to the Supplemental Appendix Section \ref{prob_setup}.
In the large $N$ limit, the problem  \eqref{max} can be reduced to solving a system of two coupled partial differential equations instead of $N$ (one for each $i$). 
The key feature of this system is the forward/backward dimension.
The optimality equation
of the representative individual (also known as a mean field) is solved backwards starting from an endpoint condition.
The relevant information over the actions of all agents is captured over the observation of $p^a(x,t)$, which represents the actual dynamics of the resource implied by individual behaviours. This information is then internalized via the pro-environmental attitudes $\bar{X}_t$ in \eqref{coupl_prob} and calibrated by the effect of adherence $g_2(a)$, yielding the optimal deforestation policy $q_t^{MFE}$.
Forest consumption thus directly affects the overall forest resources in $dt$ via the agents' heterogeneity $p_0(a)$ and consequently the spatial distribution of forest cover $p^a(x,dt)$.
The spatial distribution therefore keeps being updated forward via the probability evolution equation,
starting from an initial resource distribution $p^a(x,0)$, and agents form their forward-looking expectations based on their anticipation of what $p^a(dx,dt)$ will be like.
Lastly, rational expectations imply that for all agents there must be self-consistency between the anticipated $p^a$ and the actual (realized) $p^a$, and thus we will have to prove the existence of a fixed point. 

Let us now present the following general (i.e. independent of the choice of utility function) result, showing how for any $t \in [0,T]$  any optimal deforestation policy $q^{*}_t  $ in equilibrium changes with ATR adherence according to the following equation:

\begin{equation}
  \partial_a q^{*}_t  = \frac{1}{ u_{qq} } \mathbb{E}_t^P \int_t^T e^{-\int_t^s (\rho - \mu  + \partial_x q^{*}_v ) dv} \left (  D_s \partial_a u_{q \bar{x}} - \partial_a u_x    \right )ds \label{geneul}
\end{equation}
where the $\mathcal{F}^a_t$-measurable process $D_t = (\mu - \sigma^2/2) \bar{X}_t - \int \bar{q}^{*}_t p(da) $ is the instantaneous median forest change, net of median deforestation. 
The expectations are calculated with respect to the measure $P$, which is induced by the distribution $p^a(x,t)$ of all individual forest cover processes $X_{it}$ in equilibrium. The derivation of this Euler-type equation is based on a differentiation of the envelope of the associated Hamilton-Jacobi-Bellman equation (i.e. at the first-order condition), some straightforward derivations
(at the supremum envelope $\partial_a V_x = u_{qq} \partial_a q$, where $V$ is the value function of our problem)
and an application of the Feynman-Kac formula. 

Equation \eqref{geneul} shows how deforestation is determined by the interplay of three forces: individual adherence to ATR, the marginal value of one unit of forest cover for personal use, and the utility the individual assigns to the overall state of the environment via the spatial forest distribution. 
Furthermore, it shows how the change in deforestation due to a marginal increase in ATR is the sum of two effects discounted over time. One is always negative, and depends on each agent's  sensitivity to ATR of the marginal utility of forest cover. This is intuitive: the more valuable for an agent is the stock - not the consumption - of forest left standing, and the more this quantity is sensitive to ATR adherence, the more the agent will reduce deforestation. 
This insight maps directly to our reduced form results, and specifically to our shock coping mechanisms, validating our preference structure. The imposed signs of $g_1, g_2$ and their effect on $u_x$ and $u_{q \bar{x}}$ in \eqref{geneul} identify the effect of ATR adherence as a state-contingent ``restraint'' (i.e. a reduction in deforestation). In particular, the idiosyncratic shock difference shown in Table \ref{tab:shocks_full} is observed exactly at the margin where this mechanism is expected to be most visible. Such a behaviour can emerge only from stock-valuing preferences such as those posited by \eqref{jcrit} and \eqref{max}. 
The second term depends on the \emph{change} in median deforestation $D_t$. An increase in global deforestation reinforces the overall negative effect of ATR adherence on individual deforestation via the pro-environmental attitudes, weighed by the sensitivity to ATR of the mixed second partial derivative of forest consumption and scarcity, which is the key indicator of substitutability between the two. 
The slope of each individual deforestation policy $\partial_x q$ determines the (stochastic) discount factor which makes the present value of lifetime marginal utility of forest consumption equal its future expectation (i.e. a martingale).
\\

We now assume functional forms such that we can obtain explicit solutions. 
We choose a CRRA utility function of the form $u(z) = \frac{ z^{1-\gamma}}{1-\gamma}$, with $\gamma > 0$.
We impose $\rho > \mu (1-\gamma) - \frac{\sigma^2}{2} \gamma (1 - \gamma)$, a standard assumption that allows the problem to be well-defined at the baseline $g_1 = 1, g_2 = 0$. We then can state the following Proposition: 
\\

\textbf{Proposition 1}. \emph{An equilibrium exists for all $T$. For $T \to \infty$, the equilibrium deforestation rate $q^{MFE} := q^* (a) / X_t$  for the problem \eqref{max} is given for every $a \sim p_0(a)$ by}

\begin{eqnarray}
q^{MFE} &=& \epsilon_q  \left ( \rho - \mu ( \nu_q + \nu_{\bar{x}} ) - \frac{\sigma^2}{2} \left (  \frac{\nu_q}{\epsilon_q} + \nu_{\bar{x}} \right ) + \nu_{\bar{x}} \tilde{q}^* \right ) \label{qmfe} \\
  \tilde{q}^*  &=& \left ( \rho \overline{\epsilon}_q  - \mu \overline{(1- \epsilon_q})   - \frac{\sigma^2}{2} (1-\gamma) \overline{g}_1  - \left (\mu - \frac{\sigma^2}{2} \right )  \overline{ \epsilon_q \nu_{\bar{x}}}  \right ) \left ( 1 - \overline{\epsilon_q \nu_{\bar{x}} }) \right )^{-1}  \nonumber
\end{eqnarray}
\emph{where $\overline{\omega } = \int_A \omega p_0(da)$ for any $\omega$, $ \epsilon_q  := \epsilon_q (a)= - \frac{u_q }{ u_{qq}}\frac{1}{q}  $ is the elasticity of intertemporal substitution for forest use, $\nu_q :=\nu_q(a) =q \frac{u_q }{u}  $ is the (point) elasticity of utility with respect to forest consumption $q$, and similarly  $\nu_{\bar{x}}:= \nu_{\bar{x}}(a) = \bar{x} \frac{u_{\bar{x}}}{u} $ is the elasticity of utility with respect to the median forest cover $\bar{x}$}. 

\textbf{Proof:} See Appendix Section \ref{prop1}, which also shows the solution for the finitely-lived case, proves the existence of an equilibrium for all $t$ and discusses the finite-time evolution of the equilibrium spatial forest distribution. Note that all the expectations with respect to $p_0(da)$ exist and are bounded for all continuous distributions, since $\epsilon_q (0)$ is bounded and $a \in [0,1]$.

We can now state the key characteristics of the equilibrium deforestation policy, which allow us to better map the model to our reduced form results. 
We define as \emph{unsustainable deforestation} a deforestation policy such that $q^{MFE} > \mu - \frac{\sigma^2}{2}$ for some $a \in A$, since in each $dt$ individuals with adherence $a$ consume an amount of forest resource which is greater than its ``natural'' growth rate $\mu - \frac{\sigma^2}{2}$, and thus leading to an almost sure (in the probability sense) forest depletion. Conversely, we define as sustainable deforestation a policy such that $q^{MFE} < \mu - \frac{\sigma^2}{2}$ for all $a$. We can now state the following proposition, which summarises the key characteristics of the optimal equilibrium deforestation policy, and will help map the model to the patterns that we observe in the data.
\\           

\textbf{Proposition 2:} \emph{ (i) Local deforestation increases with decreasing individual ATR adherence and with decreasing global deforestation. 
(ii) Decreasing global ATR adherence yields increasing global deforestation rates and decreased deforestation rates for high adherence individuals.
(iii) Unsustainable deforestation policies can occur in low ATR adherence areas, and can switch to a sustainable policy as the population ATR adherence increases only in the presence of pro-environmental attitudes.}
\\

\textbf{Proof:} See Appendix Section \ref{prop2}.
\\

First, in equilibrium the extraction rate of any individual is a known function of their adherence and of the population's adherence distribution, and it falls with adherence. More adherent individuals clear less, as in the reduced form. 
The intuition behind (ii), which might seem counterintuitive, is that  when adherence declines in the population, aggregate forest clearing rises because there are fewer restrained individuals, while those who remain adherent restrain further in response to the thinning forest they observe. 
This feedback reconciles ours arrondissement-level results, where adherence loss raises clearing, with the household-level results, where adherents hold back most when the landscape around them is under pressure. 
Lastly, (iii) yields the key insight that cannot be obtained with the reduced form alone, as is that whether a population's deforestation is sustainable, in the sense of staying below the forest's natural growth rate, depends on adherence only through the ecology channel, if adherents indeed harbour pro-environmental attitudes. With the sacred channel alone, adherence changes the level of forest clearing but can never change an unsustainable deforestation activity into a sustainable one. This implication provides a useful prediction to be mapped to data.

\subsection{Zero-adherence counterfactuals and sustainable deforestation}

We bring the model to the Benin data and estimate its primitives. 
We assume infinitely-lived agents, as having a fully explicit solution of the problem simplifies identification and estimation considerably. Furthermore, the dataset on ATR beliefs does not have information over the individuals' lifecycle, and thus the time-invariant nature of the optimal deforestation rate $q^{MFE}/X_t$ given by \eqref{qmfe} is better suited for our estimation requirements. 
Since we have information on ATR adherence only at the arrondissement level, we associate each arrondissement to an agent $i$ in our model, and assume that the reported ATR adherence is representative of all individuals in the arrondissement. This choice directly maps to the model primitives $g_1$ and $g_2$, as in our data ATR adherence is the percentage of adherents and thus naturally falls between 0 and 1 as is the case for the functions $g_1$, $g_2$. The model heterogeneity in agents’ beliefs, therefore, is mapped directly to the observed heterogeneity across Benin arrondissements.

For our model estimations we use tree canopy cover area (TA) as our $X_{it}$, expressed in square kilometers. We choose TA because forest cover is a specific, much more narrow land cover type, with a lot more zeroes than tree canopy cover. More importantly, we want to look at trees in general since ATR sacralize trees as much as forests, as discussed in Section \ref{back: atr}. We therefore use NASA's Making Earth System Data Records for Use in Research Environments (MEaSUREs) Vegetation Continuous Fields Version 1 data product (VCF5KYR) created by \cite{song2018global}. 
The natural growth rate of the forest $\mu$ is estimated using data from the W-Arly-Pendjari complex, which is a protected area and thus deforestation is negligible. Using the 2013 census data, we obtain the predicted arrondissement-level TA via the closed-form solution of the dynamic equations \eqref{gbm} with each optimal policy $q^{MFE}$ evaluated at the estimated parameters shown in Table \ref{table: struc_est}, heterogeneous over the distribution $p_0(a)$. 
Section \ref{struc} in the Supplemental Appendix brings evidence of how the data supports a preference structure such that $g_1 = 1$ and $g_2 = a$, in other words, where the ecology dominates over the sacred. The section also presents in detail the different estimation methods we employ, the data and moment conditions used, our parametric identification challenges and how we overcome them, as well as the model estimates.

\begin{figure}
    \centering
    \setkeys{Gin}{width=\linewidth}
\begin{subfigure}{0.45\textwidth}
\includegraphics{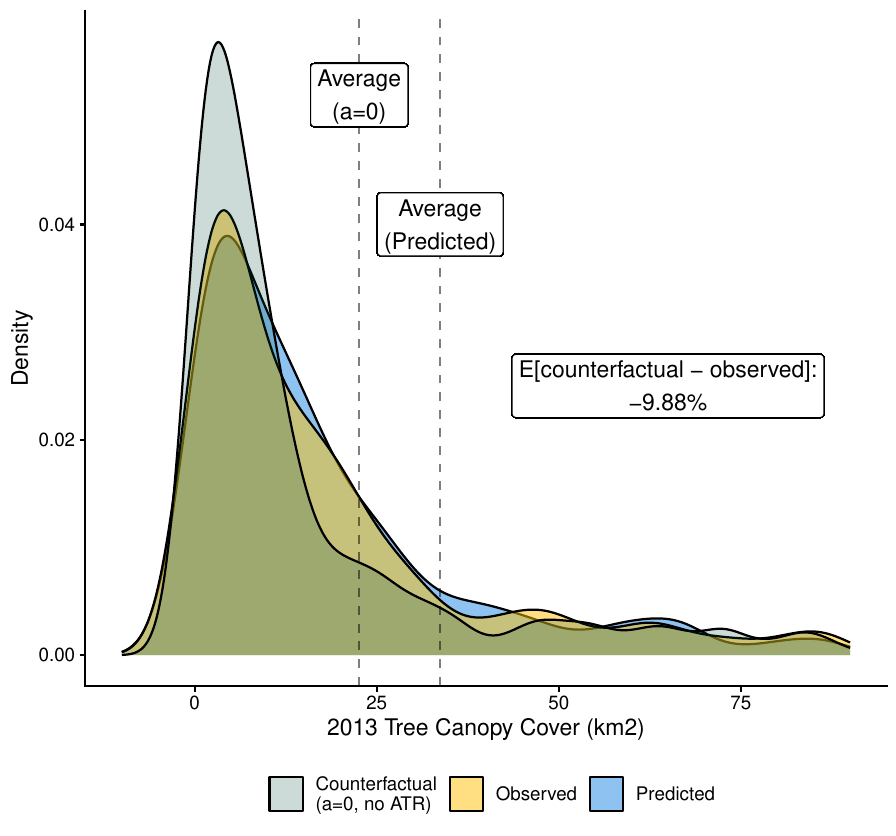}
\caption{}
\label{fig:capparatus}
\end{subfigure}
\hfil
\begin{subfigure}{0.5\textwidth}
\includegraphics{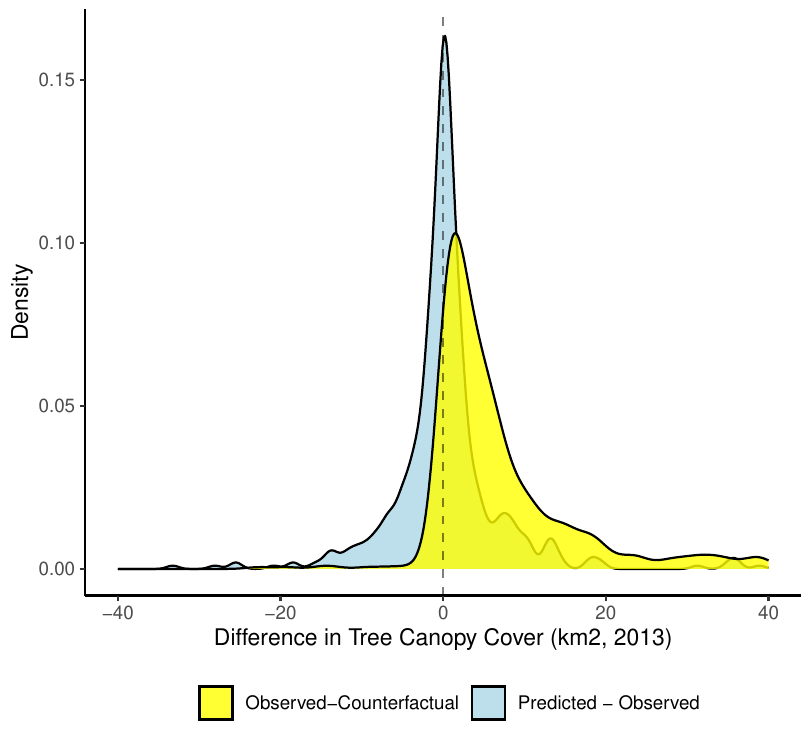}
\caption{}
\label{fig:cdiagram}
\end{subfigure}
\caption{Panel (a): Counterfactual (no ATR), observed and predicted 2013 Benin tree canopy densities. Panel (b): difference between observed and predicted, and observed and counterfactual 2013 tree canopies.}
  \label{fig:spatx_figures}
    \end{figure}

Equipped with the parameter estimates, we obtain the counterfactual canopy area distribution resulting from setting all adherence levels uniformly to zero. This distribution is obtained by setting $a=0$ in $q^{MFE}$, removing all heterogeneity in beliefs and obtaining a counterfactual no-ATR deforestation rate $\hat{q}^{MFE}_0$.
We then predict the trajectories of each dynamic equation $d X_{it}$ between 2002 and 2013 using the no-ATR deforestation rates. We then map the three distributions, counterfactual (no ATR), observed and predicted for Benin in 2013\footnote{We map these spatial distributions in Supplemental Appendix Figure \ref{fig:spatdist}.}. The left panel of Figure \ref{fig:spatx_figures} shows how the observed and predicted TA distribution match closely, and how the counterfactual distribution is clearly first-order dominated by the other two. 
The right panel of Figure \ref{fig:spatx_figures} further clarifies this point, showing the density of the difference between observed and predicted TA to be centered on zero (mean: 0.02, median: 0.082, std. dev: 10.68)

The average difference between observed and counterfactual 2013 TA is -3.32 km$^2$ per arrondissement, which corresponds to a canopy loss of 9.88\%. We find the largest counterfactual tree canopy losses in the southern departments of Mono and Kouffo, which is consistent with these departments exhibiting the highest numbers of ATR adherents. We conclude with an interesting observation: evaluated at our estimated parameters, the counterfactual no-ATR deforestation rate is not sustainable, as $\hat{q}^{MFE}_0 = 0.0451$, a yearly deforestation rate of 4.51\%.  
This rate implies an ``effective'' growth rate of \eqref{gbm} is negative\footnote{We refer to the parameter estimates in the Supplemental Appendix Table \ref{table: struc_est}, which shows how $\hat{q}^{MFE}_0 dt  > (\hat{\mu} - \hat{\sigma}^2/2) dt$ at the estimated parameters.}, which means that in absence of any ATR beliefs, the forest resources in Benin will eventually be depleted. On the other hand, there exist departments in Benin that show a sustainable (estimated) deforestation policy, all of which are high-adherence areas and are concentrated in the southwestern department of Kouffo, overlapping with the historical Adja migration corridor, and in the north-western department of Atakora, bordering the WAP protected areas.
This result is consistent with Proposition 2: the fact that deforestation switches from unsustainable to sustainable rates for arrondissements above a certain adherence threshold shows pro-environmental attitudes, the ecology, can drive the forest use policies of the adherent population.

\bibliographystyle{chicago}
\bibliography{sample}

@article{song2018global,
  title={Global land change from 1982 to 2016},
  author={Song, Xiao-Peng and Hansen, Matthew C and Stehman, Stephen V and Potapov, Peter V and Tyukavina, Alexandra and Vermote, Eric F and Townshend, John R},
  journal={Nature},
  volume={560},
  number={7720},
  pages={639--643},
  year={2018},
  publisher={Nature Publishing Group}
}

@article{hansen2013high,
  title={High-resolution global maps of 21st-century forest cover change},
  author={Hansen, Matthew C and Potapov, Peter V and Moore, Rebecca and Hancher, Matt and Turubanova, Svetlana A and Tyukavina, Alexandra and Thau, David and Stehman, Stephen V and Goetz, Scott J and Loveland, Thomas R and others},
  journal={science},
  volume={342},
  number={6160},
  pages={850--853},
  year={2013},
  publisher={American Association for the Advancement of Science}
}

@article{hyman2021pastoralists,
  title={How pastoralists weight future environmental benefits when managing natural resources},
  author={Hyman, Amanda A and Gaoue, Orou G and Tamou, Charles and Armsworth, Paul R},
  journal={Conservation Letters},
  volume={14},
  number={2},
  pages={e12770},
  year={2021},
  publisher={Wiley Online Library}
}

@article{lasry2007mean,
  title={Mean field games},
  author={Lasry, Jean-Michel and Lions, Pierre-Louis},
  journal={Japanese journal of mathematics},
  volume={2},
  number={1},
  pages={229--260},
  year={2007},
  publisher={Springer}
}

@article{gabaix2016dynamics,
  title={The dynamics of inequality},
  author={Gabaix, Xavier and Lasry, Jean-Michel and Lions, Pierre-Louis and Moll, Benjamin},
  journal={Econometrica},
  volume={84},
  number={6},
  pages={2071--2111},
  year={2016},
  publisher={Wiley Online Library}
}

@article{alesina2003fractionalization,
  title={Fractionalization},
  author={Alesina, Alberto and Devleeschauwer, Arnaud and Easterly, William and Kurlat, Sergio and Wacziarg, Romain},
  journal={Journal of Economic growth},
  volume={8},
  number={2},
  pages={155--194},
  year={2003},
  publisher={Springer}
}

@article{wren2020formalizing,
  title={Formalizing land rights can reduce forest loss: Experimental evidence from Benin},
  author={Wren-Lewis, Liam and Becerra-Valbuena, Luis and Houngbedji, Kenneth},
  journal={Science Advances},
  volume={6},
  number={26},
  pages={eabb6914},
  year={2020},
  publisher={American Association for the Advancement of Science}
}

@article{stoop2019voodoo,
  title={Voodoo, vaccines, and bed nets},
  author={Stoop, Nik and Verpoorten, Marijke and Deconinck, Koen},
  journal={Economic Development and Cultural Change},
  volume={67},
  number={3},
  pages={493--535},
  year={2019},
  publisher={The University of Chicago Press Chicago, IL}
}

@book{claffey2007christian,
  title={Christian churches in Dahomey-Benin: a study of their socio-political role},
  author={Claffey, Patrick},
  year={2007},
  publisher={Brill}
}

@book{sheridan2008african,
  title={African sacred groves: ecological dynamics \& social change},
  author={Sheridan, Michael J and Nyamweru, Celia},
  year={2008},
  publisher={Oxford: James Currey},
  note = {Last accessed 1 October 2026: \url{https://archive.org/details/africansacredgro0000unse_t7s7}}
}

@article{alonso2016voodoo,
  title={Voodoo versus fishing committees: The role of traditional and contemporary institutions in fisheries management},
  author={Alonso, Elena Briones and Houssa, Romain and Verpoorten, Marijke},
  journal={Ecological Economics},
  volume={122},
  pages={61--70},
  year={2016},
  publisher={Elsevier}
}

@article{reichel1976cosmology,
  title={Cosmology as ecological analysis: a view from the rain forest},
  author={Reichel-Dolmatoff, Gerardo},
  journal={Man},
  pages={307--318},
  year={1976},
  publisher={JSTOR}
}

@misc{aderibigbe2022palgrave,
  title={The Palgrave Handbook of African Traditional Religion},
  author={Aderibigbe, Ibigbolade S and Falola, Toyin},
  year={2022},
  publisher={Springer}
}

@book{manning2004slavery,
  title={Slavery, colonialism and economic growth in Dahomey, 1640-1960},
  author={Manning, Patrick},
  number={30},
  year={2004},
  publisher={Cambridge University Press}
}

@article{guiso2006does,
  title={Does culture affect economic outcomes?},
  author={Guiso, Luigi and Sapienza, Paola and Zingales, Luigi},
  journal={Journal of Economic perspectives},
  volume={20},
  number={2},
  pages={23--48},
  year={2006}
}

@article{alesina2019public,
  title={Public goods and ethnic diversity: Evidence from deforestation in Indonesia},
  author={Alesina, Alberto and Gennaioli, Caterina and Lovo, Stefania},
  journal={Economica},
  volume={86},
  number={341},
  pages={32--66},
  year={2019},
  publisher={Wiley Online Library}
}

@article{iyer2016new,
  title={The new economics of religion},
  author={Iyer, Sriya},
  journal={Journal of Economic Literature},
  volume={54},
  number={2},
  pages={395--441},
  year={2016}
}

@book{mbiti1990african,
  title={African religions \& philosophy},
  author={Mbiti, John S},
  year={1990},
  publisher={Heinemann}
}

@article{kahn2011policing,
  title={Policing ‘Evil’: state-sponsored witch-hunting in the People’s Republic of B{\'e}nin},
  author={Kahn, Jeffrey},
  journal={Journal of Religion in Africa},
  volume={41},
  number={1},
  pages={4--34},
  year={2011},
  publisher={Brill}
}

@article{tall1995democratie,
  title={De la d{\'e}mocratie et des cultes voduns au B{\'e}nin (On Democracy and Voodoo in Benin)},
  author={Tall, Emmanuelle Kadya},
  journal={Cahiers d'{\'e}tudes africaines},
  pages={195--208},
  year={1995},
  publisher={JSTOR}
}

@book{juhe2010forets,
  title={For{\^e}ts sacr{\'e}es et sanctuaires bois{\'e}s: des cr{\'e}ations culturelles et biologiques (Burkina Faso, Togo, B{\'e}nin)},
  author={Juh{\'e}-Beaulaton, Dominique},
  year={2010},
  publisher={KARTHALA Editions}
}

@article{nunn2011slave,
  title={The slave trade and the origins of mistrust in Africa},
  author={Nunn, Nathan and Wantchekon, Leonard},
  journal={American Economic Review},
  volume={101},
  number={7},
  pages={3221--52},
  year={2011}
}

@article{conley1999gmm,
  title={GMM estimation with cross sectional dependence},
  author={Conley, Timothy G},
  journal={Journal of econometrics},
  volume={92},
  number={1},
  pages={1--45},
  year={1999},
  publisher={Elsevier}
}

@book{falen2018african,
  title={African science: Witchcraft, vodun, and healing in Southern Benin},
  author={Falen, Douglas J},
  year={2018},
  publisher={University of Wisconsin Press}
}

@article{gershman2016witchcraft,
  title={Witchcraft beliefs and the erosion of social capital: Evidence from Sub-Saharan Africa and beyond},
  author={Gershman, Boris},
  journal={Journal of Development Economics},
  volume={120},
  pages={182--208},
  year={2016},
  publisher={Elsevier}
}

@book{berkes2017sacred,
  title={Sacred ecology},
  author={Berkes, Fikret},
  year={2017},
  publisher={Routledge}
}

@article{owen2007culture,
  title={Culture and public goods: The case of religion and the voluntary provision of environmental quality},
  author={Owen, Ann L and Videras, Julio R},
  journal={Journal of Environmental Economics and Management},
  volume={54},
  number={2},
  pages={162--180},
  year={2007},
  publisher={Elsevier}
}

@article{alesina1999public,
  title={Public goods and ethnic divisions},
  author={Alesina, Alberto and Baqir, Reza and Easterly, William},
  journal={The Quarterly journal of economics},
  volume={114},
  number={4},
  pages={1243--1284},
  year={1999},
  publisher={MIT Press}
}

@incollection{iannaccone2009economics,
  title={Economics of religion},
  author={Iannaccone, Laurence R and Bainbridge, William Sims},
  booktitle={The Routledge Companion to the Study of Religion},
  pages={475--489},
  year={2009},
  publisher={Routledge}
}

@article{alidou2019only,
  title={Only women can whisper to gods: Voodoo, menopause and women’s autonomy},
  author={Alidou, Sahawal and Verpoorten, Marijke},
  journal={World Development},
  volume={119},
  pages={40--54},
  year={2019},
  publisher={Elsevier}
}

@article{alidou2021beliefs,
  title={Beliefs and investment in child human capital: case study from Benin},
  author={Alidou, Sahawal},
  journal={The Journal of Development Studies},
  volume={57},
  number={1},
  pages={88--105},
  year={2021},
  publisher={Taylor \& Francis}
}

@article{alesina2023religion,
  title={Religion and educational mobility in Africa},
  author={Alesina, Alberto and Hohmann, Sebastian and Michalopoulos, Stelios and Papaioannou, Elias},
  journal={Nature},
  volume={618},
  number={7963},
  pages={134--143},
  year={2023},
  publisher={Nature Publishing Group UK London}
}

@article{ciscato2024astrology,
  title={Astrology and Matrimony: Social Reinforcement of Religious Beliefs on Marriage Matching in Vietnam},
  author={Ciscato, Edoardo and Do, Quoc-Anh and Nguyen, Kieu-Trang},
  year={2024},
  publisher={CESifo Working Paper}
}

@techreport{montero2025price,
  title={The Price of Faith: Economic Costs and Religious Adaptation in Sub-Saharan Africa},
  author={Montero, Eduardo and Yang, Dean and Yentzen, Triana},
  year={2025},
  institution={National Bureau of Economic Research}
}

@article{bonikowski2026nature,
  title={When nature has agency: Traditional worldviews and contemporary environmental conservation},
  author={Bonikowski, Bart and Jimenez, Juan Miguel and Nunn, Nathan and Stuhler, Oscar and Wasti, Satish},
  journal={Evolution and Human Behavior},
  volume={47},
  number={5},
  pages={106942},
  year={2026},
  publisher={Elsevier}
}

@article{becker2009weber,
  title={Was {W}eber wrong? A human capital theory of Protestant economic history},
  author={Becker, Sascha O and Woessmann, Ludger},
  journal={The Quarterly Journal of Economics},
  volume={124},
  number={2},
  pages={531--596},
  year={2009},
  publisher={MIT Press}
}

@article{nunn2017being,
  title={Why being wrong can be right: Magical warfare technologies and the persistence of false beliefs},
  author={Nunn, Nathan and Sanchez de la Sierra, Raul},
  journal={American economic review},
  volume={107},
  number={5},
  pages={582--87},
  year={2017}
}

@techreport{pewresearchcenter_2014,
  title = {Global religious diversity: Half of the most religiously diverse countries are in Asia-Pacific region},
  author = {{Pew Research Center}},
  year = {2014},
  address = {{Washington, D.C.}},
  url = {https://www.pewresearch.org/religion/wp-content/uploads/sites/7/2014/01/RestrictionsV-full-report.pdf},
  language = {en-US}
}

@techreport{world2020benin,
  title = {Benin Country Forest Note},
  author = {{World Bank}},
  year = {2020},
  address = {{Washington, D.C.}},
  url = {https://openknowledge.worldbank.org/handle/10986/34437},
  language = {en-US}
}

@article{michalopoulos2013pre,
  title={Pre-colonial ethnic institutions and contemporary African development},
  author={Michalopoulos, Stelios and Papaioannou, Elias},
  journal={Econometrica},
  volume={81},
  number={1},
  pages={113--152},
  year={2013},
  publisher={Wiley Online Library}
}

@article{murdock1967ethnographic,
  title={Ethnographic atlas: a summary},
  author={Murdock, George Peter},
  journal={Ethnology},
  volume={6},
  number={2},
  pages={109--236},
  year={1967},
  publisher={JSTOR}
}

@article{lowes2026religion,
  title={Religion in emerging and developing regions},
  author={Lowes, Sara and Marx, Benjamin and Montero, Eduardo},
  journal={Annual Review of Economics},
  volume={18},
  year={2026},
  publisher={Annual Reviews}
}

@article{mccleary2006religion,
  title={Religion and economy},
  author={McCleary, Rachel M and Barro, Robert J},
  journal={Journal of Economic perspectives},
  volume={20},
  number={2},
  pages={49--72},
  year={2006}
}

@book{idowu1973african,
  title={African traditional religion: A definition},
  author={Idowu, Emanuel B},
  year={1973},
  publisher={Orbis Books}
}

@book{bay2012wives,
  title={Wives of the leopard: Gender, politics, and culture in the Kingdom of Dahomey},
  author={Bay, Edna G},
  year={1998},
  publisher={University of Virginia Press}
}

@book{carvalho2019advances,
  title={Advances in the Economics of Religion},
  author={Carvalho, Jean-Paul and Iyer, Sriya and Rubin, Jared},
  year={2019},
  publisher={Springer}
}

@inproceedings{defourny2009accuracy,
  title={Accuracy assessment of a 300 m global land cover map: The GlobCover experience},
  author={Defourny, Pierre and Schouten, Leon and Bartalev, Sergey and Bontemps, Sophie and Caccetta, P and De Wit, AJW and Di Bella, Carlos Marcelo and G{\'e}rard, Bruno and Giri, P and Gond, Val{\'e}ry and others},
  year={2009},
booktitle= {Conference Proceedings: 33rd International Symposium on Remote Sensing of Environment, Sustaining the Millennium Development Goals}
}

@book{di2005land,
  title={Land cover classification system: classification concepts and user manual: LCCS},
  author={Di Gregorio, Antonio},
  volume={2},
  year={2005},
  publisher={Food \& Agriculture Org.}
}

@misc{esa2017land,
  title={Land cover CCI product user guide version 2. Technical Report},
  author={ESA},
  year={2017},
  publisher={UCL Geomatics Louvain, Belgium},
note   = {Available online at: \url{https://maps.elie.ucl.ac.be/CCI/viewer/download/ESACCI-LC-Ph2-PUGv2_2.0.pdf}, 
           last accessed on 16.03.2025}
}

@article{baragwanath2023collective,
  title={Collective property rights lead to secondary forest growth in the Brazilian Amazon},
  author={Baragwanath, Kathryn and Bayi, Ella and Shinde, Nilesh},
  journal={Proceedings of the National Academy of Sciences},
  volume={120},
  number={22},
  pages={e2221346120},
  year={2023},
  publisher={National Academy of Sciences}
}

@book{mbiti1970concepts,
  title={Concepts of god in Africa},
  author={Mbiti, John S},
  year={1970},
publisher={New York : Praeger Publishers}
}

@book{food1990major,
  title={Major Significance of" Minor" Forest Products: The Local Use and Value of Forests in the West African Humid Forest Zone},
  author={ Koppell, Carla RS and Falconer, Julia},
  year={1990},
publisher={Food and Agriculture Organization of the United Nations},
note={\url{https://www.fao.org/4/t9450e/t9450e00.htm}}
}

@article{taylor2016lynn,
  title={Lynn White Jr. and the greening-of-religion hypothesis},
  author={Taylor, Bron and Van Wieren, Gretel and Zaleha, Bernard Daley},
  journal={Conservation Biology},
  volume={30},
  number={5},
  pages={1000--1009},
  year={2016},
  publisher={Wiley Online Library}
}

@techreport{le2022social,
  title={The social consequences of traditional religion in contemporary Africa},
  author={Le Rossignol, Etienne and Lowes, Sara and Nunn, Nathan},
  year={2022},
  institution={National Bureau of Economic Research}
}

@article{asamoah2009african,
  title={African Traditional Religion, Pentecostalism and the Clash of Spiritualties in Ghana},
  author={Asamoah-Gyadu, J Kwabena},
  journal={Fundamentalism and the Media},
  pages={161--178},
  year={2009},
  publisher={Bloomsbury Academic New York}
}

@article{camilla2005nouveaux,
  title={Les nouveaux r{\'e}seaux {\'e}vang{\'e}liques et l’{\'E}tat: le cas du B{\'e}nin},
  author={Strandsbjerg, Camilla},
  journal={Laurent Fourchard, Andr{\'e} Mary, Entreprises religieuses transnationales en Afrique de l’Ouest, Paris, Karthala},
  pages={223--241},
  year={2005}
}

@incollection{mayrargue20059,
  title={9. Dynamiques transnationales et mobilisations pentec{\^o}tistes dans l’espace public b{\'e}ninois},
  author={Mayrargue, C{\'e}dric},
  booktitle={Entreprises religieuses transnationales en Afrique de l'Ouest},
  pages={243--265},
  year={2005},
  publisher={Karthala}
}

@article{lee2022valid,
  title={Valid t-ratio Inference for IV},
  author={Lee, David S and McCrary, Justin and Moreira, Marcelo J and Porter, Jack},
  journal={American economic review},
  volume={112},
  number={10},
  pages={3260--3290},
  year={2022},
  publisher={American Economic Association 2014 Broadway, Suite 305, Nashville, TN 37203}
}

@article{ordonez2025crops,
  title={When crops fail, forests follow: Agricultural shocks and deforestation in Zambia},
  author={Ord{\'o}{\~n}ez, Pablo J and Hadunka, Protensia and Del Rossi, Gemma and Baylis, Kathy},
  journal={Proceedings of the National Academy of Sciences},
  volume={122},
  number={40},
  pages={e2427156122},
  year={2025},
  publisher={National Academy of Sciences}
}

@article{borusyak2023nonrandom,
  title={Nonrandom exposure to exogenous shocks},
  author={Borusyak, Kirill and Hull, Peter},
  journal={Econometrica},
  volume={91},
  number={6},
  pages={2155--2185},
  year={2023},
  publisher={Wiley Online Library}
}

@book{butinda2023importance,
  title={On the importance of African traditional religion for economic behavior},
  author={Butinda, Lewis Dunia and Lameke, Aimable Amani and Nunn, Nathan and Posch, Max and de la Sierra, Raul Sanchez},
  year={2023},
  publisher={National Bureau of Economic Research}
}

@article{burgess2012political,
  title={The political economy of deforestation in the tropics},
  author={Burgess, Robin and Hansen, Matthew and Olken, Benjamin A and Potapov, Peter and Sieber, Stefanie},
  journal={The Quarterly journal of economics},
  volume={127},
  number={4},
  pages={1707--1754},
  year={2012},
  publisher={MIT Press}
}

@article{ferraro2020conditional,
  title={Conditional cash transfers to alleviate poverty also reduced deforestation in Indonesia},
  author={Ferraro, Paul J and Simorangkir, Rhita},
  journal={Science Advances},
  volume={6},
  number={24},
  pages={eaaz1298},
  year={2020},
  publisher={American Association for the Advancement of Science}
}

@article{pfaff1999drives,
  title={What drives deforestation in the Brazilian Amazon?: Evidence from satellite and socioeconomic data},
  author={Pfaff, Alexander SP},
  journal={Journal of environmental economics and management},
  volume={37},
  number={1},
  pages={26--43},
  year={1999},
  publisher={Elsevier}
}

@article{veestraeten2004conditional,
  title={The conditional probability density function for a reflected Brownian motion},
  author={Veestraeten, Dirk},
  journal={Computational Economics},
  volume={24},
  number={2},
  pages={185--207},
  year={2004},
  publisher={Springer}
}

@article{ives2024role,
  title={The role of religion in shaping the values of nature},
  author={Ives, Christopher D and Kidwell, Jeremy and Anderson, Christopher B and Arias-Ar{\'e}valo, Paola and Gould, Rachelle K and Kenter, Jasper and Murali, Ranjini},
  journal={Ecology and Society},
  volume={29},
  number={2},
  pages={10},
  year={2024},
  publisher={Resilience Alliance}
}

@article{amanze2024african,
  title={African Approaches to the Protection and Conservation of the Environment: The Role of African Traditional Religions},
  author={Amanze, James N},
  journal={Religion and Development},
  volume={2},
  number={3},
  pages={445--462},
  year={2024},
  publisher={Brill Sch{\"o}ningh}
}

@article{gratz2011paroles,
  title={`{P}aroles de vie': {C}hristian radio producers in the {R}epublic of {B}enin},
  author={Gr{\"a}tz, Tilo},
  journal={Journal of African Media Studies},
  volume={3},
  number={2},
  pages={161--188},
  year={2011},
  publisher={Intellect Ltd.}
}

@article{hackett1998charismatic,
  title={Charismatic/Pentecostal appropriation of media technologies in Nigeria and Ghana},
  author={Hackett, Rosalind IJ},
  journal={Journal of religion in Africa},
  pages={258--277},
  year={1998},
  publisher={JSTOR}
}

@techreport{ipbes2019intergovernmental,
  title={Summary for Policymakers of the Methodological Assessment Report on the Diverse Values and Valuation of Nature of the Intergovernmental Science-Policy Platform on Biodiversity and Ecosystem Services.},
  author={{IPBES}},
  journal={Intergovernmental science-policy platform on biodiversity and ecosystem services},
  year={2022},
url = {https://zenodo.org/records/7410287}
}

@book{cornevin1981republique,
  title     = {La R{\'e}publique populaire du B{\'e}nin: des origines dahom{\'e}ennes {\`a} nos jours},
  author    = {Cornevin, Robert},
  year      = {1981},
  publisher = {Maisonneuve \& Larose : Acad{\'e}mie des sciences d'outre-mer},
  address   = {Paris}
}

@book{law1991slave,
  title={The Slave Coast of West Africa, 1550-1750: the impact of the Atlantic slave trade on an African society},
  author={Law, Robin},
  year={1991},
  publisher={Oxford University Press}
}

@article{klein2011hyde,
  title={The HYDE 3.1 spatially explicit database of human-induced global land-use change over the past 12,000 years},
  author={Klein Goldewijk, Kees and Beusen, Arthur and Van Drecht, Gerard and De Vos, Martine},
  journal={Global Ecology and Biogeography},
  volume={20},
  number={1},
  pages={73--86},
  year={2011},
  publisher={Wiley Online Library}
}

@inproceedings{natta2003riparian,
  title={Riparian forests and biodiversity conservation in Benin (West Africa)},
  author={Natta, AK and Sinsin, B and Van der Maesen, LJG},
  booktitle={Proceedings of the XII World Forestry Congress, Qu{\'e}bec City (Canada)},
  year={2003}
}

@article{goodman2019geoquery,
  title={GeoQuery: Integrating HPC systems and public web-based geospatial data tools},
  author={Goodman, Seth and BenYishay, Ariel and Lv, Zhonghui and Runfola, Daniel},
  journal={Computers \& geosciences},
  volume={122},
  pages={103--112},
  year={2019},
  publisher={Elsevier}
}

@article{van2024global,
  title={A global time series of traffic volumes on extra-urban roads},
  author={van Strien, Maarten J and Gr{\^e}t-Regamey, Adrienne},
  journal={Scientific data},
  volume={11},
  number={1},
  pages={470},
  year={2024},
  publisher={Nature Publishing Group UK London}
}

@article{mermoz2018impacts,
  title={Impacts of the forest definitions adopted by African countries on carbon conservation},
  author={Mermoz, St{\'e}phane and Bouvet, Alexandre and Toan, Thuy Le and Herold, Martin},
  journal={Environmental Research Letters},
  volume={13},
  number={10},
  pages={104014},
  year={2018},
  publisher={IOP Publishing}
}

@book{schroeder2010secondary,
  title={Secondary forests in West Africa: a challenge and opportunity for management.},
  author={Schroeder, Jobst-Michael and Oke, David O and Onyekwelu, Jonathan C and Yirdaw, Eshetu},
  volume={25},
  year={2010},
  publisher={International Union of Forestry Research Organizations}
}

@article{lugo2010tolerance,
  title={Tolerance and tension: Islam and Christianity in sub-Saharan Africa},
  author={Lugo, Luis and Cooperman, Alan},
  journal={Washington, DC, Pew Research Center},
  volume={147},
  year={2010}
}

\clearpage 

\addcontentsline{toc}{section}{Supplemental Appendix}

\startcontents[appendix]
\printcontents[appendix]{l}{1}{
    \section*{\centering Sacred Ecology\\ Supplemental Appendix}
}

\setcounter{section}{0}
\setcounter{figure}{0}
\setcounter{table}{0}

\counterwithin{figure}{section}
\counterwithin{table}{section}

\renewcommand{\thesection}{S\Alph{section}}
\renewcommand{\thefigure}{\thesection\arabic{figure}}
\renewcommand{\thetable}{\thesection\arabic{table}}

\renewcommand{\theHsection}{online.\Alph{section}}
\renewcommand{\theHfigure}{online.\Alph{section}.\arabic{figure}}
\renewcommand{\theHtable}{online.\Alph{section}.\arabic{table}}

\section{Tables}

\begin{table}[h]
\centering
\begin{threeparttable}
\caption{ESA-CCI-LC dataset: Land Cover classifications used in the study}
\label{lctype}
\begin{tabular}{lll}
\toprule
\multicolumn{2}{c}{Land Cover (LC) classification}  & LC type \\
\midrule
\multirow{8}{*}{Forests} & Tree cover, broadleaved, evergreen, closed to open ($\geq15\%$)  & 50 \\
& Tree cover, broadleaved, deciduous, closed to open ($\geq15\%$) & 60, 61, 62 \\
 & Tree cover, needleleaved, evergreen, closed to open ($\geq15\%$) & 70, 71, 72 \\
  & Tree cover, needleleaved, deciduous, closed to open ($\geq15\%$) & 80, 81, 82 \\
   & Tree cover, mixed leaf type (broadleaved and needleleaved) & 90 \\
    & Mosaic tree and shrub ($>50\%$) / herbaceous cover ($< 50\%$) & 100 \\
     & Tree cover, flooded, fresh or brakish water & 160 \\
      & Tree cover, flooded, saline water & 170 \\
\hline
Savanna      & Shrubland & 120, 121, 122\\
\bottomrule
\end{tabular}

\begin{tablenotes}
\footnotesize
\item Global land cover (LC) data from the European Space Agency (ESA) Climate Change Initiative (CCI) programme. The data is binary as each pixel represents one type of LC classification. A total of 37 original LC types are presented in the dataset based on the LC Classification System developed by \cite{di2005land} for FAO. The LC types used in the study are Forests, Agriculture and Savanna. For further details refer to Section 3 in the main paper and \cite{esa2017land}.
\end{tablenotes}

\end{threeparttable}
\end{table}

\begin{table}[htb]
\centering
\begin{threeparttable}
\caption{Summary statistics, EHCVM cluster panel}
\label{tab:sumstats_lsms}
\begin{tabular}{l rr rr}
\toprule
 & \multicolumn{2}{c}{Pooled} & \multicolumn{2}{c}{Mean by survey year} \\
\cmidrule(lr){2-3} \cmidrule(lr){4-5}
Variable & Mean & SD & 2018 & 2021 \\
\midrule
\multicolumn{5}{l}{\textit{Panel A: Landscape outcomes and rainfall (cluster $\times$ wave)}} \\
\quad ESA forest--savanna share, 10 km (\%) & 27.95 & 31.00 & 27.54 & 28.38 \\
\quad Hansen loss over recall window, 10 km (\% of land) & 1.49 & 2.13 & 1.72 & 1.25 \\
\quad Apr--Oct rainfall, survey year (mm) & 1,071.3 & 191.1 & 1,091.3 & 1,050.3 \\
\quad Apr--Oct rainfall, recall window (mm) & 1,070.0 & 144.7 & 1,057.7 & 1,083.0 \\
\addlinespace
\multicolumn{5}{l}{\textit{Panel B: Shocks reported, 3-year recall (household)}} \\
\quad Idiosyncratic shocks (count) & 0.60 & 0.73 & 0.76 & 0.33 \\
\quad Climate shocks (count) & 0.66 & 0.78 & 0.71 & 0.58 \\
\addlinespace
\multicolumn{5}{l}{\textit{Panel C: Losses caused by shocks (household)}} \\
\quad Any loss (count) & 1.45 & 1.16 & 1.89 & 0.72 \\
\quad Agricultural production losses (count) & 1.27 & 1.12 & 1.65 & 0.62 \\
\quad Income and asset losses (count) & 1.38 & 1.14 & 1.80 & 0.68 \\
\quad Livestock losses (count) & 1.00 & 1.13 & 1.38 & 0.37 \\
\addlinespace
\multicolumn{5}{l}{\textit{Panel D: Household head religion and controls (household)}} \\
\quad ATR head (0/1) & 0.21 & 0.41 & 0.22 & 0.20 \\
\quad Christian head (0/1) & 0.41 & 0.49 & 0.40 & 0.42 \\
\quad Muslim head (0/1) & 0.34 & 0.47 & 0.34 & 0.34 \\
\quad Household size & 6.07 & 3.19 & 6.08 & 6.05 \\
\quad Share of plots owned & 0.66 & 0.46 & 0.64 & 0.70 \\
\quad Plot area (ha) & 12.84 & 265.40 & 16.80 & 6.23 \\
\quad Assets (CFA) & 310,053 & 1,447,631 & 287,392 & 347,926 \\
\quad Salaried income (daily CFA) & 3,911 & 29,668 & 3,418 & 4,736 \\
\addlinespace
\midrule
\quad Household-years & 5,997 & & 3,752 & 2,245 \\
\quad Clusters & 457 & & 452 & 427 \\
\bottomrule
\end{tabular}
\begin{tablenotes}[flushleft]\footnotesize
\item \textit{Notes:} Estimation sample of Table~\ref{tab:shocks_full}: the 2018 and 2021 EHCVM waves, farming households with reported religion, after dropping two households with a missing control and 33 clusters with a single household-year. Panel A is measured at the cluster--wave level (879 cells) using the cluster's GPS location: ESA-CCI forest--savanna share of land within a 10 km buffer in the survey year; Hansen stand-replacing loss over the wave's own three-year recall window (2016--18 for 2018, 2019--21 for 2021) as a share of land area in the 10 km grid cell; CHIRPS v3 April--October rainfall in the survey year and averaged over the recall window. Panels B--D are household reports from module S14 (shocks in the past three years and their consequences) and the household roster. Idiosyncratic shocks: illness or death of a household member, divorce, end of transfers, bankruptcy, loss of salaried employment, theft. Climate shocks: drought, flood, fire, crop or livestock disease. Loss counts are the number of reported shocks that reduced the item. 
\end{tablenotes}
\end{threeparttable}
\end{table}

\begin{table}[!htb]
\centering
\caption{Where the corridor's ATR outflow went}
\label{tab:destination}

\begin{threeparttable}

\small
\setlength{\tabcolsep}{5pt}
\renewcommand{\arraystretch}{1.00}

\begin{tabular}{lccccc}
\toprule
& \multicolumn{2}{c}{REP $\times$ 2002}
& \multicolumn{2}{c}{REP $\times$ 2013}
& \shortstack{Share of\\2013 outflow} \\
\cmidrule(lr){2-3}\cmidrule(lr){4-5}
Destination
& Coef.
& $t$
& Coef.
& $t$
& \\
\midrule

ATR
& -1.02$^{**}$
& (-2.5)
& -3.96$^{***}$
& (-7.3)
& -- \\

\midrule

Other Christianity
& 0.45$^{**}$
& (2.3)
& 1.35$^{***}$
& (4.8)
& 34\% \\

Catholic
& 0.55$^{**}$
& (2.1)
& 1.17$^{***}$
& (3.8)
& 30\% \\

Protestant
& 0.09
& (0.8)
& 0.52$^{***}$
& (3.6)
& 13\% \\

Islam
& -0.22
& (-1.1)
& -0.15
& (-0.7)
& -4\% \\

Other religion / none
& 0.16
& (0.5)
& 1.06$^{***}$
& (3.0)
& 27\% \\

\bottomrule
\end{tabular}

\vspace{3pt}

\begin{tablenotes}[flushleft]
\footnotesize
\item \textit{Notes:}
Each row reports a separate regression in which the dependent variable is
the arrondissement share of the indicated religious category, measured in
percentage points.
The regressors of interest are relative Adja proximity (REP$_i$)
interacted with indicators for 2002 and 2013, with 1992 as the omitted
reference year.
All specifications use the full sample and include arrondissement fixed
effects, homeland $\times$ year fixed effects, time varying and geographic controls.
The final column reports each destination's 2013 coefficient as a share
of the estimated 2013 ATR outflow.
``Other Christianity'' comprises Christians who identify as neither
Catholic nor Protestant, including evangelical, Pentecostal, Born-Again,
and related Charismatic denominations.
$t$-statistics are based on standard errors clustered by arrondissement.
$^{*}p<0.10$; $^{**}p<0.05$; $^{***}p<0.01$.
\end{tablenotes}

\end{threeparttable}
\end{table}

\begin{table}[!htb]
\centering
\caption{Alternative temporal profiles of the national shifter: South sub-sample}
\label{tab:shifter_esa_atr_south}

\begin{threeparttable}

\small
\setlength{\tabcolsep}{3.0pt}
\renewcommand{\arraystretch}{1.05}

\begin{tabular}{
@{}
p{3.4cm}
c
c
c
c
c
c
@{}
}
\toprule
National series $W_t$
& \makecell{$W_t$ path\\(92/02/13)}
& \makecell{2013/2002\\ratio}
& \makecell{REP $\times$ 2013\\coef.\ ($t$)}
& \makecell{Effective\\$F$}
& \makecell{2SLS\\$\hat{\beta}$}
& \makecell{Sargan\\$p$} \\
\midrule

Pentecostal radio stations
& 0 / 1 / 3
& 3.0
& -5.12 (-7.4)
& 55.3
& \makecell{0.165$^{***}$\\(0.049)}
& -- \\

Other Christianity share
& 0 / 4.96 / 11.6
& 2.3
& -5.16 (-7.3)
& 52.8
& \makecell{0.169$^{***}$\\(0.051)}
& -- \\

ATR outflow since 1992
& 0 / 12.2 / 23.6
& 1.9
& -5.08 (-7.1)
& 49.9
& \makecell{0.172$^{***}$\\(0.053)}
& -- \\

Linear time trend
& 0 / 10 / 21
& 2.1
& -5.13 (-7.2)
& 51.3
& \makecell{0.171$^{***}$\\(0.052)}
& -- \\

Post-1992 dummy
& 0 / 1 / 1
& 1.0
& -3.35 (-5.8)
& 33.3
& \makecell{0.191$^{***}$\\(0.066)}
& -- \\

Unrestricted year coefficients
& --
& 2.7
& -5.15 (-7.2)
& 30.3
& \makecell{0.167$^{***}$\\(0.050)}
& 0.385 \\

\bottomrule
\end{tabular}

\begin{tablenotes}[flushleft]
\footnotesize
\item \textit{Notes:}
$N=1{,}166$.
Each row instruments ATR share with
REP$_i \times \widetilde{W}_t$, where
$\widetilde{W}_t=(W_t-W_{1992})/(W_{2013}-W_{1992})$,
so each national series is normalized to zero in 1992 and one in 2013.
The coefficient on REP $\times$ 2013 therefore gives the implied
1992--2013 change in ATR share per unit of REP and is comparable across
candidate shifters; the associated $t$-statistic is in parentheses.
The 2013/2002 ratio describes the temporal profile imposed by each
series. 
The final row instead instruments ATR share with REP interacted separately
with the 2002 and 2013 indicators, leaving the first-stage time profile
unrestricted.
All specifications include arrondissement fixed effects,
homeland $\times$ year fixed effects, time-varying and
geographic controls.
Time-varying controls include log population density, nighttime-light
intensity, informal-employment share, mean wealth index, mean temperature,
precipitation, and annual average daily traffic on the nearest road.
Geographic controls comprise latitude, longitude, and FAO-GAEZ soil
suitability for maize, cassava, and cotton, each interacted with census-year
indicators.
Standard errors, clustered by arrondissement, are reported in parentheses
beneath the 2SLS estimates.
$^{*}p<0.10$; $^{**}p<0.05$; $^{***}p<0.01$.
\end{tablenotes}

\end{threeparttable}
\end{table}

\begin{table}[!htbp]
\centering
\caption{Migration composition and the PFR land reform}
\label{tab:mig_pfr_combined}

\begin{threeparttable}

\small
\setlength{\tabcolsep}{4.2pt}
\renewcommand{\arraystretch}{1.05}

\begin{tabular}{lccccc}
\toprule
& \makecell{Baseline}
& \makecell{Born-here\\control}
& \makecell{Stability\\interaction}
& \makecell{PFR\\control}
& \makecell{PFR\\interaction} \\
& (1) & (2) & (3) & (4) & (5) \\
\midrule

\multicolumn{6}{c}{
\textit{Panel A: Dependent variable: forest--savanna cover (\%)}
} \\[2pt]

ATR share (\%)
& 0.185$^{***}$
& 0.190$^{***}$
& 0.207$^{***}$
& 0.182$^{***}$
& 0.184$^{***}$ \\

& (0.056)
& (0.061)
& (0.064)
& (0.056)
& (0.057) \\[2pt]

Born-here share
&
& -2.122
&
&
\\

&
& (2.737)
&
&
\\[2pt]

ATR share $\times$ Stable
&
&
& -0.027
&
\\

&
&
& (0.034)
&
\\[2pt]

PFR Dummy
&
&
&
& -0.563
& -0.745 \\

&
&
&
& (0.397)
& (0.746) \\[2pt]

ATR share $\times$ PFR Dummy
&
&
&
&
& 0.008 \\

&
&
&
&
& (0.023) \\[2pt]

Observations
& 1,637
& 1,637
& 1,635
& 1,637
& 1,637 \\

First-stage $F$: ATR share
& 61.4
& 56.4
& 30.0
& 61.0
& 30.6 \\

First-stage $F$: interaction
&
&
& 37.2
&
& 104.3 \\

\midrule

\multicolumn{6}{c}{
\textit{Panel B: Dependent variable: subsequent forest loss (\%)}
} \\[2pt]

ATR share (\%)
& -0.068$^{***}$
& -0.072$^{***}$
& -0.061$^{***}$
& -0.068$^{***}$
& -0.064$^{***}$ \\

& (0.015)
& (0.017)
& (0.018)
& (0.016)
& (0.015) \\[2pt]

Born-here share
&
& 1.589$^{**}$
&
&
\\

&
& (0.654)
&
&
\\[2pt]

ATR  $\times$ Stable
&
&
& -0.009
&
\\

&
&
& (0.016)
&
\\[2pt]

PFR Dummy
&
&
&
& 0.113
& -0.110 \\

&
&
&
& (0.102)
& (0.153) \\[2pt]

ATR $\times$ PFR Dummy
&
&
&
&
& 0.010 \\

&
&
&
&
& (0.008) \\[2pt]

Observations
& 1,084
& 1,084
& 1,082
& 1,084
& 1,084 \\

First-stage $F$: ATR share
& 74.8
& 70.2
& 24.7
& 74.3
& 37.7 \\

First-stage $F$: interaction
&
&
& 28.9
&
& 108.9 \\

\bottomrule
\end{tabular}

\begin{tablenotes}[flushleft]
\footnotesize
\item \textit{Notes:}
All columns report 2SLS estimates using the full sample.
Column (1) reports the baseline specification.
Column (2) additionally controls for the contemporaneous born-here share.
In column (3), Stable is an indicator for above-median born-here share in
1992, a predetermined measure of residential stability; ATR share and
ATR share $\times$ Stable are instrumented with the baseline instrument
and its interaction with Stable.
PFR dummy indicates arrondissements containing at least one
village treated by the randomized Plans Fonciers Ruraux land-formalization
programme, rolled out from 2009 and therefore equal to one for treated
arrondissements in 2013.
Column (4) controls directly for PFR treatment.
In column (5), ATR share $\times$ PFR is additionally instrumented with
the baseline instrument interacted with PFR treatment.
All specifications include time varying and geographic controls, arrondissement and homeland $\times$ year fixed
effects.
The forest-loss outcome is available for the 2002 and 2013 waves only.
First-stage $F$ statistics are reported for ATR and, where applicable,
the endogenous interaction.
Standard errors, clustered by arrondissement, are in parentheses.
$^{*}p<0.10$; $^{**}p<0.05$; $^{***}p<0.01$.
\end{tablenotes}

\end{threeparttable}
\end{table}

\begin{table}[!htbp]
\centering
\caption{Instrument construction: alternative specifications}
\label{tab:spec_curve}

\begin{threeparttable}

\small
\setlength{\tabcolsep}{3.0pt}
\renewcommand{\arraystretch}{1.08}

\begin{tabular}{lcccccc}
\toprule
& \multicolumn{2}{c}{Cover, full}
& \multicolumn{2}{c}{Cover, south}
& \multicolumn{2}{c}{Loss, full} \\
\cmidrule(lr){2-3}
\cmidrule(lr){4-5}
\cmidrule(lr){6-7}
Specification
& Estimate & Eff.\ $F$
& Estimate & Eff.\ $F$
& Estimate & Eff.\ $F$ \\
\midrule

\multicolumn{7}{l}{\textit{Alternative exposure measures}} \\[2pt]

Baseline 2SLS
& \makecell{0.185$^{***}$\\(0.056)}
& 61.4
& \makecell{0.158$^{***}$\\(0.045)}
& 59.1
& \makecell{-0.068$^{***}$\\(0.015)}
& 74.8 \\

5 km distance floor
& \makecell{0.179$^{***}$\\(0.052)}
& 70.2
& \makecell{0.141$^{***}$\\(0.037)}
& 100.1
& \makecell{-0.066$^{***}$\\(0.014)}
& 73.3 \\

10 km distance floor
& \makecell{0.183$^{***}$\\(0.054)}
& 62.3
& \makecell{0.134$^{***}$\\(0.035)}
& 89.8
& \makecell{-0.060$^{***}$\\(0.013)}
& 71.2 \\

Linear distances (per 100 km)
& \makecell{0.247$^{***}$\\(0.073)}
& 45.1
& \makecell{0.175$^{***}$\\(0.045)}
& 69.6
& \makecell{-0.048$^{***}$\\(0.011)}
& 62.9 \\

Distance to nearest waypoint
& \makecell{0.137$^{***}$\\(0.040)}
& 141.9
& \makecell{0.114$^{***}$\\(0.034)}
& 142.4
& \makecell{-0.047$^{***}$\\(0.011)}
& 81.7 \\

Absolute Adja proximity ($-\ln d_{\text{Adja}}$)
& \makecell{0.198$^{***}$\\(0.051)}
& 35.4
& \makecell{0.205$^{***}$\\(0.052)}
& 27.5
& \makecell{-0.082$^{***}$\\(0.022)}
& 45.1 \\

\midrule

\multicolumn{7}{l}{\textit{Dropping individual Yoruba corridor branches}} \\[2pt]

Drop If\`e--K\'etou--Sak\'et\'e branch
& \makecell{0.231$^{***}$\\(0.059)}
& 52.1
& \makecell{0.207$^{***}$\\(0.050)}
& 40.5
& \makecell{-0.076$^{***}$\\(0.018)}
& 63.2 \\

Drop Oyo--Sav\`e branch
& \makecell{0.175$^{***}$\\(0.056)}
& 52.6
& \makecell{0.157$^{***}$\\(0.050)}
& 44.5
& \makecell{-0.077$^{***}$\\(0.017)}
& 66.5 \\

Drop Egba--Dassa branch
& \makecell{0.168$^{***}$\\(0.059)}
& 53.0
& \makecell{0.143$^{***}$\\(0.046)}
& 50.8
& \makecell{-0.076$^{***}$\\(0.017)}
& 66.5 \\

\midrule

\multicolumn{7}{l}{\textit{Alternative endogenous variable}} \\[2pt]

Vodun share
& \makecell{0.122$^{***}$\\(0.038)}
& 52.4
& --
& --
& \makecell{-0.056$^{***}$\\(0.013)}
& 50.7 \\

\bottomrule
\end{tabular}

\begin{tablenotes}[flushleft]
\footnotesize
\item \textit{Notes:}
Each row re-estimates the preferred 2SLS specification using an alternative
exposure measure interacted with the national Pentecostal broadcaster count.
The baseline measure is
REP $=\ln d_{\text{Yoruba}}-\ln d_{\text{Adja}}$, where distances are measured
from the arrondissement centroid to the nearest corridor polyline.
The 5 km and 10 km distance-floor specifications replace $d$ with
$\max(d,5)$ and $\max(d,10)$ km, respectively.
Linear distance uses
$(d_{\text{Yoruba}}-d_{\text{Adja}})/100$.
Distance to nearest waypoint replaces distance to the corridor polyline with
distance to the nearest historical waypoint.
Absolute Adja proximity uses $-\ln d_{\text{Adja}}$.
The next three specifications sequentially exclude one Yoruba branch from
the reference corridor.
The final row replaces ATR share with Vodun share as the endogenous variable.
Because the census separately identifies Vodun only in 2002 and 2013,
this specification uses the two-wave panel for both outcomes.
All specifications include arrondissement and homeland $\times$ year fixed
effects, time-varying controls, and geographic controls.
Time-varying controls include log population density, nighttime-light
intensity, informal-employment share, mean wealth index, mean temperature,
precipitation, and annual average daily traffic on the nearest road.
Geographic controls comprise latitude, longitude, and FAO-GAEZ soil
suitability for maize, cassava, and cotton, each interacted with census-year
indicators.
Eff.\ $F$ denotes the effective first-stage $F$-statistic.
Standard errors, clustered by arrondissement, are in parentheses.
$^{*}p<0.10$; $^{**}p<0.05$; $^{***}p<0.01$.
\end{tablenotes}

\end{threeparttable}
\end{table}

\begin{table}[!htbp]
\centering
\caption{Heterogeneity analysis}
\label{tab:het_combined}

\begin{threeparttable}

\small
\setlength{\tabcolsep}{4.0pt}
\renewcommand{\arraystretch}{1.06}

\begin{tabular}{lcccc}
\toprule
& \makecell{Fractional-\\ization}
& Drought
& Poverty
& Illiteracy \\
& (1) & (2) & (3) & (4) \\
\midrule

\multicolumn{5}{c}{
\textit{Panel A: Dependent variable: forest--savanna cover (\%)}
} \\[2pt]

ATR share (\%)
& 0.183$^{***}$
& 0.218$^{***}$
& 0.189$^{***}$
& 0.198$^{***}$ \\

& (0.057)
& (0.079)
& (0.056)
& (0.059) \\[2pt]

ATR share $\times$ High
& 0.051
& -0.055
& -0.032
& -0.032 \\

& (0.061)
& (0.048)
& (0.029)
& (0.033) \\[2pt]

Observations
& 1,635
& 1,637
& 1,635
& 1,635 \\

Effective $F$: ATR share
& 31.2
& 39.3
& 31.2
& 30.9 \\

Effective $F$: interaction
& 9.5
& 54.4
& 39.4
& 33.1 \\

\midrule

\multicolumn{5}{c}{
\textit{Panel B: Dependent variable: subsequent forest loss (\%)}
} \\[2pt]

ATR share (\%)
& -0.067$^{***}$
& -0.063$^{***}$
& -0.062$^{***}$
& -0.059$^{***}$ \\

& (0.017)
& (0.021)
& (0.016)
& (0.018) \\[2pt]

ATR share $\times$ High
& -0.009
& -0.009
& -0.022
& -0.020 \\

& (0.026)
& (0.019)
& (0.014)
& (0.015) \\[2pt]

Observations
& 1,082
& 1,084
& 1,082
& 1,082 \\

Effective $F$: ATR share
& 30.1
& 37.4
& 38.5
& 30.4 \\

Effective $F$: interaction
& 14.3
& 42.5
& 46.1
& 31.1 \\

\bottomrule
\end{tabular}

\begin{tablenotes}[flushleft]
\scriptsize
\item \textit{Notes:}
Each column reports 2SLS estimates in which ATR share and its interaction
with the indicated moderator are instrumented by the baseline instrument
and its interaction.
High is an indicator for arrondissements above the median of the relevant
moderator; for Poverty, High denotes arrondissements below the median
1992 mean wealth index, while the contemporaneous wealth index remains
included among the controls.
Fractionalization, drought, poverty, and illiteracy are measured at
baseline.
All specifications include arrondissement and homeland $\times$ year fixed
effects, time-varying controls, and geographic controls.
Time-varying controls include log population density, nighttime-light
intensity, informal-employment share, mean wealth index, mean temperature,
precipitation, and annual average daily traffic on the nearest road.
Geographic controls comprise latitude, longitude, and FAO-GAEZ soil
suitability for maize, cassava, and cotton, each interacted with census-year
indicators.
The forest-loss outcome measures Hansen forest loss in the year following
each census and is available for the 2002 and 2013 waves only.
Conditional effective first-stage $F$-statistics are reported separately
for ATR share and the endogenous interaction.
Standard errors, clustered by arrondissement, are in parentheses.
$^{*}p<0.10$; $^{**}p<0.05$; $^{***}p<0.01$.
\end{tablenotes}

\end{threeparttable}
\end{table}

\begin{table}[htb]
\centering
\begin{threeparttable}
\caption{Household balance by religion of the household head}
\label{tab:balance_lsms}
\small
\begin{tabular}{l rr r rr}
\toprule
 & \multicolumn{2}{c}{Mean} & & \multicolumn{2}{c}{$p$-value} \\
\cmidrule(lr){2-3}\cmidrule(lr){5-6}
Variable & Non-ATR & ATR & Diff. & Raw & Within cluster \\
\midrule
\multicolumn{5}{l}{\textit{Panel A: Shocks reported, 3-year recall}} \\
\quad Idiosyncratic shocks (count) & 0.60 & 0.62 & 0.03 & 0.281 & 0.859 \\
\quad Climate shocks (count) & 0.61 & 0.83 & 0.22 & 0.000 & 0.188 \\
\quad All shocks (count) & 1.76 & 2.04 & 0.28 & 0.000 & 0.303 \\
\multicolumn{5}{l}{\textit{Panel B: Losses caused by shocks}} \\
\quad Any loss (count) & 1.40 & 1.63 & 0.23 & 0.000 & 0.010 \\
\quad Agricultural production losses & 1.22 & 1.40 & 0.17 & 0.001 & 0.026 \\
\quad Income losses & 1.33 & 1.54 & 0.21 & 0.000 & 0.034 \\
\quad Asset losses & 1.26 & 1.49 & 0.23 & 0.000 & 0.014 \\
\quad Livestock losses & 0.97 & 1.11 & 0.14 & 0.005 & 0.078 \\
\multicolumn{5}{l}{\textit{Panel C: Household controls}} \\
\quad Household size & 6.28 & 5.28 & -1.00 & 0.000 & 0.358 \\
\quad Share of plots owned & 0.68 & 0.60 & -0.08 & 0.000 & 0.009 \\
\quad Plot area (sqm) & 97,775 & 239,181 & 141,405 & 0.283 & 0.337 \\
\quad Assets (CFA)  & 320,619 & 304,528 & -16,091 & 0.752 & 0.533 \\
\quad Salaried income (CFA) & 4,541 & 2,140 & -2,401 & 0.000 & 0.001 \\
\midrule
Household-years & 4,747 & 1,285 & & & \\
\bottomrule
\end{tabular}
\begin{tablenotes}[flushleft]\footnotesize
\item \textit{Notes:} Household-level means by religion of the household head, pooling the 2018 and 2021 EHCVM waves. Diff.\ is the ATR minus non-ATR difference. Raw $p$-values are from a regression of each variable on the ATR-head indicator; within-cluster $p$-values add survey-cluster fixed effects. Standard errors clustered by survey cluster in both cases. Shocks and losses are reported over the three years preceding each wave.
\end{tablenotes}
\end{threeparttable}
\end{table}

\begin{table}[!htbp]
\centering
\caption{Losses instrumented by idiosyncratic shocks: 2SLS}
\label{tab:shocks_iv}
\begin{threeparttable}
\small
\setlength{\tabcolsep}{8pt}
\begin{tabular}{lcc}
\toprule
 & Forest--savanna cover (\%) & Forest loss (\% of land) \\
 & (1) & (2) \\
\midrule
\multicolumn{3}{l}{\textit{Panel A: Second stage}} \\[2pt]
Any loss $\times$ ATR head ($\beta_3$) & 0.284$^{***}$ & -0.069$^{*}$ \\
 & (0.082) & (0.042) \\
Effect for ATR households ($\beta_1+\beta_3$) & 0.232$^{***}$ & -0.094$^{**}$ \\
 & (0.065) & (0.039) \\
Any loss ($\beta_1$) & -0.052 & -0.024 \\
 & (0.065) & (0.025) \\
ATR head ($\beta_2$) & -0.001 & 0.004 \\
 & (0.047) & (0.026) \\
\midrule
\multicolumn{3}{l}{\textit{Panel B: First stage, dependent variable: any loss}} \\[2pt]
Idiosyncratic shocks & 0.455$^{***}$ & 0.457$^{***}$ \\
 & (0.024) & (0.025) \\
Idiosyncratic shocks $\times$ ATR head & 0.067$^{*}$ & 0.062 \\
 & (0.040) & (0.040) \\
Effective $F$ & 214.1 & 210.9 \\
\midrule
\multicolumn{3}{l}{\textit{Panel C: First stage, dependent variable: any loss $\times$ ATR head}} \\[2pt]
Idiosyncratic shocks & -0.049$^{***}$ & -0.046$^{***}$ \\
 & (0.007) & (0.007) \\
Idiosyncratic shocks $\times$ ATR head & 0.692$^{***}$ & 0.686$^{***}$ \\
 & (0.038) & (0.038) \\
Effective $F$ & 189.8 & 184.5 \\
\midrule
Mean of dependent variable & 30.01 & 1.28 \\
Observations & 5,997 & 5,997 \\
\bottomrule
\end{tabular}
\begin{tablenotes}[flushleft]\scriptsize
\item \textit{Notes:} 2SLS estimates of equation~\eqref{sox2} in which the count of reported losses (any loss) and its interaction with the ATR-head indicator are instrumented by the count of idiosyncratic shocks and its interaction with the ATR-head indicator. Both counts are centred at their sample means (1.45 losses, 0.60 idiosyncratic shocks), so that $\beta_2$ is the ATR versus non-ATR difference for a household reporting the average number of losses; $\beta_1$, $\beta_3$ and the first stages are unaffected by centring. Both columns include survey-cluster and year fixed effects, household controls (household size, share of plots owned, plot area, assets, salary), survey-wave $\times$ year indicators, the cluster's latitude and longitude interacted with year indicators, and a cubic spline (3 d.f.) in cluster rainfall over the survey year (column 1) or the three-year recall window (column 2). Effect for ATR households is $\beta_1+\beta_3$ with a delta-method standard error. Effective $F$ is the conditional first-stage effective $F$-statistic for each endogenous term. Standard errors clustered by survey cluster in parentheses. $^{*}p<0.10$; $^{**}p<0.05$; $^{***}p<0.01$.
\end{tablenotes}
\end{threeparttable}
\end{table}

\section{Figures}

 \begin{figure}
     \centering
    \includegraphics[scale=0.6]{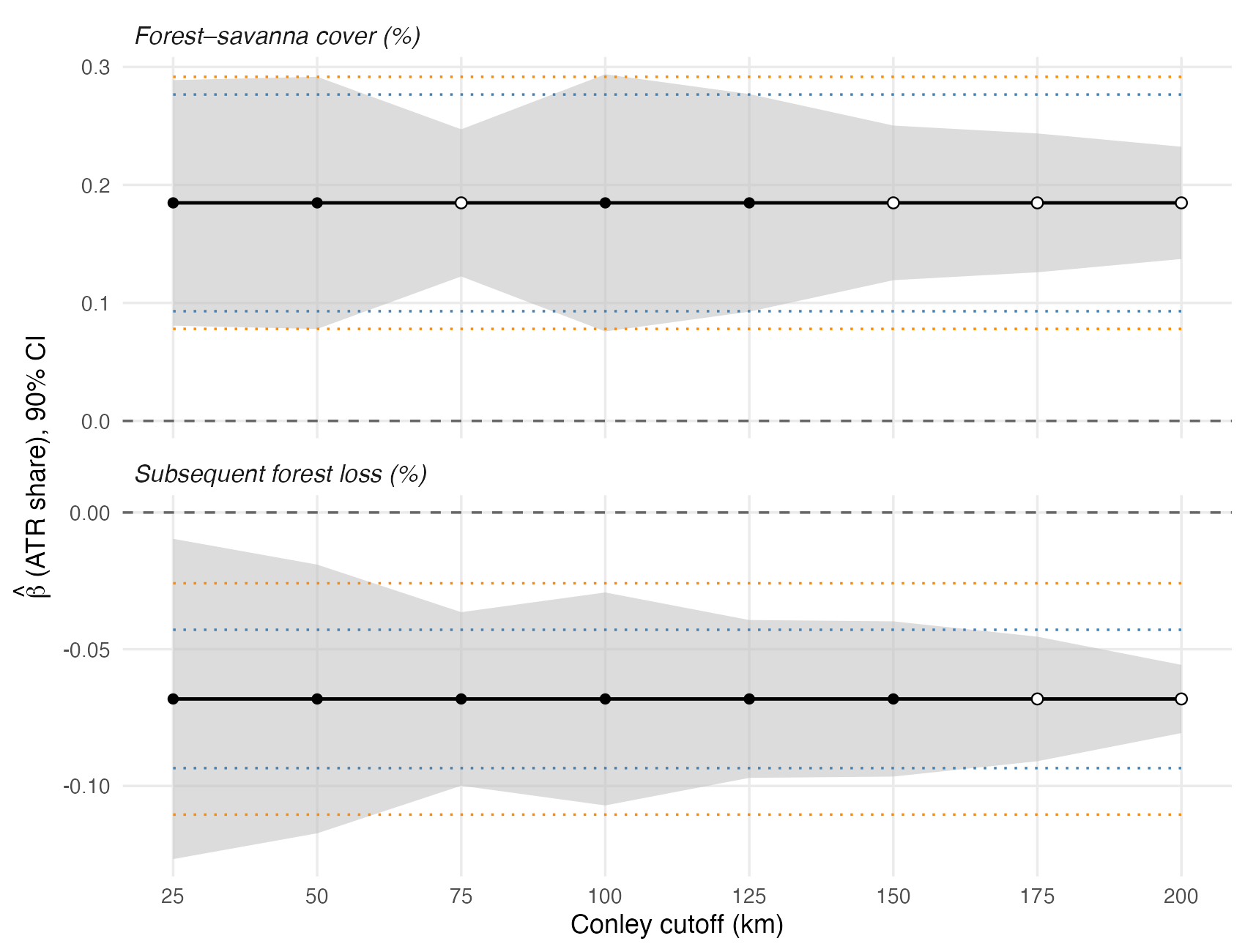}
     \caption{Robustness of the 2SLS estimates to computing Conley standard errors for spatial bandwidth between 25-200 km.  90\% confidence intervals. Blue dashed line: arrondissement-level error clustering. Orange dashed line: administrative level 2 error clustering.}
     \label{fig:conley}
 \end{figure}

\begin{figure}
    \centering
    \includegraphics[width=0.9\linewidth]{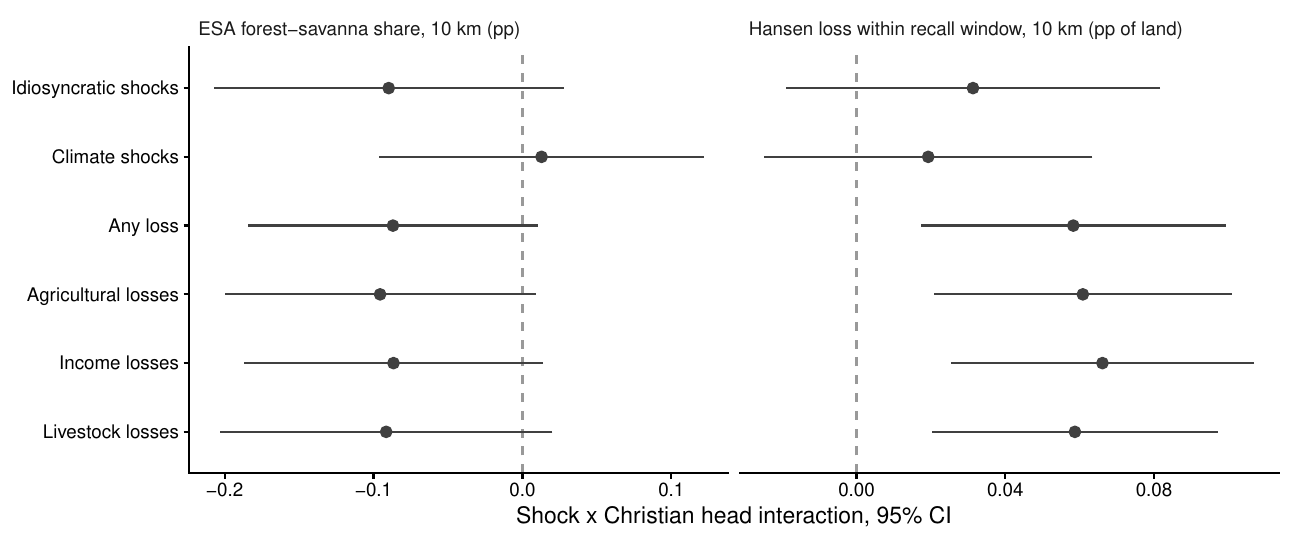}
    \caption{Estimation of Equation \eqref{sox} by using an indicator of whether each household head is Christian. Christian households resort more to deforestation as a response to idiosyncratic shocks and livelihood losses. Left panel: ESA forest/savanna cover measured at each respective survey year, right panel: Hansen forest loss for each survey year's three-year shock recall window.}
    \label{fig:christians_coping}
\end{figure}

\begin{figure}
    \centering
    \includegraphics[width=0.9\linewidth]{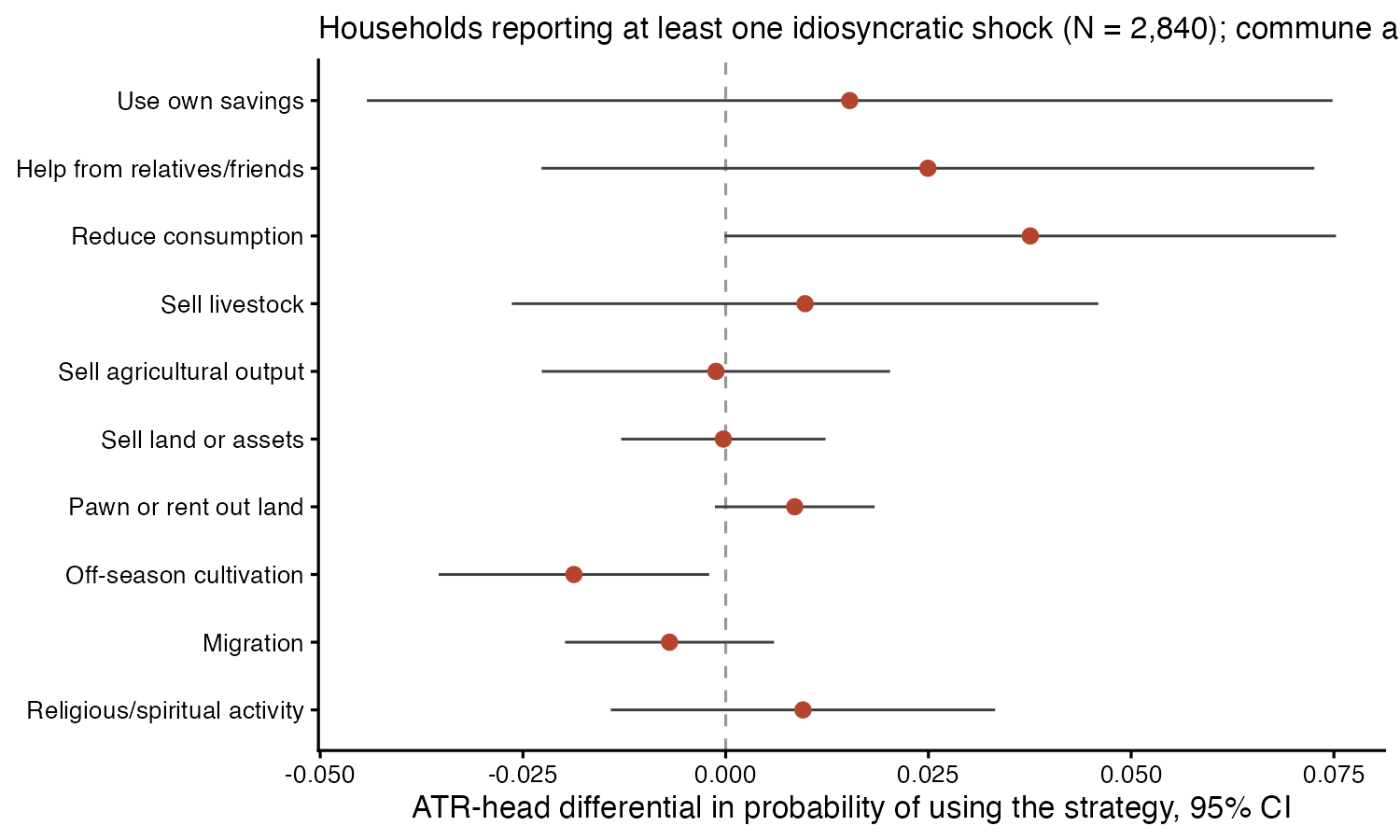}
    \caption{Each row is the coefficient of an ATR-head indicator in a separate linear probability regression of whether the household used the strategy after any of its three most severe shocks, among households reporting at least one idiosyncratic shock. All regressions control for household size, share of plots owned, plot area, assets and salary, and include commune and wave fixed effects. Standard errors clustered by survey cluster in parentheses. }
    \label{fig:coping_atr}
\end{figure}


\begin{figure}
\centering
\includegraphics[scale=2]{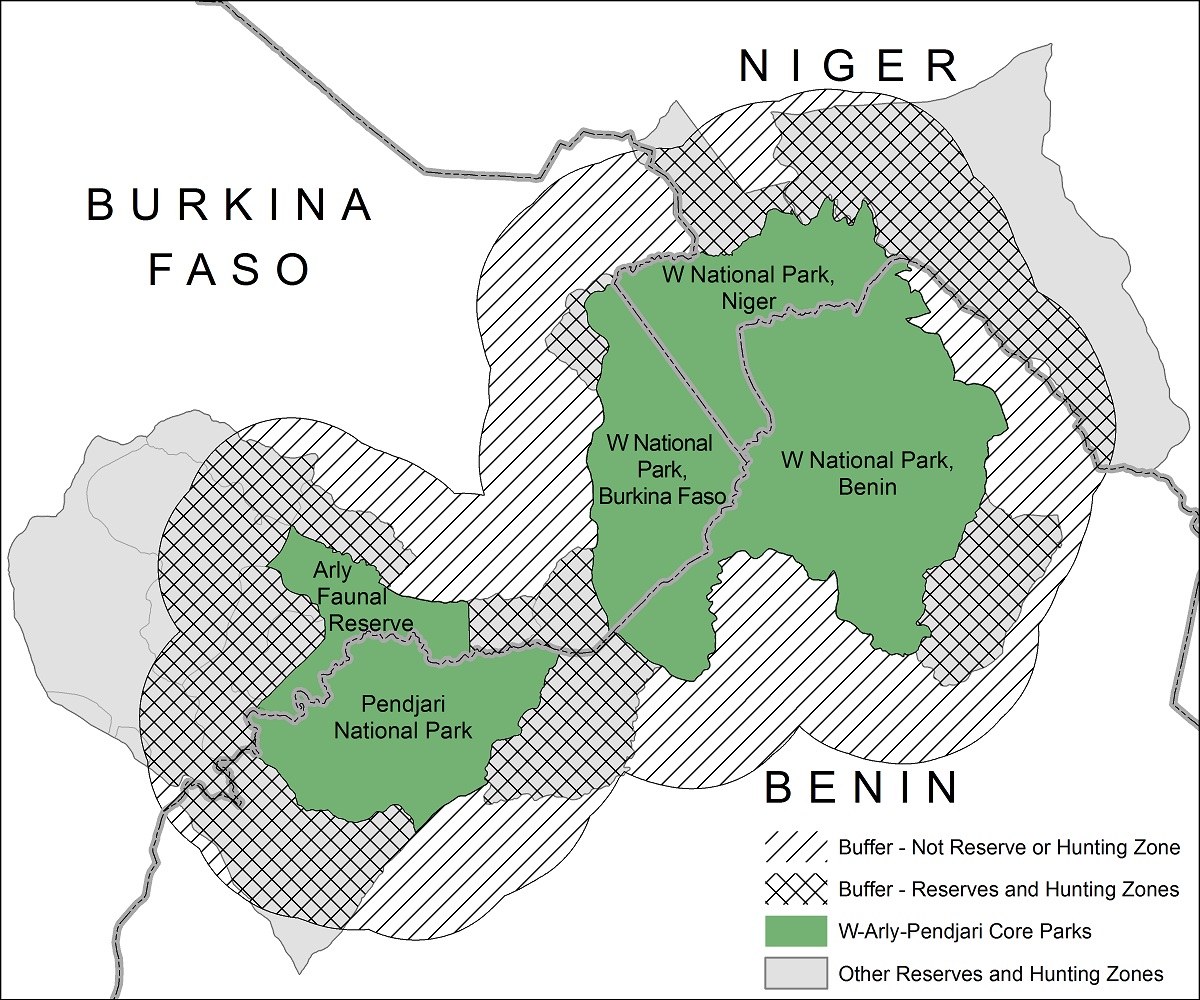}
\caption{W-Arly-Pendjari complex and surrounding buffer zones. \\ 
Source: USGS EROS, \url{https://eros.usgs.gov/westafrica/case-study/w-arly-pendjari-transboundary-biosphere-reserve}}
\label{fig:wap}
\end{figure} 
\FloatBarrier


\section{Model Setup}  \label{prob_setup}
\label{finite}

Let us first show in detail the reason behind the particular form of coupling we have assumed, the geometric mean $ \exp\mathbb{E}^p [\log X^q_{it}]$, as with it one can explicitly show the dynamics of the (geometric) average forest cover. 

We consider a continuum of individuals, or equivalently a large $N$, yielding a mean-field game. 
Mean-field games are a branch of game theory pioneered by 
\cite{lasry2007mean}.
To our knowledge, our application of this powerful framework is novel, both in its approach to heterogeneity and pro-environmental beliefs. 
Our application of this powerful framework introduces two novel features, particularly underexplored in environmental and natural resource economics: first, it successfully couples heterogeneity in beliefs and attitudes with managed ecosystems by means of strategic decision making.  
Second, it establishes a framework for analyzing the feedback of global environmental dynamics on natural resource policies and their sustainability, which remains eminently tractable and estimable whilst maintaining the essence of the complexity of the problem. 
Since each individual $i$ has her own level of exogenous adherence to ATR beliefs $a_i \in A$, the forest cover spatial distribution $p(x,t)$ will depend jointly on the distribution of ATR beliefs, i.e. it should be defined as $p(x,a,t)$. 
This density is equipped with an initial distribution $ p_0(x,a,0)$, which is the initial joint distribution of forest cover and ATR adherence. 
We thus augment our filtration $\mathcal{F}_t$ to $\mathcal{F}_t^a$, which includes the information available to all individuals at time 0 over the initial state of both forest and adherence.

It must therefore be that the beliefs have a marginal distribution
$$
\int_x p(dx, a, t) = \int_x p_0(dx, a,0) := p_0(a),
$$
which is the distribution of ATR beliefs among the population. This allows to separate forest cover and beliefs in the joint distribution: 
\begin{equation}
p(dx,da, t) = p^a(dx,t) p_0(da) \label{dec}
\end{equation}
where $p^a$ is a probability measure in $x$ for $p_0$-almost any $a$. 

Using \eqref{dec}, we can write the interaction term between agents as

\begin{equation}
\bar{X}_t = \exp \int_{[0,\infty) \times A} \log x \ p( dx, da, t)   =  \exp \int_{[0,\infty)} \int_A \log x \ p^a( dx,t) p_0(da) . \label{geomcont}
\end{equation}
Note that for a discrete number $j=1,\dots,n$ of observations $x_j$, $p^a(x,t)$ is exactly the empirical measure associated to the realizations $x_j$ and \eqref{geomcont} becomes the familiar expression for the geometric mean $(\Pi_j x_j)^{1/n}$. 



First, a straightforward application of It\^o's lemma yields 

$$
d \log{X_t} =\left ( \mu - \frac{\sigma^2}{2} - q_t\right) dt + \sigma d W_t
$$
where $\bar{W}_t$ is a $\mathcal{F}^a_t$-adapted standard Brownian motion. 
Since all agents face idiosyncratic and independent Brownian motions and $\mu, \sigma$ are assumed to be the same across the population, and expectations of a Brownian motion are equal to zero, the dynamics of the median forest cover $\bar{X}_t$ are given therefore by

\begin{equation}
d \bar{X}_t := d \exp \left ( \mathbb{E} [\log{X_{it}^q} |\mathcal{F}_t^a] \right ) = \left [ \bar{X}_t \left(\mu - \frac{\sigma^2}{2} \right)  - \int_A  \bar{q}_t p( da) \right]   dt. \label{geomsdem}
\end{equation}
where $\bar{q}_t = \int q p^a(dx,t)$ is the median deforestation policy for individuals with any adherence $a$.
The dynamics implied by \eqref{geomsdem} show clearly how the average state of the forest depends not only on the ecological parameters $\mu$ and $\sigma$, but also on the heterogeneity in beliefs within the population, which shape global deforestation.
We now conjecture individual deforestation policies to be linear in $\bar{X}$, i.e. $q_{it} := q(X_{it},t) = q_{it} X_{it}$ and thus assume $q_{it} \in [0,1]$ to be the average fraction of forest consumed by an individual at each $t$. This conjecture will be shown to be correct later - getting slightly ahead of ourselves, we begin with this conjecture as it is well-known that a single-agent (or representative) agent's problem with power utility and a geometric Brownian motion yields optimal policies linear in the state variable. Equipped with \eqref{geomsdem}, we can write our problem as shown in Eq. (8) of the paper.

\section{Proof of Proposition 1} \label{prop1}

Let us now focus on the problem of finding an equilibrium. In the limit of a large amount of individuals, and using \eqref{dec}, the $\mathcal{F}^a_t$-measurable deforestation policy in order to be a mean-field equilibrium must solve the following coupled backward/forward system of Hamilton-Jacobi-Bellman and Kolmogorov forward equations given by
\begin{eqnarray}
-V_t + \rho V &=& \sup_{q \in Q} \left \{ u(q^{g_1(a)} \bar{x}^{g_2(a)})  - q V_x \right \} + \mu x V_x  + \frac{\sigma^2}{2}x^2 V_{xx} \label{mfgsys1}\\
\partial_t p^a(x,t)  &=&  -\partial_x \left [(\mu x - q) p^a(x,t) \right ] + \frac{\sigma^2}{2} \partial_{xx} x^2 p^a(x,t) \label{mfgsys2}\end{eqnarray}
with the coupling condition 
\begin{equation}
    \bar{x} = \exp \int_{[0,\infty)} \int_A \log x \ p( dx,t)p_0(da),\label{coupl}
\end{equation}
under the decomposition \eqref{dec} and the boundary conditions for the HJB (backwards, so with a final condition) $V(x,T) = b(T; a)$ and Kolmogorov (forward, so with an initial condition) $p(x,0,a) = p_0^a(x_0,0)p_a(a)$.

The key feature of this system is the forward/backward dimension, essential characteristic of mean-field games. The Hamilton-Jacobi-Bellman equation (the optimality equation) is obtained backwards, as for all Bellman equations, starting from an endpoint condition. On the other hand, the Kolmogorov forward equation \eqref{mfgsys2} (the probability equation) starts from an initial resource distribution $p(x,0)$ which is known to the agents, and transports probability forward by means of the individuals' optimal decisions $q$ as well as its diffusive part. Starting from this distribution, they observe an endogenously time-evolving distribution $p(x,t)$ for all $t$, which identifies an aggregate forest cover level $\bar{x}$ via the coupling \eqref{coupl}. They then optimize proportionally to the ``forest rent'' $V_x$, since the optimal extraction policy $q^*$ conditional on the observation of $p$ for all times $t$ satisfies the first-order condition $q^* = u_q^{-1} (V_x)$
Since $u$ is increasing and weakly concave in $q$ for all $a \in A$, the optimal policy $q^*$ is decreasing in the resource rent associated to the ``average'' forest stock $V_x$, increasing in the utility associated to scarcity $\bar{x}$, and increasing when $g_1$ increases.

Using the condition \eqref{geomsdem}, we have thus obtained a closed-form expression for the dynamics of $\bar{X}_t$ and thus can incorporate the coupling condition of $\bar{X}$ given by \eqref{coupl} together with the controlled state process in the extended state space  $(X_t,\bar{X}_t,t)$, subject to the SDE 
$$
dX_t = (\mu X_t - q) dt + \sigma X_t dW_t
$$
and the ODE \eqref{geomsdem}. 

The HJB equation in the extended state space $(x,\bar{x},t) \in \mathbb{R}^+\times \mathbb{R}^+ \times [0,T]$ becomes
\begin{equation}
-V_t+ \rho V =\sup_{q \in Q} \left \{ \frac{1}{1-\gamma} \left [ q^{g_1(a)} \bar{x}^{g_2(a)} \right ]^{1-\gamma}   - q V_x(x,t) \right \} + \mu x V_x(x,t) +  \bar{\mu}^a \bar{x} V_{\bar{x}} + \frac{\sigma^2}{2}x^2 V_{xx} \label{hjb}
\end{equation}
with boundary condition $V(x,\bar{x},T) = b(x,\bar{x},T; a)$. Note that this HJB equation is random, as it depends on the $\mathcal{F}^a_t$-measurable distribution of beliefs $p_0(a)$.


The supremum is achieved by the optimal policy

\begin{equation}q^*(x, \bar{x},a, t) = \left ( \frac{V_x}{g_1  \bar{x}^{\nu_{\bar{x}}}} \right )^{- \epsilon_q }\label{optq}
\end{equation} 
where $ \epsilon_q$ is the elasticity of intertemporal substitution for forest use given by 
$$
\epsilon_q (a)= - \frac{u_q }{ u_{qq}}\frac{1}{q} = (1-g_1(a) (1-\gamma) )^{-1} 
$$
and $\nu_q :=\nu_q(a) =q \frac{u_q }{u}  = g_1(a)(1 - \gamma) $ is the point elasticity of utility with respect to $q$, and similarly for $\bar{x}$, $\nu_{\bar{x}}:= \bar{x} \frac{u_{\bar{x}} }{u} = g_2(a) (1-\gamma)$.
Inserting \eqref{optq} in the HJB yields the nonlinear PDE

\begin{equation}
    -V_t + \rho V = \frac{\bar{x}^{\nu_{\bar{x} }} }{1-\gamma} \left [\frac{ V_x}{g_1 \bar{x}^{\nu_{\bar{x} }} }\right ]^{\nu_{q} \epsilon_q} - \left [ \frac{V_x}{g_1 \bar{x}^{\nu_{\bar{x}}}  } \right ]^{\epsilon_q} V_x + \mu x V_x + \bar{\mu}^a \bar{x} V_{\bar{x}} + \frac{\sigma^2}{2} x^2 V_{xx} ,\label{hjbcrra}
\end{equation}.

We now assume \emph{for simplicity} a bequest function of the form $b = h^{-1/\epsilon_q} \frac{X_T^{\nu_q}\bar{X}_T^{\nu_{\bar{x}}} }{1-\gamma}$, where $h:= h(a), h' > 0, h: [0,1] \to [0,1]$. Individuals thus want to leave at end of life $T$ an amount of forest (both local and global) which depends on their own ATR adherence. 

Using as \emph{ansatz} for the value function a time-separable guess $V(x,\bar{x},t) =f(t) \frac{x^{g_1 (1-\gamma)}  \bar{x}^{g_2(1-\gamma)}}{1-\gamma} $ for a continuous function $f: \mathbb{R}^+ \to [0,1]$, differentiable at least once, that exploits the homothetic and multiplicative form of the objective function. 
This \emph{ansatz} implies that if an appropriate $f(t)$ is found, then the optimal deforestation policy is given by $q^* = f(t)^{-\epsilon_q} x$ for $p_0$-almost any $a$, which means that this form will apply for each individual, fixed any $\bar{q}$. This validates our conjecture of the median deforestation policy being linear in $\bar{X}_t$. 
After some lengthy computations we obtain that the function $\phi(t)$ needs to solve the ODE 
$$
f'(t) = -f(t) ^{\nu_q \epsilon_q} (1-\nu_q) - f(t) \left [ \mu \nu_q + \left ( \mu - \frac{\sigma^2}{2} - \int_A \bar{q}_t p(da) \right ) \nu_{\bar{x}} + \frac{\sigma^2}{2} \frac{\nu_q}{\epsilon_q} - \rho 
 \right ]
$$
with boundary condition $f(T) = h^{-1/\epsilon_q} $. This nonlinear ordinary differential equation has a closed form solution:
\begin{eqnarray}
f(t) &=&  \left [ \frac{\exp \left ( - \epsilon_q   (T - t) \right )  ( 1 - \nu_q + h C )}{C} \right ]^{- \frac{1}{\epsilon_q}} \label{foft}\\
     C&=&\mu \nu_q + \left ( \mu - \frac{\sigma^2}{2} - \int_A \bar{q}_t p(da) \right ) \nu_{\bar{x}} + \frac{\sigma^2}{2} \frac{\nu_q}{\epsilon_q} - \rho, \label{copt}
\end{eqnarray}
which in turn yields the optimal individual deforestation rate, conditional on observing $\bar{q}$ and for any $a$, given by 
    
\begin{equation}
    \frac{q^a_t}{X_t} = \frac{\exp \left ( - \epsilon_q C  (T - t) \right )  ( 1 - \nu_q + h C ) }{C}
    \label{optex} 
\end{equation}
where $C$ is again given by \eqref{copt}.
The function $f(t)$, which is pinned down using the final boundary condition on $V(x,\bar{x},T)$, depends clearly on the individual adherence $a$, the bequest $h$ and continuously on a combination of all model parameters as well as the  mean extraction integrated over the whole beliefs distribution. Note that for $a=0$ and $p(a) = 0$ one reverts to a standard consumption problem under uncertainty. 

However, \eqref{optex} is not yet a solution of our problem, because it depends on the median deforestation policy $\int_A \bar{q}_t p_0(da)$. What is then now required is to prove the existence of a mean-field equilibrium, and thus solve a fixed point problem. We need to then prove self-consistency of the mean-field equilibrium solution $q^{MFE}$.
Let $q^*_t$ be a $\mathcal{F}^a_t$-measurable extraction policy for our problem, and assume it to be admissible, such that $\mathbb{E} \int_0^T | (q_t^*)^2 | dt < \infty$. Let $\bar{X_t}$ be a $\mathcal{F}^a_t$-measurable random variable given by $\bar{X_t} = \exp (\mathbb{E} \log X^*_t) $, where $X^*_t$ is the forest cover associated with the policy $q^*_t$. The policy $q^*_t$ is a mean-field equilibrium if $q^*_t$ is optimal for each agent upon this choice of $\bar{X_t}$, under the consistency condition that $\bar{X}_t = \exp  \int  \log x^* p(dx^*,da,t)$ for all $t \in [0,T]$. This implies that the equilibrium is associated with a continuum of optimal choices, weighed by each individual adherence, in a framework where essentially each agent in the continuum faces an independently distributed copy of the same optimization problem, based on the intial draw from the beliefs distribution. In other words, the equilibrium happens at the point for which $  \int_A q^a_t/x \  p_0(da) = \int_A  \bar{q}_t   p_0(da)$ for all $t \in [0,T]$.

This is the case if there exists a function for the population forest consumption averaged over the whole beliefs distribution $\bar{q}^a_t: [0,T ]\to [0,1]$ that solves the fixed point relation given by 

\begin{equation}
    \int_A \bar{q}_t p_0(da) := q^a_t =  \int_A  \frac{\exp \left ( - \epsilon_q C  (T - t) \right )  ( 1 - \nu_q + h C ) }{C} p_0(da). \label{eqmf}
\end{equation}
If this fixed point exists, there exists a mean-field equilibrium where each individual consumes an amount of forest inversely proportional to their \emph{own} individual beliefs $a$ and to the collective average actions $\bar{q}^a_t$ of the population, which are defined by the fixed point in \eqref{eqmf}. 


Assuming the right-hand side exists and is bounded, which is equivalent to assuming a reasonably well-behaving distribution of beliefs $p(a)$, we study the map $Q_A (q^a_t)_t:Q[(0,T)] \to Q ([0,T)]$ given by

$$
Q_A (q^a)_t := \int_A f(t, q^a_t, a)  p_0(da)
$$
where $f(t, q^a, a)  = \frac{\exp \left ( - \epsilon_q C  (T - t) \right )  ( 1 - \nu_q + h C ) }{C}$, which is the optimal policy for \emph{any} median policy $q^a$. This map is a contraction, since it is defined in $[0,T] \to [0,1]$ and for any $q_1^a, q_2^a \in Q^A[(0,T)]$ in which $q_i^a = \int_A q_i p(da)$ (i.e. the dependence on $a$ has been integrated out) one has

\begin{eqnarray*}
| Q_A (q_1^a)_t -  Q_A (q_2^a)_t | & = & \sup_{t \in [0,T]} \left |   \int_A f(t, q^a_1, a)^{- \frac{1}{\epsilon_q}} p_0(da) - \int_A f(t, q^a_2, a)^{- \frac{1}{\epsilon_q}} p_0(da) \right |  \\
& = & \sup_{t \in [0,T]} \left |  \int_A   \left [  f(t, q^a_1, a)^{- \frac{1}{\epsilon_q}} -f(t, q^a_2, a)^{- \frac{1}{\epsilon_q}} \right ]  p_0(da) \right |  \\
&\leq & \int_A\left [1-\nu_q + h  \left ( \mu  \nu_q + \left (\mu- \frac{\sigma^2}{2} \right )\nu_{\bar{x}} + \frac{\sigma^2}{2} \frac{\nu_q}{\epsilon_q} \right ) \right ] \times  \\
& &  \times \frac{h | q_1^a - q^a_2 |  \exp \left ( - \epsilon_q \nu_{\bar{x}} | q_1^a - q^a_2 |   T \right )    }{\nu_{\bar{x}}}  p_0(da) \\
& \leq & \bar{Q} \  || q_1^a - q_2^a ||,
\end{eqnarray*}
for some constant $\bar{Q}$ in which $a$ has been integrated out, since $a$ is a random variable with bounded support $A \in [0,1]$ and all terms $\nu_q, \nu_{\bar{x}}, \epsilon_q, 1/\epsilon_q$ are bounded for all realisations of $a \in A$. Note also that $\epsilon_q \to 1$ as $a \to 0$. 
One then can apply Banach's fixed point theorem which guarantees the existence of a solution $\bar{q}^*_t := \int_A \bar{q}^*_t p_0(da)$
There thus exists an equilibrium deforestation rate for all times $t \in [0,T]$ and adherences $a \in p_0(a)$,  $q^{MFE}_t$, given by 

\begin{eqnarray}
       q^{MFE}_t &=&  \frac{\exp \left ( - \epsilon_q C^{\bar{q}^*_t }  (T - t) \right )  ( 1 - \nu_q + h C^{\bar{q}^*_t } ) }{C^{\bar{q}^*_t }} \\
        C^{\bar{q}^*_t }  &=&\mu \nu_q + \left ( \mu - \frac{\sigma^2}{2} - \bar{q}^*_t  \right ) \nu_{\bar{x}} + \frac{\sigma^2}{2} \frac{\nu_q}{\epsilon_q} - \rho,  \label{qmfegeneq}
\end{eqnarray}
and $\bar{q}^*_t $ solves \eqref{eqmf}.\\ 

Consider now the infinite horizon case $T \to \infty$. 
The HJB equation \eqref{hjbcrra} is then solved by setting $V_t = 0$, and we have a stationary solution $q^a_t = q^a$ by using the \emph{ansatz}

$$
V(x,\bar{x}) = B \frac{x^{g_1 (1-\gamma)} \bar{x}^{g_2 (1-\gamma)}}{1-\gamma}.
$$
By similar computations as before, we can obtain the individual deforestation policy, conditional on the median policy $\int_A \bar{q} p_0(da) $ given by 

\begin{equation}
q^a = \epsilon_q  \left ( \rho - \mu ( \nu_q + \nu_{\bar{x}} ) - \frac{\sigma^2}{2} \left (  \frac{\nu_q}{\epsilon_q} + \nu_{\bar{x}} \right ) + \nu_{\bar{x}} \bar{q}^a \right )X_t \nonumber. \label{qmfegen}
\end{equation}
The equilibrium is achieved by imposing the self-consistency condition 

$$
\int_{ A } q^a p_0(da) = \int_A \bar{q}^a p_0(da) 
$$
Integrating the optimal policy over all adherence levels $a$, one obtains the equilibrium condition
$$
\int_A \bar{q} p_0(da) = \int_A \epsilon_q  \left ( \rho - \mu ( \nu_q + \nu_{\bar{x}} ) - \frac{\sigma^2}{2} \left (  \frac{\nu_q}{\epsilon_q} + \nu_{\bar{x}} \right ) + \nu_{\bar{x}} \int_A \bar{q} p_0(da) \right ) p_0(da).
$$
One can then use the law of iterated expectations, solve for the equilibrium median policy, and obtain Equation \eqref{qmfe} in Proposition 1.

\begin{figure}
    \centering
    \includegraphics[width=\linewidth]{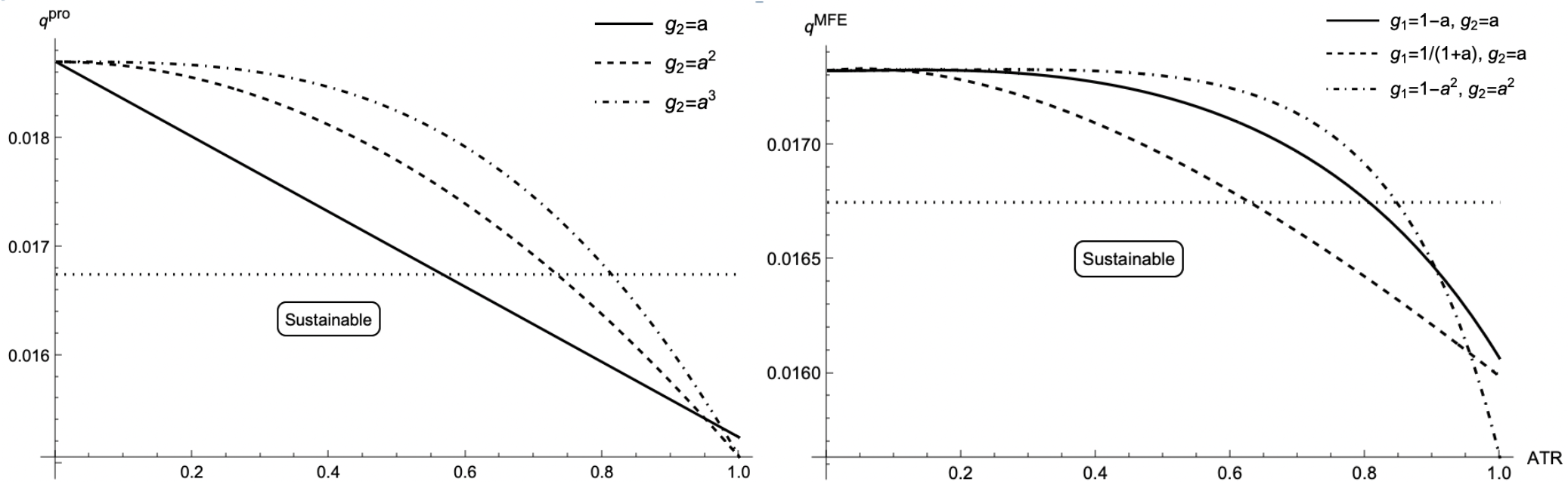}
    \caption{Optimal deforestation rates and ATR adherence. Left panel: model with only pro-environmental attitudes ($g_1 = 1$) for different choices of $g_2$ ($\gamma = 1.5$). Right panel: model with adherence affecting both individual consumption and pro-environmental attitudes, for different choices of $g_1$ and $g_2$ ($\gamma = 2.2$). Full and dot-dashed lines: local forest consumption and global scarcity are perfect substitutes. Other parameters: $\rho = 0.02, \mu = 0.018, \sigma = 0.05$. Beliefs distribution $p_o(a) \sim \text{Beta} (0.55, 2.55)$, estimated from the Benin census data. }
    \label{mod_fig_1}
\end{figure}

Figure \ref{mod_fig_1} in the paper shows the shape of the stationary deforestation rate $q^{MFE}$. The left-hand panel is for a model with only pro-environmental attitudes, where $g_1 = 0$ for all $a$, and ATR adherence affects individual utility only via global scarcity. The panel plots the equilibrium deforestation rate $q^{pro}$ for three different choices of $g_2$, linear ($g_1 = a$), quadratic and cubic, in order to uncover patterns that can then be matched with the data. The interpretation of $g_2$ is straightforward: the more its convexity increases, and the more individuals will only care about the state of the environment only for ``high'' levels of ATR adherence. This is directly reflected in the shape of the deforestation policy. The right-hand panel shows the model with both $g_1, g_2 \geq 0 $, such that individuals care both about their own local forest, via sacralization of their lived environment, and the (average) global state of the forest across the entire population. 
The panel shows that if local forest consumption and global scarcity are perfect substitutes (i.e. $g_1 + g_2 = 1$), then the optimal deforestation policy is nonlinear even for linear choices for $g_1, g_2$, and the two mechanisms of valuing local vs. global environment balance each other except for high levels of ATR adherence. For both scenarios, Figure \ref{mod_fig_1} shows that for high levels of adherence deforestation can switch from unsustainable (above the dashed line, equal to $\mu - \sigma^2/2$) to sustainable, such that one observes a net forest growth even with active deforestation. This switch is not possible if individuals' adherence affects only local forest use and there are no pro-environmental attitudes at play. 

For reference, it is useful to show the more compact expression for the optimal deforestation rate when ATR influences forest utility only via pro-environmental attitudes, i.e. when $g_1 = 1$ everywhere and assume $g_2(a) = a^k$, where $k \in \mathbb{R}^+$. This form of $g_2$ allows to flexibly specify the shape of the ATR-environment nexus, such that the higher is $k$, the more individuals care about the environment only for high levels of adherence. We then have

\begin{eqnarray}
q^{\text{pro}} & =&   q^*_{a=0}  + \frac{\nu_{\bar{x}}}{ \gamma }\left (   \mu  - \frac{\sigma^2}{2}  - \frac{ \gamma  q^*_{a=0}  - (\mu  - \frac{\sigma^2}{2}) (1-\gamma) \langle \bar{a}^k \rangle   }{1-(1-\gamma)  \langle \bar{a}^k \rangle }\right )  \label{qpro}\\
q^*_a &= & \epsilon_q \left [ \rho - \mu \nu_q - \frac{\sigma^2}{2} \frac{\nu_q}{\epsilon_q} \right ] \label{qnoint}
\end{eqnarray}
where the quantity $q^*_{a=0}$ is the optimal deforestation policy \eqref{qnoint} when $a=0$ everywhere, without any effect played by both ATR adherence and pro-environmental beliefs, $\langle \bar{a}^k \rangle := \int_{A} a^k \ p_0(da)$ is the $k$-th raw moment of the ATR adherence distribution across the population and $q^*_a$ is the optimal deforestation policy without pro-environmental beliefs, i.e. when $g_2 = 0$ everywhere, there are no interactions and each individual chooses its deforestation policy based only on observing local forest stocks.
Note that $\epsilon_q (0) = \gamma$, so that in absence of ATR beliefs the elasticity of intertemporal substitution of deforestation is the inverse of risk aversion. 
Here $q^*_0$ is essentially an upper bound, and since $g_2' > 0, \nu_{\bar{x}} < 0  \ \forall \ a \in A$, the deforestation rate with only pro-environmental beliefs is thus a \emph{reduction} in deforestation with respect to the ``no ATR, no interactions'' rate. 
Figure \ref{mod_fig_1} presents an illustration of the optimal deforestation rates under different functional choices. 

Once equipped with $q^{MFE}$, one can integrate in time the equilibrium Kolmogorov equation to obtain the (transition) forest spatial density .
We can now use the linearity in $x$ of \eqref{qmfegen} for both finite and infinite time and the decomposition \eqref{dec} to obtain the measure $p(x,a,t)$. Using the Kolmogorov forward equation at the optimal control $q^{MFE}_t$ for the joint distribution $p(x,a,t)$ we obtain

$$
\partial_t p(x,a, t)  =  -\partial_x \left [ \mu -  \frac{\exp \left ( - \epsilon_q C^{\bar{q}^*_t }  (T - t) \right )  ( 1 - \nu_q + h C^{\bar{q}^*_t } ) }{C^{\bar{q}^*_t }}  \right ] x \  p(x,a, t)  + \frac{\sigma^2}{2} \partial_{xx} x^2 p (x,a,t).
$$
where $C^{\bar{q}^*_t }$ is given by \eqref{qmfegeneq} and is evaluated at the equilibrium median deforestation rate $\bar{q}^*_t $. 
Since $p_0(da)$ is a probability measure, we can apply the dominated convergence theorem and integrate the PDE over the domain $A$, using \eqref{dec} and the time invariance of $p_0$, which yields

$$
p_0(a) \partial_t p^a(x,t) =  - p_0(a)\partial_x \left [\mu -\frac{\exp \left ( - \epsilon_q C^{\bar{q}^*_t }  (T - t) \right )  ( 1 - \nu_q + h C^{\bar{q}^*_t } ) }{C^{\bar{q}^*_t }}  \right ]x \ p^a(x,t) + \frac{\sigma^2}{2} p_0(a) \partial_{xx} x^2 p^a(x,t).
$$
Integrating in time and using very well-known results, one obtains that solving the above PDE under the normalization condition $\int p^a(dx,t)=1$ yields a transition density given by 
\begin{eqnarray}
p(dx,da,t) &=& \frac{p_0(da) }{dx \sigma \sqrt{2 \pi t}}\exp \left (- \frac{1}{2 \sigma^2 t} \left [\log \left ( dx \right )  - \left ( \mu  - \frac{\sigma^2}{2} - \right . \right . \right .  
\nonumber\\
    & & - \left .  \left .  \left . \int_0^t \frac{\exp \left ( - \epsilon_q C^{\bar{q}^*_s }  (T - s) \right )  ( 1 - \nu_q + h C^{\bar{q}^*_s } ) }{C^{\bar{q}^*_s }} ds \right ) t \right ]^2 \right ),\nonumber \label{mfetpd_2}
\end{eqnarray}
with starting condition $p(x,a,0) = p_0(a)p_0(x_0,0)$. The reader can now verify that the same applies without the time integral for the infinite horizon case $q^{MFE}$ shown in Proposition 1, and that the cross-sectional spatial forest density evolves in time according to the log-normal transition density. 

As a more explicit example, the forest cover starting in year $t_0$ for any individual $i$ with adherence $a$ and only pro-environmental beliefs ($g_1 = 0$ for all $a$, assuming an infinite horizon and $g_2 = a^k$ for $k$ positive integer) evolves according to the transition density 

\begin{eqnarray*}
    p^a(x,t | x_0, t_0 ) &=& \frac{1}{x \sigma \sqrt{2 \pi (t-t_0) }}\exp \left (- \frac{1}{2 \sigma^2 (t-t_0) } \left [\log \left ( \frac{x}{x_0} \right )  - \left ( \mu  - \frac{\sigma^2}{2} - \right . \right .  \right . \\
    & & \left . \left . \left . - q^*_{0}  + \frac{1}{ \gamma } \left [ \nu_{\bar{x}}\left (   \mu  - \frac{\sigma^2}{2}  - \frac{ \gamma  q^*_{0}  - (\mu  - \frac{\sigma^2}{2}) (1-\gamma) \langle \bar{a}^k \rangle   }{1-(1-\gamma)  \langle \bar{a}^k \rangle }\right )  \right ]\right )( t -t_0) \right ]^2 \right )
\end{eqnarray*}
where $\bar{a}^k = \int_A a^K  p_0(da)$ is the $k$-th raw moment of the beliefs distribution. 

An Ito process with \eqref{mfetpd_2} as transition density, however, has no stationary distribution, as the cross-sectional variance explodes in time unless deforestation is unsustainable, i.e. $\mu - \sigma^2/2- q^{MFE}_t < 0 $, and in such a case it is a degenerate one $\delta(x=0)$ - no more forest.
We therefore exploit the fact that there is a maximum forest area available to the agents $S$, identifiable as the total overall geographical area available for the population.
We therefore augment \eqref{geomsdem} with a reflection term $- dL_t$, where $L_t$ is a nonnegative, right-continuous and nondecreasing process that occurs both instantaneously as well as with infinitesimal magnitude (so that the sample path   s of $X_t$ are always continuous) whenever $X_t = S$, $\mathcal{F}^a_t$-measurable for all $t\geq 0$ and thus adapted to $\mathcal{F}^a_t$ generated by
the Brownian motion process together with the adherence distribution. 
The term $dL_t$ is of finite variation, and the measure of the time spent on the boundary has Lebesgue measure zero with probability one.
This term does not affect the optimal deforestation policy, as it is in feedback form, linear in $X_t$ and the boundary condition of the HJB equation is one for a final time $T$, as well as a natural one $V(0,t) = 0$ which is easily satisfied by our policy and is unaffected by the upper bound. 
Furthermore, we have by construction $dtdL = dWdL = dLdL  = 0$ and thus the policy remains the one shown in Proposition 1. 
It does however affect the transition density: we need to solve the KFE \eqref{mfgsys2} now with an added Robin condition that the conditional probability mass of at the boundary is zero. 
We evaluate the limit of a very useful result by \cite{veestraeten2004conditional}, using the boundary condition given by 

$$
\lim_{x \to S} \frac{\sigma^2}{2} \partial_x x p^a(x, t| x_0, t_0) - \mu^* x  p^a(x, t| x_0, t_0) = 0
$$
where we define 
$$
\mu^* := \mu  - \frac{\sigma^2}{2} -  \int_0^t \frac{\exp \left ( - \epsilon_q C^{\bar{q}^*_s }  (T - s) \right )  ( 1 - \nu_q + h C^{\bar{q}^*_s } ) }{C^{\bar{q}^*_s }} ds,
$$
and obtain the equilibrium transition density with reflections at $S$ as 

\begin{eqnarray}
  p^a(x,t | x_0, t_0) &=& \frac{1}{ x \sigma  \sqrt{2 \pi t}} \exp \left [ -\frac{(\ln x - \ln x_0 - \mu^* (t -t_0))^2}{2 \sigma^2 (t -t_0)} \right ] +\nonumber \\   
  & & + \frac{1}{x \sigma \sqrt{2 \pi t}}\exp \left [ -\frac{2 \mu^* }{\sigma^2 } (\ln S + \ln x_0  )   \right ] \exp \left  [ 
 - \frac{(\ln x - \ln x_0 + 2 \ln S - \mu^* (t -t_0))^2 }{2 \sigma^2 (t -t_0) }\right ] \nonumber + \\
 & & + \frac{2 \mu^*}{\sigma^2 x } \exp \left [ - \frac{2 \mu^*}{\sigma^2  } ( \ln S - \ln x) \right ] \Phi \left ( \frac{\ln x + \ln x_0 - 2 \ln S + \mu (t -t_0)}{
 \sigma (t -t_0) 
 }\right ) 
\end{eqnarray}
where $\Phi(z)$ is the Gaussian CDF.
The stationary distribution $p^a(x)$ is obtained setting $\partial_t p^a =0$ in the KFE, using $\mu^* \to \mu^s := \mu- \sigma^2/2 - q^{MFE}$, and after defining $y := \ln x$ is obtained by solving 
$$
0 = - \mu^s \partial_y y p^a + \frac{\sigma^2}{2}  \partial_{yy} y^2 p^a
$$
 with the boundary condition $p^a (S) = 0$, and is a power law-type distribution in $x$ (exponential in $y$) given by

\begin{equation}
p^a(y) = C \frac{\sigma^2}{2 \mu^s} \left ( S^{\frac{2 \mu^s}{\sigma^2}} - e^{\frac{2 \mu^s}{\sigma^2} y } \right )     
\end{equation}
where $C$ is a normalization constant. We note that a very similar result would be obtained by imposing instead a lower reflection boundary $S$, i.e. a minimum forest size (one tree, for example), or individual births and deaths, one of many stabilizing forces that can be found in the Appendix of \cite{gabaix2016dynamics} that generate power law-type stationary distributions.

\section{Proof of Proposition 2} \label{prop2}

Assume $a = 0$, and hence $g_1 = 1, g_2  = 0$. It's then clear that $q_0^* \geq 0$ since we assumed $\rho > \mu (1-\gamma) - \frac{\sigma^2}{2} \gamma (1 - \gamma)$ for the problem to be well-posed. Because of our parametric assumption, then it's easily seen that $\partial_a q^*_a < 0$ for all $a$. Then $q^*_1 = \rho < q^*_0$ and therefore $q^*_a \geq 0 $ for all $a \in A$. Since $g_2' \geq 0 $ by assumption, then if $\gamma < 1$ $\partial_a \nu_{\bar{x}} < 0 $ and one can see that $ \frac{ \gamma  q^*_{0}  - (\mu  - \frac{\sigma^2}{2}) (1-\gamma) \langle a^k \rangle  }{1-(1-\gamma)  \langle a^k \rangle } > \mu - \frac{\sigma^2}{2}$, and viceversa for $\gamma< 1$, so it follows that $\partial_a q^{pro} \leq 0$ for all $a \in A$.  Note this is exactly what \eqref{geneul} implies, in a simplified scenario and for our particular choice of $u$. One can make the same argument using the same parametric restriction, albeit for a more complicated expression, for the general policy given by Eq. 
\eqref{qmfe} in the paper. \\

Let us now show the impact on individual deforestation policies of a change in the beliefs distribution in the population. Let $\mu_a$ be a parameter (or a combination of parameters) that regulates the first-order stochastic dominance between different parametrizations of the beliefs distribution, i.e. 
\begin{equation}
\mu_a \text{ s.t. } \frac{d}{d \mu_a} \int_A a \times  p_0^{(\mu_a)}(da) >0. \label{stdom}
\end{equation}
Firstly, since $p_0(a)$ is a probability measure it is Lebesgue-integrable and so is its cumulative density $P_0(a)$. Parametrizing both as $p_0(a; \mu_a)$ and $P_0(a; \mu_a)$ and assuming that $\partial_{\mu_a} p_0, \partial_{\mu_a} P_0$ exist and are bounded (a simple requirement that holds for most non-degenerate parametrized distributions), for any bounded an continuous function $f(a)$ one can apply the bounded convergence theorem and show that 

$$
 \frac{d}{  d \mu_a} \int_A f(a) p_0(da; \mu_a) = \int_A f(a) \partial_{\mu_a} p_0(da; \mu_a).
$$

Integrating by parts and using Fubini's theorem we obtain
\begin{equation}
\int_A f(\bar{a}) \partial_{\mu_a} p_0(d\bar{a}; \mu_a) = f(a) \partial_{\mu_a} P_0(a; \mu_a) |^{\bar{A}}_{0} - \int_A f' (a)\partial_{\mu_a} P_0(da; \mu_a), \label{pr2}
\end{equation}
where $\bar{A}$ is the upper boundary of the domain of the beliefs distribution. 
For distributions in a bounded domain, which applies for example such as the uniform and Beta distributions, one has 

$$
\int_A f(\bar{a}) \partial_{\mu_a} p_0(d\bar{a}; \mu_a) =  f(\bar{A})\partial_{\mu_a} P_0(\bar{A}; \mu_a) - \int_A f' (a) \partial_{\mu_a} P_0(da; \mu_a).
$$
Furthermore, by the definition \eqref{stdom} of first-order stochastic dominance, an increase in the parameter that ranks the beliefs distributions in such order must generate a shift to the right of the beliefs CDF, hence $\partial_{\mu_a} P_0(da) \leq 0 $. 
Hence, the first term on the right-hand side of the previous expression is zero at its maximum, achieved at $\bar{A}$. 
It immediately follows that $\int_A f(\bar{a}) \partial_{\mu_a} p_0(d\bar{a}; \mu_a) = f(0)\partial_{\mu_a} P_0(0; \mu_a) - \int_A f' (a) \partial_{\mu_a} P_0(da; \mu_a)$. Now applying this result to our functional forms, one can see that for \eqref{qpro} we have $f = g_2$, and $g_2(0) = 0$, so it's easily checked that $d q^{pro}/d \mu_a  \geq 0$, implying that since global deforestation increases with decreasing global adherence, pro-environmental beliefs imply that individuals react to increasing scarcity with reduced individual deforestation policies. For the general form of the solution given by Equation \eqref{qmfe} in the paper, one needs to apply the same result to each term, noting that $g_1(0) = 1$ and so $\epsilon_q ( 0 ) = \gamma^{-1}$, and $\nu_{\bar{x}} (0) = 0$, and it's easily checked that $d q^{MFE} / d \mu_a \geq 0 $ and $d q^{MFE} / d \mu_a =0  $ for $a=0$ (i.e. when there are no pro-environmental beliefs. 

Regarding the switch from unsustainable to sustainable, one first sees immediately that without pro-environmental beliefs ($g_2 = 0$) the optimal deforestation policy can only either be sustainable or unsustainable regardless of adherence, since for $\gamma < 1$ in order for the problem to be well-defined one must have the forest growth rate $\mu$ low enough to satisfy the parametric assumption, and hence since $\partial_a q^*_a \leq 0 $ for all $a$ deforestation will always be unsustainable. Viceversa for sustainable deforestation, which occurs if $\gamma > 1$. Now, if one introduces back pro-environmental beliefs and $g_2 \neq 0$, 
Then, noting that $\nu_{\bar{x}} (0) = 0$ and that for $\gamma > 1$ we have $\nu_{\bar{x}} < 0$ for all $a$, then one can see immediately that for $\gamma > 1 $ there exist $\mu, \sigma, \mu_a$ such that  $q^{pro} < \mu - \frac{\sigma^2}{2} $ for some $a \in A$. Then since $g_1, g_2$ are continuous functions of $a$, because of the intermediate value theorem there exists an $\underline{a} \in A$ such that $q^{pro}(\underline{a})$ switches from unsustainable to sustainable. \\

Proposition 2 describes the key characteristics of the equilibrium deforestation policy. The first characteristic, as also evidenced by the reduced form results, implies that a greater individual adherence yields a lower deforestation rate. 
One can observe in Equation \eqref{qmfe} in the paper that the equilibrium individual deforestation rate is determined by the interplay of individual adherence to ATR and the median forest cover across the population, which is determined continuously by the median deforestation rate $ \tilde{q}^*$, whose heterogeneity is driven by the global distribution of ATR beliefs. 
A decrease in first-order stochastic dominance for the ATR distribution, therefore, yields a global average increase in deforestation, since there are ``less adherents'' across the population with less pro-environmental concerns, which yields a higher aggregate deforestation and lower aggregate forest cover. 
In turn, individuals with nonzero adherence which have not been personally affected by the change in ATR population distribution (i.e. it wasn't them who changed adherence) observe this change and as a response decrease in $dt$  their own deforestation policy based on their own preferences (i.e. their own adherence $a$), the ``average`` deforestation policy across all levels of adherence (the mean field) adjusts and the system remains in equilibrium.

\section{Model Estimation: beliefs, preferences and optimal deforestation} \label{struc}

In this section we fit the model to the data and estimate its parameters. 
In all that follows we will assume infinitely-lived agents, as having a fully explicit solution of the problem simplifies identification and estimation considerably. Furthermore, the dataset on ATR beliefs does not have information over the individuals' lifecycle, and thus the time-invariant nature of the optimal deforestation rate $q^{MFE}/X_t$ given by \eqref{qmfe} is better suited for our estimation requirements. 
We first estimate the beliefs distribution $p_0(a)$ from the Benin census data. Since we have information on ATR adherence only at the arrondissement level, we associate each arrondissement to an agent $i$ in our model, and assume that the reported ATR adherence is representative of all individuals in the arrondissement. This choice directly maps to the model primitives $g_1$ and $g_2$, as in our data ATR adherence is the percentage of adherents and thus naturally falls between 0 and 1 as is the case for the functions $g_1, g_2$. 
The model heterogeneity in agents' beliefs, therefore, is mapped directly to the observed heterogeneity across Benin arrondissements.

For our estimations we use tree canopy cover area (TA) as our $X_{it}$, expressed in square kilometers. We choose TA because forest cover is a specific, much more narrow LC type, with a lot more zeroes than tree canopy cover. More importantly, we want to look at trees in general since ATR sacralize trees as much as forests, as discussed in Section \ref{back: atr}. We therefore use NASA's Making Earth System Data Records for Use in Research Environments (MEaSUREs) Vegetation Continuous Fields Version 1 data product (VCF5KYR) created by \cite{song2018global} and is used regularly in studies of deforestation \citep{baragwanath2023collective}. The data provides historical annual global fractional vegetation cover (FVC) during the time of the local peak growing season, for the time period 1982 - 2016. 
Each pixel is characterized by the percentage of tree canopy cover ($\geq 5$ metres in height) at  5.6 km $\times$ 5.6 km spatial resolution. 
Unlike the ESA-CCI-LC dataset which is binary, VCF5KYR is continuous, providing an estimated percentage of tree cover in a given pixel.
We aggregate this quantity and obtain the total tree canopy area for each arrondissement, which will be the spatial unit associated to each representative agent $i$ in our model.

\begin{figure}
    \centering
    \includegraphics[width=0.5\linewidth]{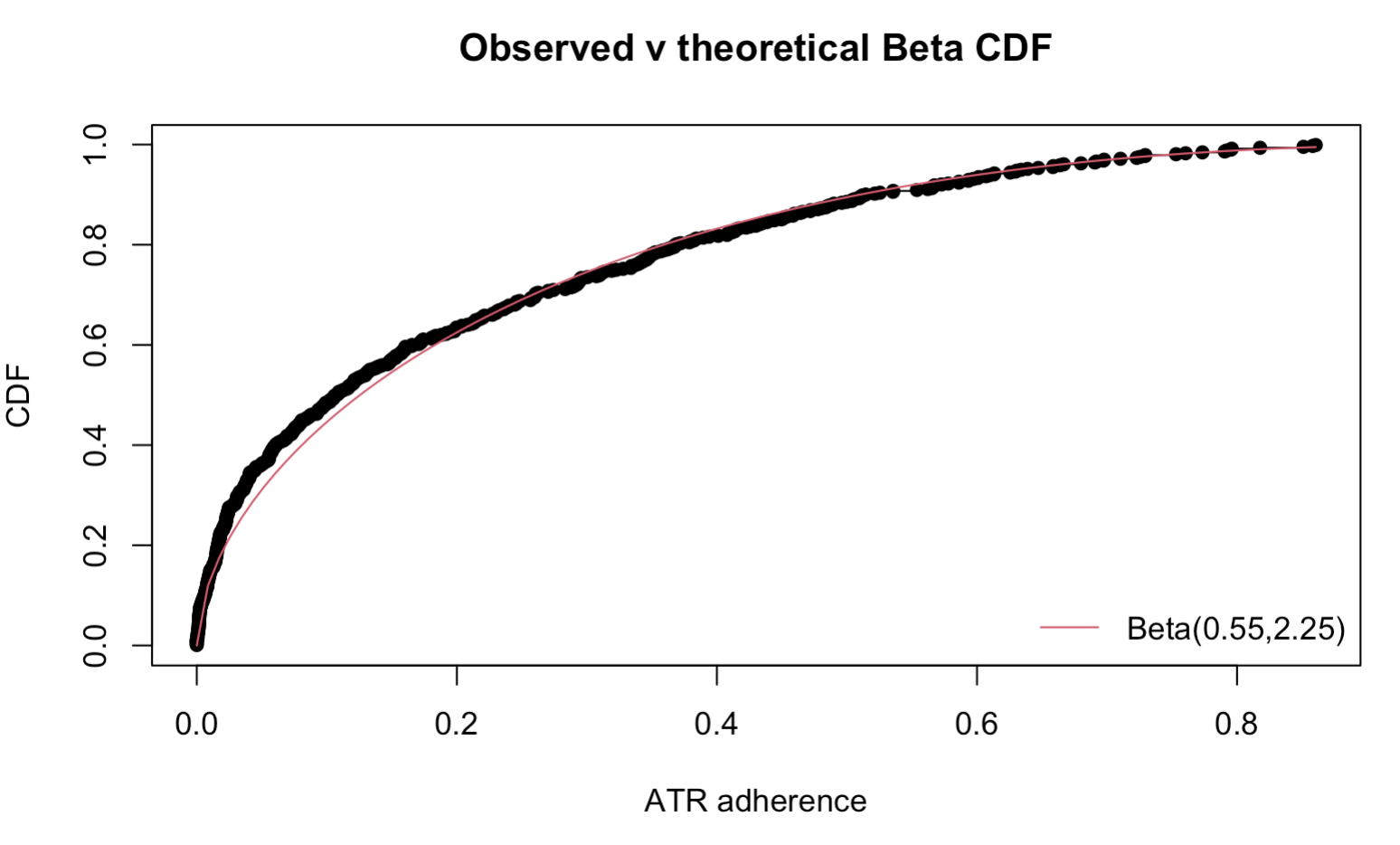}
    \includegraphics[width=0.5\linewidth]{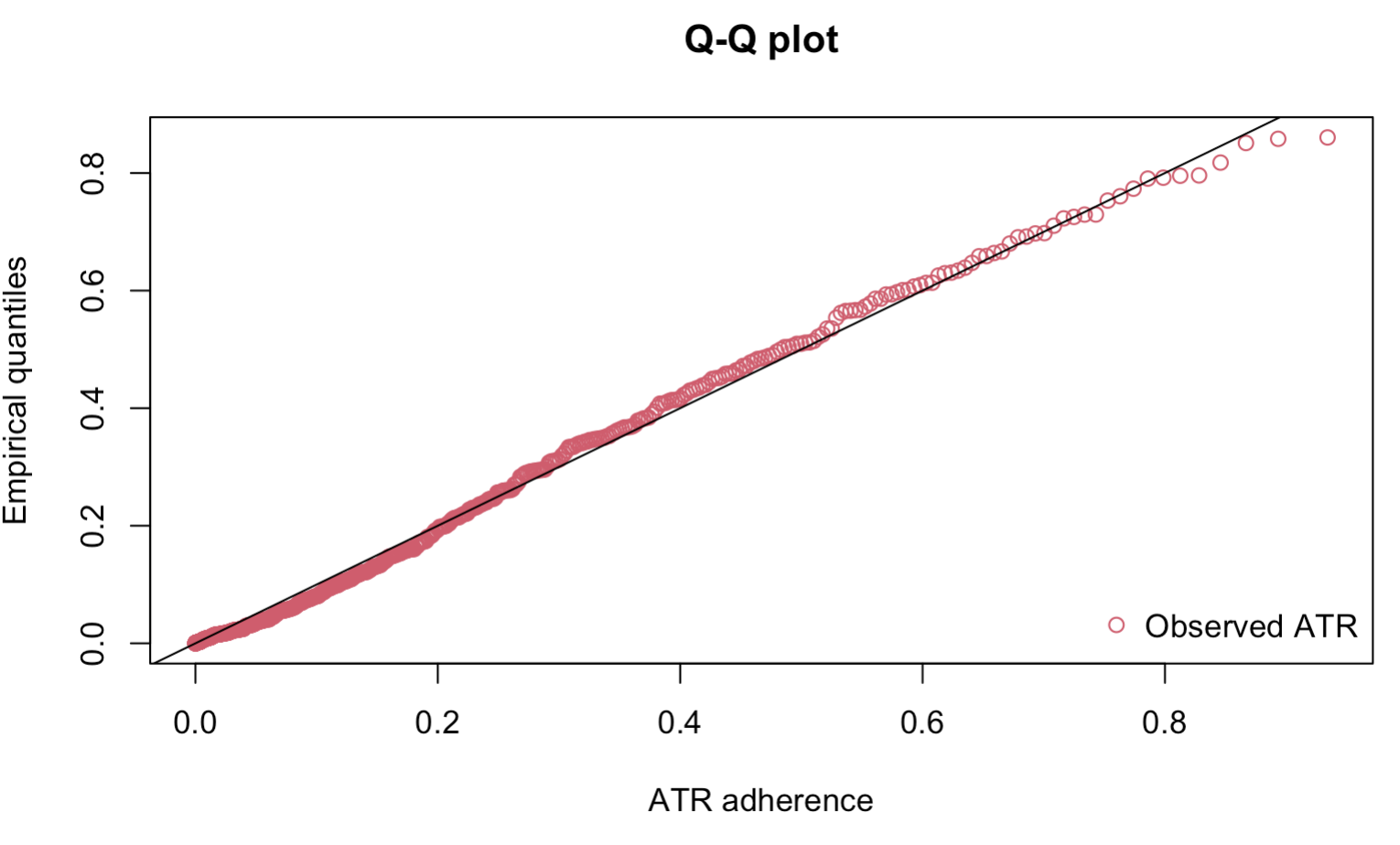}
    \includegraphics[width=0.5\linewidth]{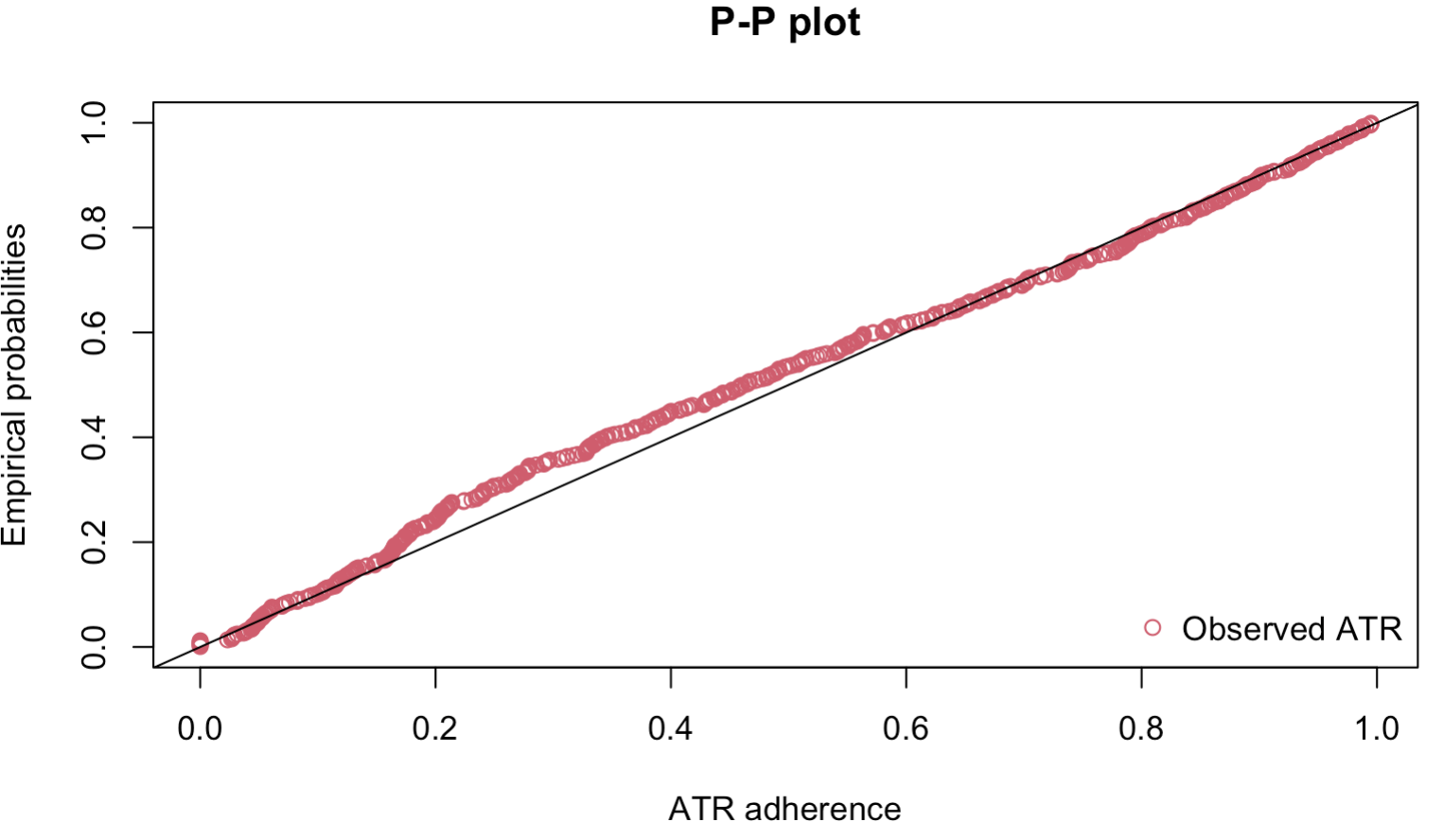}
    \caption{Maximum likelihood estimation of $p_0(a)$: Observed v theoretical Beta CDF (top), Empirical v theoretical quantiles (middle) and estimated probabilities (bottom).}
    \label{beta_fit}
\end{figure}

We first estimate the beliefs distribution $p_0(a)$ from the data. We first focus on the 2013 cross-section of our data, and normalise adherence in terms of the ``representative adherent", shown by the percentage of ATR adherence for each arrondissement in Benin. Given that our adherence measure is between 0 and 1, a natural candidate for its distribution is the Beta distribution, parametrised with the positive parameters $\alpha, \beta$ such that the distribution mean is given by $\frac{\alpha }{\alpha + \beta}$. 
We start with an initial guess of $\alpha = \hat{\mu}_a ( 1-\hat{\mu}_a) /\hat{s}_a^2 - 1$ and $\beta = (1- \hat{\mu}_a )(\hat{\mu}_a ( 1 - \hat{\mu}_a )/\hat{s}_a^2) -1 $, where $\hat{\mu}_a, \hat{s}_a$ are the mean and standard deviation of the observed ATR beliefs in the data, and estimate the parameters via maximum likelihood. We obtain that a Beta distribution  with $\alpha = 0.553$ (s.d. 0.028) and $\beta =2.251$ (s.d. 0.151) is an excellent fit for our data (see Figure \ref{beta_fit}).

A first issue we encounter in identifying the model's parameters is how to disentangle the ``uncontrolled'' drift parameter $\mu$ from the deforestation policy $q_t$, function of all model parameters, in the evolution equation of the tree canopy. Estimating the drift from individual trajectories in \eqref{gbm} for TA in each arrondissement $i$ or alternatively estimating it from the country-level geometric mean (the median), whilst perfectly feasible computationally, does not allow one to identify separately $\mu$ from $q^{MFE}$ or $ \tilde{q}^*$.
Since we assume $\mu$ to be constant across individuals, one would like to look in our data for an area where $q_t =0$ for all times, in order to exploit ``untouched'' areas to capture the natural rate of canopy growth. 
Benin, however, is a narrow, key-shaped country, measuring about 2.9 latitude degrees at its widest point (around 325 km) and is highly densely populated, making the search for such an area difficult.
As a solution to this challenge, we focus on regions where it is plausible to assume that human forest consumption is zero, or at least negligible. We turn our analysis towards the W-Arly-Pendjari (WAP) Complex which is a transnational park shared between the Republic of Niger, Burkina Faso and the Republic of Benin and has been a part of the UNESCO World Heritage List since 1996, shown in Appendix Figure \ref{fig:wap}. 
The park covers a major expanse of intact Sudano-Sahelian savannah, with vegetation types including grasslands, shrub lands, wooded savannah and extensive gallery forests.\footnote{For additional information refer to \url{https://whc.unesco.org/en/list/749/}}
We first obtain the canopy data for areas of WAP falling within Benin corresponding to the W and Pendjari national parks.
We then estimate the parameters based on the construction of a likelihood function derived from the transition probability density of the discretely sampled data, based on the uncontrolled version of \eqref{gbm}. Given that we only have access to yearly data, we bootstrap the maximum likelihood estimator (3000 repetitions) to get robust confidence bands for our estimate. 
We obtain an estimate of $\hat{\mu}=0.0482$ (s.d. 0.004),  i.e. a annual ``untouched'' growth rate of 4.82\%, for the forests in the WAP area, which we consider to be a reasonable estimate of the natural growth rate of the forest resources $\mu$ for Benin. Similarly, we estimate $\sigma^2$ by an equivalent procedure over the entire canopy cover of Benin since the diffusion coefficient is assumed to be constant across arrondissements. We obtain an estimate of $\hat{\sigma} = 0.258$ (s.d. 0.032), corresponding to a yearly variance of 6.5\%. We assume $\rho = 0.0487$, an estimate provided by  \cite{hyman2021pastoralists} for ecological (forest) goods obtained with surveys of Fulani people in Benin, and thus perfectly suited for our purposes. This estimate is similar to the 4.27\% average Benin Central Bank discount rate between 2010 and 2024.\footnote{Source: Central Bank of West African States (BCEAO).}

We estimate the ``untouched'' growth parameter $\mu$ as follows. We first obtain the forest cover data for areas of WAP falling within Benin corresponding to the W and Pendjari national parks. The forest cover grid cells that we keep for our estimates yield the Core Park area shown in Figure \ref{fig:wap}. 
Suppose $p(X_{it}, t |X_{i t-1}, t-1; \Theta)$ is the transition probability density of the area. 
The Markov property of Eq. \eqref{gbm} of the paper implies a log-likelihood function for the discrete sample given by $l_{TD}(\Theta)=\log (p(X_{it}, t |X_{i t-1}, t-1;\Theta))$ where $\Theta = [\mu, \sigma]$. 
The resulting estimator will be consistent, asymptotically normally distributed and asymptotically efficient under the usual regularity conditions for maximum likelihood estimation in dynamic models.
To perform exact ML estimation, one needs a closed form expression for $l_{TD}(\Theta)$. It is a well-known result that an uncontrolled geometric Brownian motion, via solving its corresponding Kolmogorov equation, admits a closed form, log-normal transition density:
\begin{eqnarray}
p^{WAP}(x,t+dt | x_0, t; \Theta) &=& \frac{1}{x \sigma \sqrt{2 \pi dt}}\exp \left (- \frac{1}{2 \sigma^2 dt} \left [\log \left ( \frac{x}{x_0} \right )  - \left ( \mu  - \frac{\sigma^2}{2} \right) t \right ]^2 \right ).\label{lnormtpd} 
\end{eqnarray}
Equipped with \eqref{lnormtpd}, the likelihood function can be easily constructed using the $n$ available observations in $T$ as
\begin{equation}
L_n(\Theta)=\prod_{i=1,t=1}^{n,T} p^{WAP}(X_{it+\Delta t},t+\Delta t | X_{it},t ; \Theta)p^{WAP}\left(X_{i,0}\right). 
\end{equation}
Setting all $p^{WAP}(X_{i,0}) =1$ as typically done (its weight decays quickly with observations), the log-likelihood to be maximized is
\begin{eqnarray}
\ell_n(\Theta)  &=&\log L_n(\Theta)=\sum_{i=1}^n\sum_{t=1}^T \ell_i(\Theta)  \\
& =&  \sum_{i=1}^n \sum_{t=1}^T\log p^{WAP}\left( X_{it+\Delta t},t+\Delta t | X_{it},t ; \Theta\right) .    
\end{eqnarray}

\begin{figure}
    \centering
    \includegraphics[width=0.49\linewidth]{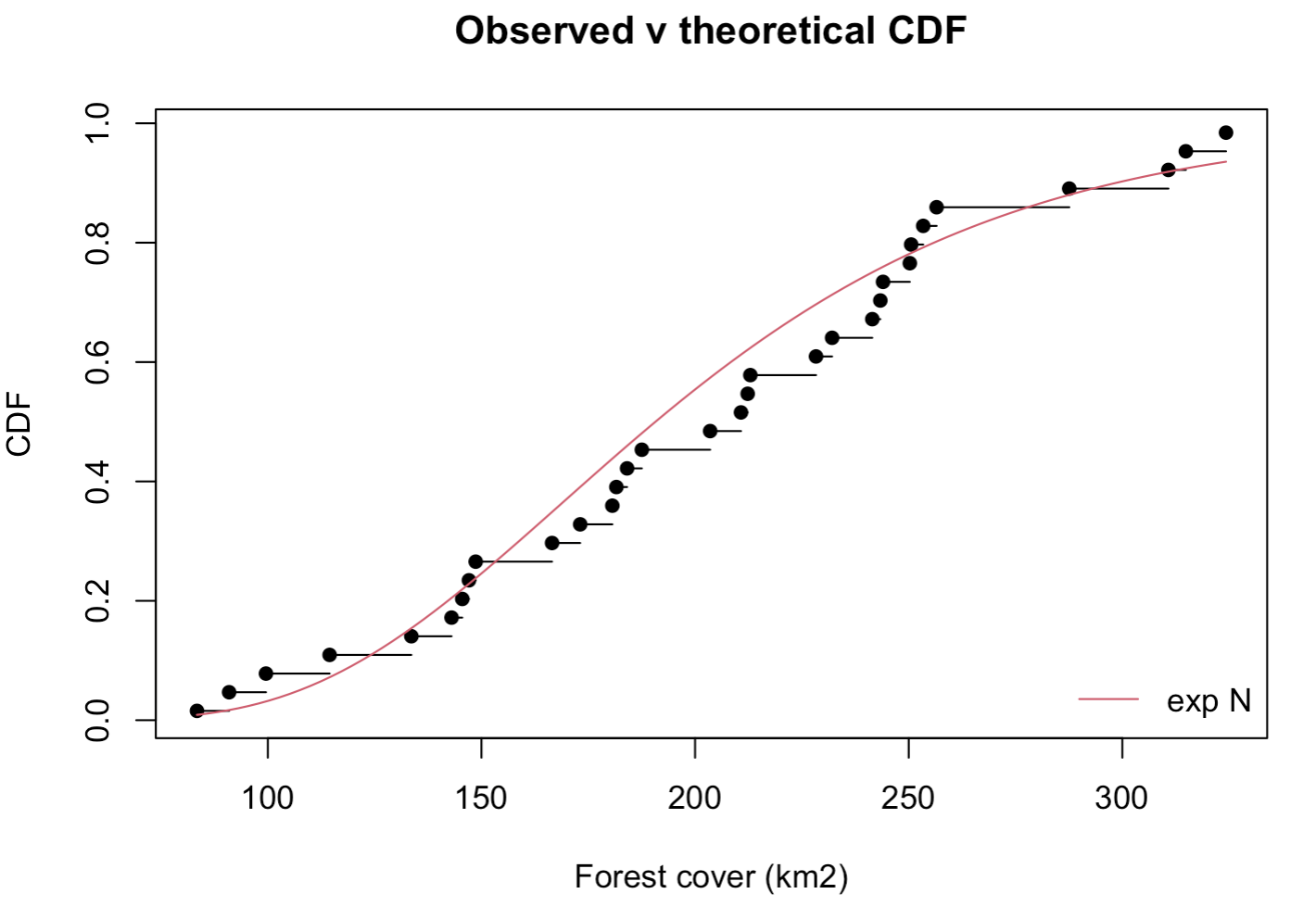}
     \includegraphics[width=0.49\linewidth]{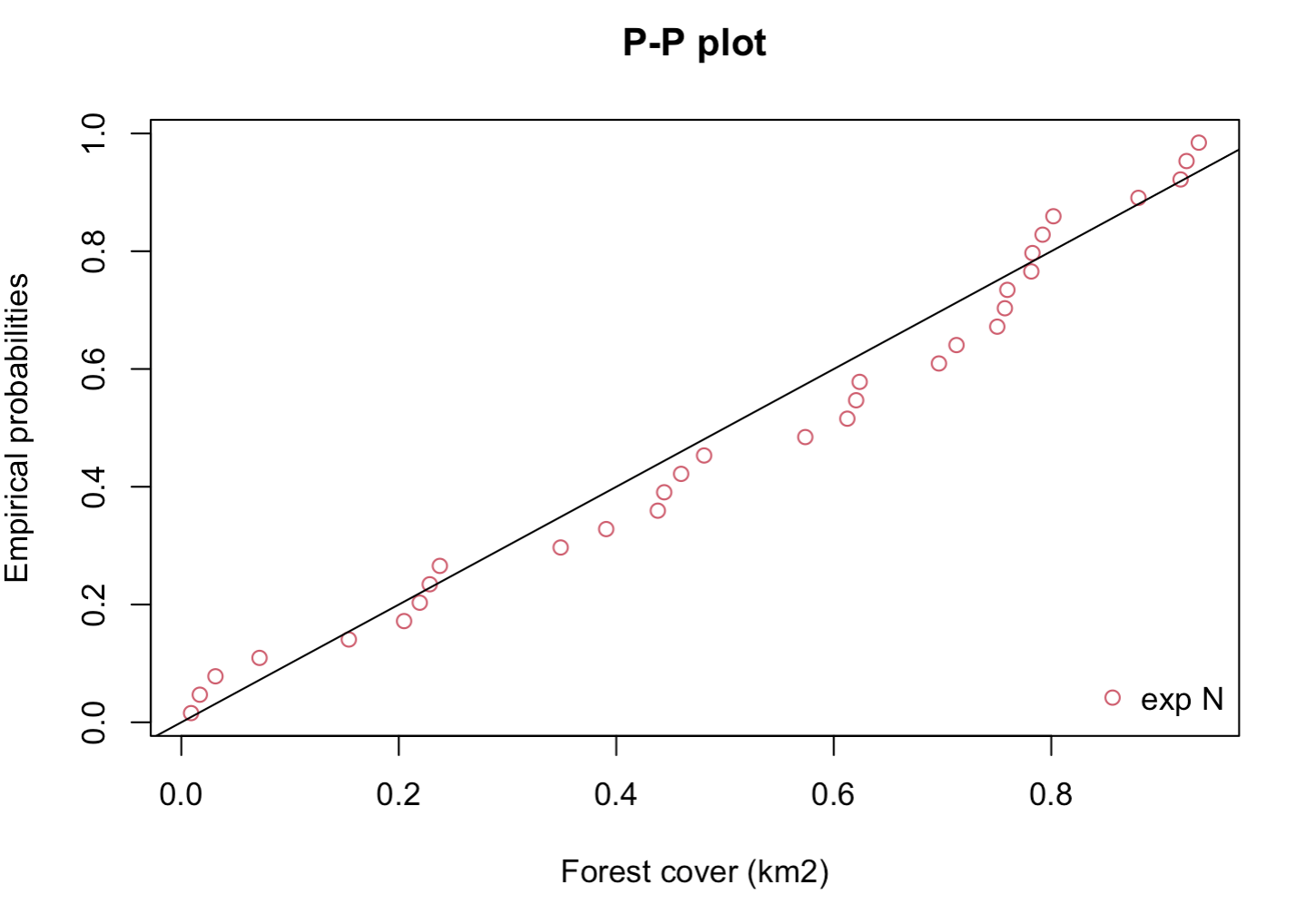}
    \caption{Maximum likelihood estimation of $p(X_{it}|X_{i t-1}, \theta)$ as the log-normal transition density \eqref{lnormtpd}  }
    \label{wap_lnorm}
\end{figure}

The discretization of the time grid has to be equivalent to the yearly frequency of the data, and thus we fix $\Delta t=1$. Estimation is obtained by bootstrapped maximum likelihood, and we obtain $\mu =  0.0482$ with a standard deviation of $ 0.003$. 
The data seems to be in good agreement with a log-normal transition density fit, as shown by Figure \ref{wap_lnorm}.
In order to obtain $\sigma^2$ we use the entirety of Benin's forest cover, and exclude the top 1\%-sized not to have our estimates driven by outliers. We obtain a ML estimate of the drift of 0.018 (std. dev. 0.047), statistically not significant, which is a reassuring result given that using the entirety of Benin data yields a drift estimate net of deforestation and thus not allowing to estimate $\mu$, as described in the main text. The estimate of $\sigma$ is not statistically different if estimated on WAP data (0.346 with std. dev. 0.045, the lower bound of a 95\% CI is 0.26), but is noisier as it's evaluated over a substantially smaller area.

\begin{figure}
\centering
\includegraphics[scale=2]{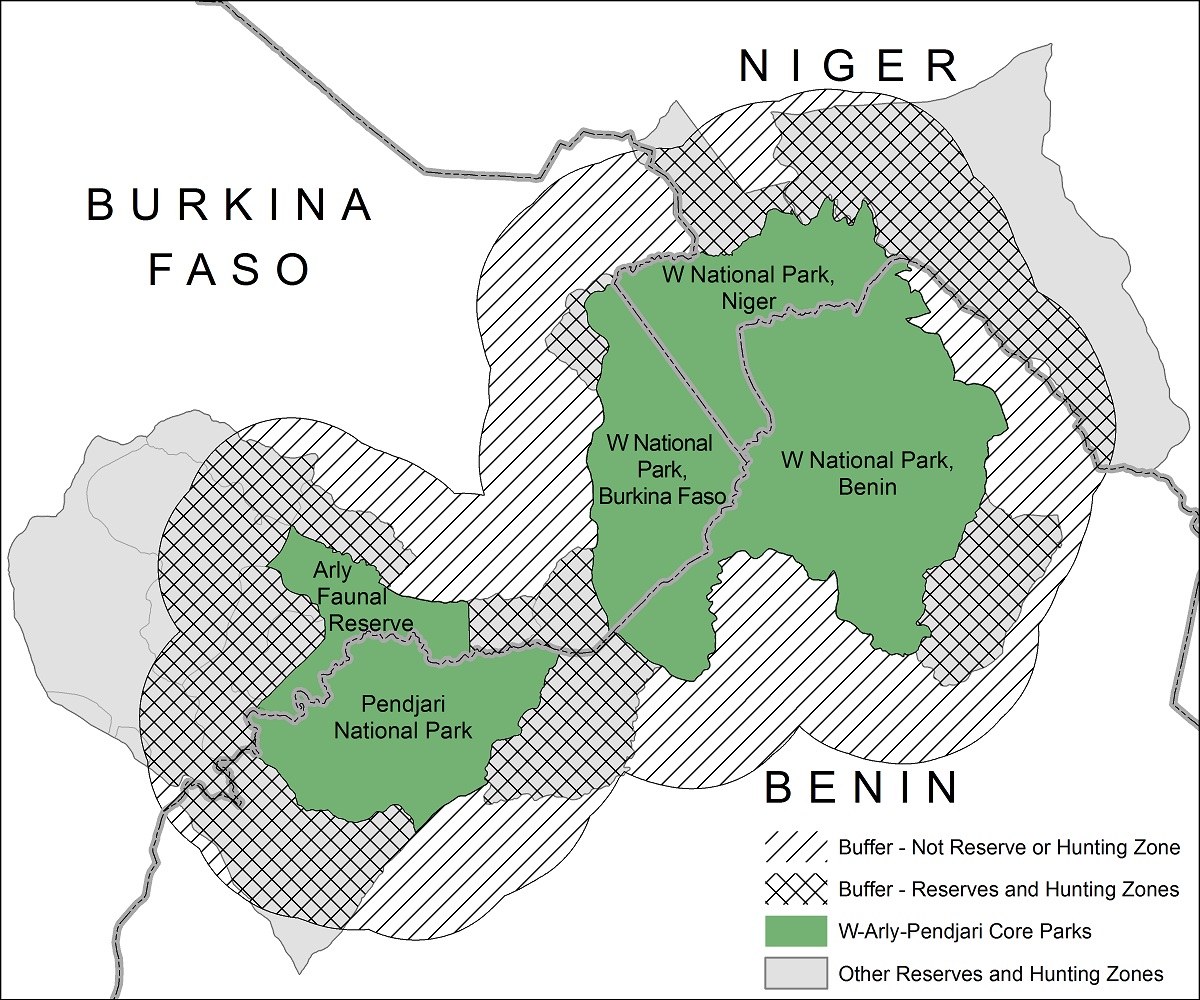}
\caption{W-Arly-Pendjari complex and surrounding buffer zones. \\ 
Source: USGS EROS, \url{https://eros.usgs.gov/westafrica/case-study/w-arly-pendjari-transboundary-biosphere-reserve}}
\label{fig:wap}
\end{figure}

\begin{figure}
    \centering
    \includegraphics[width=0.7\linewidth]{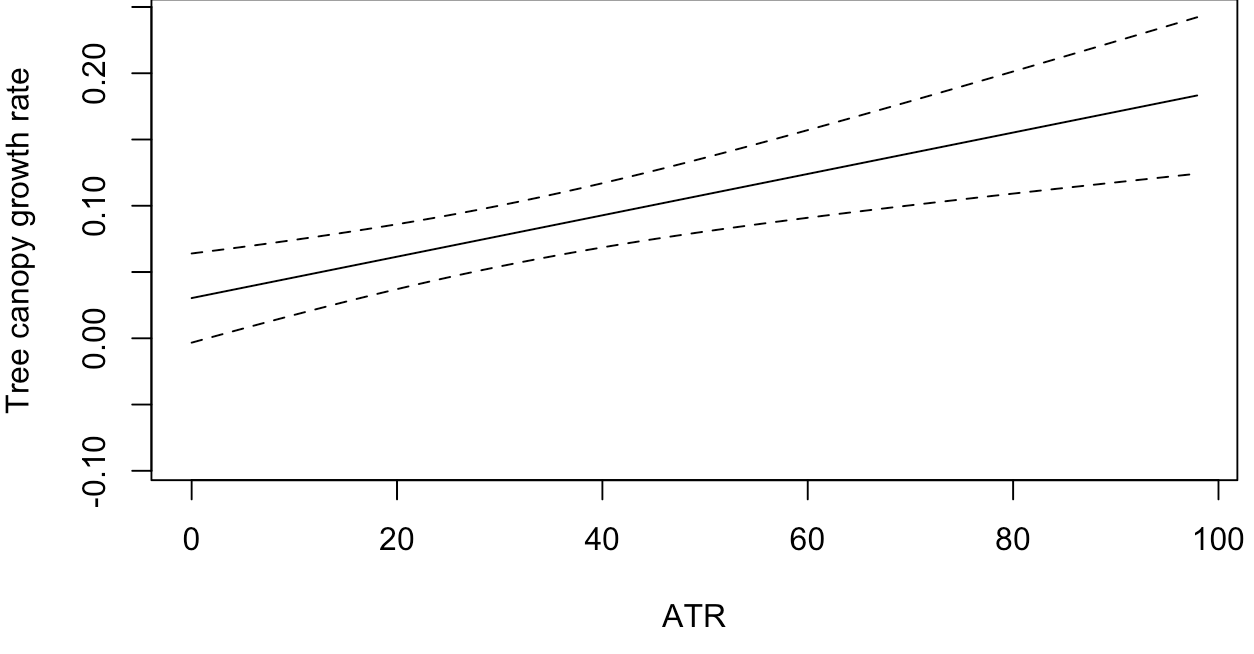}
    \caption{ Estimated nonparametric relationship between $\Delta X_i/X_i $ and $a_{i}$.}
    \label{fig:nonpar_tccrate_atr}
\end{figure}

Having identified $\mu$, we can now obtain from the data the optimal forest consumption policy $q^{MFE}$ implied by our framework.
We exploit variation for ATR adherence across the arrondissements and the three census years available, and we use the linearity in $X_{it}$ of the optimal policy \eqref{qmfe}. The form of each dynamic equation \eqref{gbm}, together with a time discretization $\Delta t$, imply that the equilibrium deforestation rate $q^{MFE}$ can be recovered as the following conditional moment condition:

\begin{equation}
    \mathbb{E}_{p^a}[ X_{i t+\Delta t} - X_{it} - \mathbb{E}_{p_0} [ (  \hat{\mu} - q^{MFE}(a_i) )  X_{it}] \Delta t | \mathcal{F}^a_t ]  = 0 .\label{momcond}
    \end{equation}
We note that using \eqref{momcond} to estimate the parameters is internally consistent, as it incorporates naturally the error structure of \eqref{gbm} in the moment condition, unlike simply matching $q^{MFE}$ with the data ``plus an error term'' (by nonlinear least squares, for example).
Furthermore, the term $\mathbb{E}_{p_0}[  \hat{\mu} - q^{MFE}(a)]$ is time-invariant. From the data across all arrondissements $i$ and given the annual nature of the TCC dataset such that $\Delta t = 1$, the moment condition \eqref{momcond} implies that $\frac{X_{it+1} - X_{it}}{X_{it}}$ can identify $q^{MFE}$ unconditionally, where $X_{it}$ is the TA for arrondissement $i$ in year $t$ and $a_i$ is arrondissement-level ATR adherence. For each of the three census years ($t=1992,2002,2013$), we use the following year's TA ($t+1=1993,2003,2014)$ and calculate the yearly growth rate of tree canopy cover area necessary for computing \eqref{momcond}. The moments of $g_2$ with respect to the distribution $p_0$ are recovered by numerical integration, using the Beta PDF at the estimated parameters $\hat{\alpha}_a, \hat{\beta}_a$. 
Figure \ref{mod_fig_1} shows how a choice of $g_1=1$ and $g_2 = a$ yields deforestation rates linear in the canopy stock. 
In order to validate our modeling choices for the point estimation, Figure \ref{fig:nonpar_tccrate_atr} shows the estimated non-parametric relationship between ATR adherence and tree canopy growth implied by our model $\Delta X_i = f(a_{i}) + \epsilon_{i}$ for the three stacked observed years $t = \{1992, 2002, 2013\}$. 
The linearity validates our choice of $k=1$, even though changing the functional shape of $g_2(a) = a^k$ via $k$ does not affect our results significantly as we have shown in the main text. We allow for different year intercepts (year dummies), due to the census sampling different individuals across each of the 10-year intervals that separates each wave, and estimate the function $g$ using straightforward local linear regression methods, a bootstrapped error band and a bandwidth that uses Kullback-Leibler cross-validation. We therefore  focus our estimation effort to the modeling choice of $g_1 = 1$.

We can then estimate $\gamma$, the CRRA parameter which drives the  mean-field interactions in our model, using a two-step generalized method of moments method over the condition induced by the optimal extraction \eqref{qmfe}. 
Estimation of the moment condition shown in Equation \eqref{momcond} of the paper is done using $ \mathcal{F}^a_t = \sigma\{X_{it},\Delta t \}$ as the $\sigma$-field generated by the $\Delta t$-sampled observations until $X_{it}$. We rewrite the moment condition as

$$
 \mathbb{E} \left [ \left . \frac{ X_{i t+\Delta t} - X_{it}}{X_{it}} - \hat{\mu} - q^{MFE} (a_i)  \right | \mathcal{F}_t^a \right ] = 0 ,\label{uncond_mom}
$$
which identifies the required parameter $\gamma$ over the sampled discretized processes in $\Delta t$. 
However, it is well-known that the conditional moments of a discretized diffusion process do not necessarily coincide with the conditional moments of its continuous-time equivalent. 
What we want to obtain is to simply condition on the $a_i$, which is the entire sample variation on ATR adherence. In other words, we want to switch from conditional to unconditional moments. This is not a trivial change: however, we exploit the fact that $q^{MFE}$ is stationary and only dependent on $a_i$.
We therefore exploit space (the measured ATR adherence across arrondissements and years) rather than time for identification.  
We stack each increment of $ \frac{ X_{i t+\Delta t} - X_{it}}{X_{it}}$, over the 3 survey years, write it as $\Delta X_{i}/X_i$ and can then write 
\begin{equation}
 \mathbb{E} \left [ \Delta X_{i }/X_i  - \hat{\mu} - q^{MFE} (a_i)  \right ] = 0. \label{uncond_mom}
\end{equation}
Equation \eqref{uncond_mom} is therefore an unconditional moment condition.
By exploiting the spatial variation over the sample periods of ATR, we bypass this difficulty and obtain consistent estimates of $\gamma$.
Estimation is then done over the condition \eqref{uncond_mom} using a standard two-step GMM procedure.  
The estimation is straightforward, uses the Brent method with an optimal weighting matrix for computing standard errors, which converges easily and yields an estimate of $\hat{\gamma} = 2.272  $ (std. dev. $ 0.856$) for a linear choice $g_2(a) = a$.
Increasing $k$ yields estimates of $\gamma$ that are slightly increasing in magnitude but not statistically different than $k=1$, and yields negligible changes in root mean squared error. 

\begin{figure}
    \centering
    \includegraphics[width=0.9\linewidth]{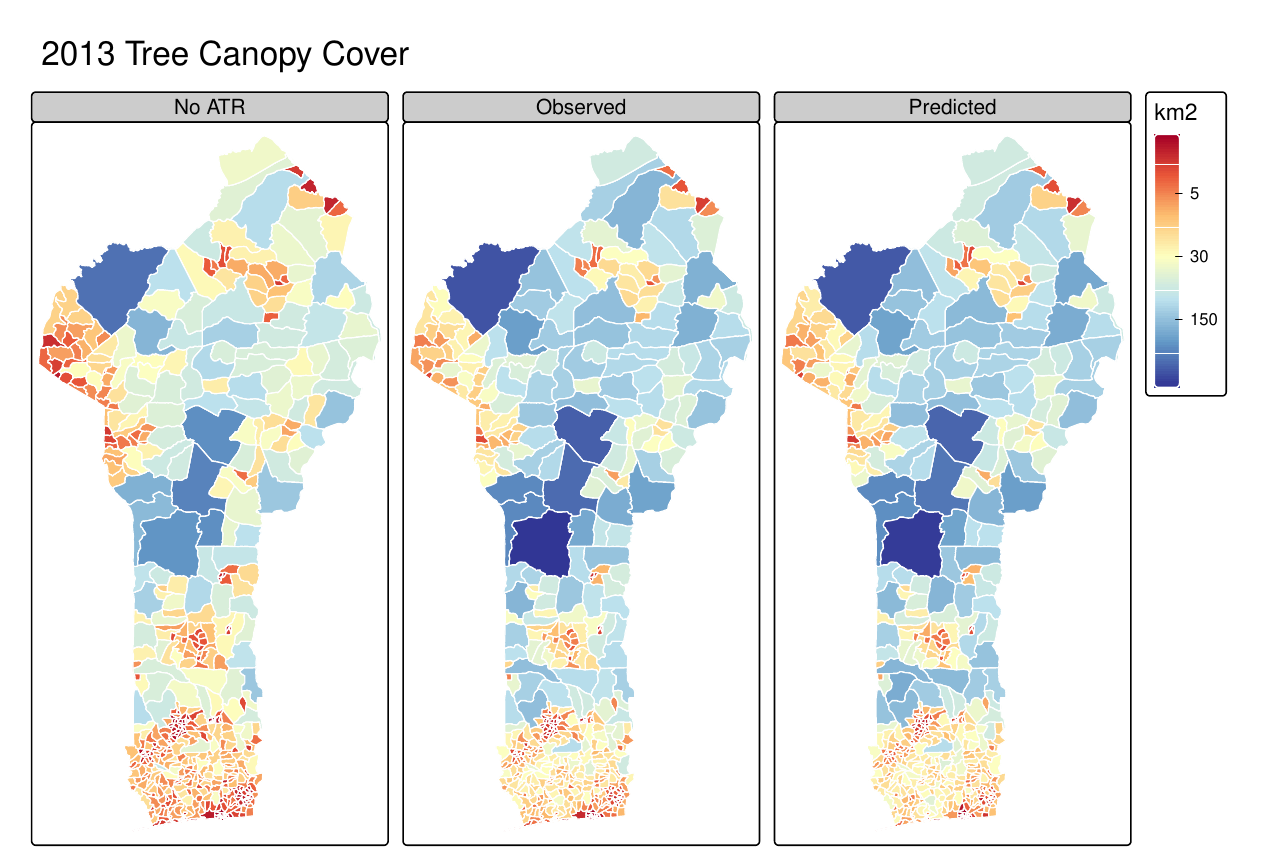}
    \caption{Counterfactual (no ATR), observed and predicted spatial tree canopy distributions in Benin for 2013.}
    \label{fig:spatdist}
\end{figure}

All the previous results are reported in Table \ref{table: struc_est}. Figure \ref{fig:spatdist} maps for Benin the observed, predicted, and zero-ATR counterfactual tree canopy cover spatial distribution over the different arrondissements.

\begin{table}\centering 
\begin{threeparttable}
  \caption{Model parameter estimates}
    \label{table: struc_est}
\begin{tabular}{>{\small}lcccc} 
\\[-1.8ex]\hline 
\hline \\[-1.8ex] 
\\[-1.8ex]&  Parameter & Estimate & \text{Std. Dev.} & \text{Method} \\ 
\hline \\[-1.8ex] 
&$\mu$ & 0.0482&  0.003 & MLE on TPD (only WAP data, 1982-2016) \\

  & $\sigma$ & 0.258 & 0.032 & MLE on TPD (1982-2016)  \\

&$\alpha_a$ &0.553 & 0.028 & MLE  \\
 
&$\beta_a$ &  2.251 & 0.151 & MLE  \\
&$\gamma$ & 2.272 & 0.856  & GMM ($g_2 = a$) \\
  \hline 
\hline \\[-1.8ex] 
\end{tabular} 
\begin{tablenotes}
\footnotesize
\item 
MLE: maximum likelihood estimation. TPD: transition probability density. WAP: W-Arly-Pendjari complex. GMM: generalized method of moments. \end{tablenotes}
\end{threeparttable}
\end{table}

We also obtain an essentially unchanged estimate of 2.35 (std. dev. 0.86) when estimating $\gamma$ ($k=1$) using the moment condition shown in Eq. \eqref{momcond} of the paper, together with the condition 
$$
\mathbb{E} \left [ \left (\frac{ X_{i t+\Delta t} -  X_{it} }{X_{it} } -  \mathbb{E}_{p_0} [ (  \hat{\mu} - q^{MFE}(a_i) ) \Delta t \right )^2  | \mathcal{F}^a_t  \right ] - \sigma^2 X_{it} \Delta t = 0,
$$
with the p-value of the J over-id test failing to reject with more than 99\% confidence. These two robustness results dampen potential concerns that the estimation of $\gamma$ might be over-sensitive to the choice of $g_2$ or of moment condition.

\end{document}